\documentclass[11pt]{article}
\usepackage{hyperref}
\usepackage{graphicx}
\usepackage{relsize,setspace}  
\usepackage{url}               

\usepackage[dvipsnames]{xcolor}
\usepackage{dcolumn}
\usepackage[ruled,vlined]{algorithm2e}
\usepackage{algorithmic}
\usepackage{xcolor}
\usepackage{graphicx,float}
\usepackage{booktabs}
\usepackage{amsmath,amsfonts,amssymb,mathtools}
\usepackage{makecell}
\usepackage{enumerate}
\usepackage{authblk}
\usepackage{bbm}
\usepackage{apacite}
\usepackage{multirow}
\usepackage{makecell}
\usepackage{cleveref}

\usepackage{fancyhdr}
\DeclareUnicodeCharacter{2212}{-}

\usepackage{subfig}
\usepackage{tikz}
\graphicspath{ {plot/} }
\usetikzlibrary{arrows.meta,shapes,shapes.arrows,
shapes.geometric,shapes.multipart,decorations.pathmorphing,
positioning,shapes.swigs,}

\newtheorem{theorem}{Theorem}
\newcommand\numberthis{\addtocounter{equation}{1}\tag{\theequation}}
\newcommand{\indep}{\perp \!\!\! \perp}

\providecommand{\keywords}[1]
{
  \small	
  \textbf{\text{Keywords: }} #1
}

\usepackage{appendix}
\usepackage{titletoc}
\newcommand\DoToC{%
  \startcontents
  \printcontents{}{2}{\textbf{Contents}\vskip3pt\hrule\vskip5pt}
  \vskip3pt\hrule\vskip5pt
}

\newtheorem{assumption}{Assumption}

\newtheorem{Lemma}{Lemma}[section]

\allowdisplaybreaks

\usepackage{float}
\DeclareGraphicsExtensions{.pdf, .jpg, .png}
\setkeys{Gin}{width=0.85\textwidth}

\title{The Multiplicative Instrumental Variable Model}
\author[1]{Jiewen Liu
}
\author[2]{Chan Park
}
\author[3]{Yonghoon Lee
}
\author[1]{Yunshu Zhang
}
\author[4]{\\ Mengxin Yu
}
\author[5]{James M. Robins 
}
\author[3,1]{Eric J. Tchetgen Tchetgen
\thanks{
Address for correspondence: Eric J. Tchetgen Tchetgen, 407 Academic Research Building, 265 South 37th Street, Philadelphia, PA 19104. Email: ett@wharton.upenn.edu
}}
\affil[1]{Department of Biostatistics, Perelman School of Medicine, University of Pennsylvania}
\affil[2]{Department of Statistics, University of Illinois Urbana-Champaign}
\affil[3]{Department of Statistics and Data Science, Wharton School, University of Pennsylvania}
\affil[4]{Department of Statistics and Data Science, Washington University in St. Louis}
\affil[5]{Department of Biostatistics, Harvard T.H. Chan School of Public Health}

\newcommand{\qq}{\quad \quad \quad \quad}
\newcommand{\mb}[1]{\mathbb{#1}}
\newcommand{\mc}[1]{\mathcal{#1}}

\newcommand{\ba}[1]{\left( #1 \right)}
\newcommand{\bb}[1]{\left\{ #1 \right\}}
\newcommand{\bc}[1]{\left[ #1 \right]}

\newcommand\norm[1]{{\left\lVert#1\right\rVert}_{P,2}}

\date{}
\begin{document}
\doublespacing
\vspace{-20mm}
\maketitle
\vspace{-8mm}
\begin{abstract}
The instrumental variable (IV) design is a common approach to address hidden confounding bias.   For validity, an IV must impact the outcome only through its association with the treatment. In addition, IV identification has required a homogeneity condition such as monotonicity or no unmeasured common effect modifier between the additive effect of the treatment on the outcome, and that of the IV on the treatment. In this work, we introduce \emph{the Multiplicative Instrumental Variable Model} (MIV), which encodes a condition of no multiplicative interaction between the instrument and an unmeasured confounder in the treatment propensity score model. Thus, the MIV provides a novel formalization of the core IV independence condition interpreted as independent mechanisms of action, by which the instrument and hidden confounders influence treatment uptake, respectively. As we formally establish, MIV provides nonparametric identification of the population average treatment effect on the treated (ATT) via a single-arm version of the classical Wald ratio IV estimand, for which we propose a novel class of estimators that are multiply robust and semiparametric efficient. Finally, we illustrate the methods in extended simulations and an application on the causal impact of a job training program on subsequent earnings.
\end{abstract}


\allowdisplaybreaks
\section{Introduction}
\label{intro}
Observational studies are widely used in epidemiology, clinical research, and the social sciences to estimate the causal effect of various treatments, exposures or interventions of interest. In practice, researchers typically make a concerted effort to control, to the extent feasible, as many common causes of the treatment and outcome of interest, also known as \emph{confounders}, in an effort to recover a valid estimate of the causal effect of interest. However, in observational studies, it is rarely possible to rule out confounding by hidden factors, leading to biased causal estimates. Even in well-designed randomized controlled trials, unmeasured confounding of treatment uptake can operate through selective non-adherence of patients to their randomly assigned treatment, potentially compromising study results \cite{dunn2005estimating}. Among existing methods to address unmeasured confounding, the instrumental variable (IV) method has emerged as a popular and widely used technique across disciplines, offering a framework that under certain conditions can account for unmeasured confounding and consistently estimate a causal effect \cite{wright1928tariff,greenland2000introduction,bollen2012instrumental,imbens2014instrumental}.

Generally speaking, a valid IV satisfies three assumptions: (i) relevance for the treatment, (ii) independence from unmeasured confounders, and (iii) no direct effect on the outcome other than through the treatment. See \Cref{sec:ns} for formal definitions of these assumptions. However, although sufficient for partial identification or for testing the causal null hypothesis, these three conditions alone are well-known to be insufficient for point identification, therefore requiring a ``fourth'' IV assumption. The type of identifiable causal effect depends on the specific form of this fourth IV condition. For instance, when both IV and treatment are binary, the monotonicity condition identifies the so-called local average treatment effect (LATE) also known as the complier average treatment effect \cite{permutt1989simultaneous,baker1994paired, imbens1994identification}. Alternatively, a fourth condition that there is no unmeasured common effect modifier between the additive effect of the treatment on the outcome and that of the IV on the treatment is known to identify the population average treatment effect (ATE) \cite{wang2018bounded,cui2021semiparametric, qiu2021optimal}. In contrast, a fourth condition that the additive treatment effect on the treated is independent of the IV conditional on covariates is also known to identify the conditional average treatment effect for the treated (ATT) given covariates \cite{robins1994correcting}, and therefore the marginal ATT is also identified. 

Finally, \citeA{tchetgen2024nudge} recently introduced sufficient conditions for identification of the so-called nudge average treatment effect (NATE), defined as the average treatment effect for the subgroup of units for whom the treatment is manipulable by the instrument, a subgroup of \emph{nudge-able} individuals which may consist of both defiers and compliers. Their fourth condition identifying the NATE states that among the nudge-able, heterogeneity in the causal effect induced by a hidden confounder is uncorrelated with corresponding heterogeneity in the share of compliers.  Notably, in the absence of measured covariates, the ATE, ATT, LATE and NATE are uniquely identified by the Wald ratio under their respective fourth condition, while in the presence of measured covariates, each estimand is recovered as a certain weighted average of the conditional Wald ratio given covariates, with weights corresponding to the covariate distribution in the subpopulation to which the causal effect applies, i.e. in the overall population, or the treated, or the compliers, or the nudge-able, respectively.  Additional alternative fourth identifying conditions for the binary IV model can be found in \citeA{tchetgen2013alternative} and \citeA{liu2020identification}. 

In this paper, we introduce a novel IV condition, which together with conditions (i)-(iii), point identifies the treatment-free counterfactual mean for the treated, and therefore the ATT, via a \emph{single-arm Wald ratio}, defined as a standard Wald ratio for the observed outcome restricted to the untreated group. In contrast to the constant treatment effect condition of \citeA{robins1994correcting}, our approach allows the ATT to be an unrestricted function of the instrument, and instead assumes a multiplicative model for the treatment uptake mechanism with respect to the IV and unmeasured confounders. Specifically, our assumption rules out any multiplicative interaction between the IV and an unmeasured confounder in the treatment model, which we refer to as a \emph{multiplicative IV model} (MIV). Intuitively, the MIV essentially encodes the core IV independence condition (ii) interpreted as an assumption of independent mechanisms of action, by which the instrument and hidden confounders influence treatment uptake, respectively. This formalization will later be made concrete through the lens of a so-called Generalized Latent Index Model (GLIM), a generalization of a certain latent index model (LIM) originally due to \citeA{heckman1979sample}. Under Heckman's LIM, one models a unit's treatment selection with a latent index crossing a random threshold, where the latent index represents a form of expected net utility of selecting into treatment. Interestingly, \citeA{vytlacil2002independence} established a fundamental equivalence between Heckman's LIM and the LATE model of \citeA{imbens1994identification}, therefore offering an alternative formulation of the LATE identifying model. Later, we use the GLIM as a platform to motivate the MIV, and further, to formally relate treatment selection IV models for the most prominent causal effects in the literature, mainly the population ATE, the LATE, the ATT and the NATE. Beyond establishing nonparametric identification of the ATT under the MIV model, we characterize the semiparametric efficiency bound for estimating the single-arm Wald ratio functional identifying the ATT under a nonparametric model for the observed data distribution, and construct a multiply robust semiparametric estimator which, combined with flexible machine learning estimators of nuisance functions, is shown to attain the efficiency bound provided the nuisance functions can be estimated at sufficiently fast convergence rates. 

\section{Framework and Identification}
\label{sec:iden}
\subsection{Causal Structure and Notation}
\label{sec:ns}
Suppose that one has observed an independent and identically distributed (i.i.d.) sample of size $N$ on data $O=(Y,A,Z,X)$; where $Y$ denotes an outcome of interest; $(A,Z) \in \{0,1\}^2$ denotes a binary treatment of interest and an instrumental variable for the effect of $A$ on $Y$; $X \in \mathcal{X} \subseteq  \mathbb{R}^d$ denotes a $d$-dimensional vector of measured covariates. Suppose further that for each unit, there exists an unmeasured confounder, $U$, of the causal effect of $A$ on $Y$. Let $Y^{a,z}$ denote the potential outcome that one would have observed had, possibly contrary to fact, the treatment and instrument been set to $A = a$ and $Z=z$. A similar definition applies to $Y^a \in \mathbb{R}$ (had only the treatment been set to $A=a$) and $A^z \in \{0,1\}$. Throughout, we make the assumption of consistency, that is, $Y=Y^{A}=Y^{A,Z}$ and $A=A^Z$ almost surely. We also assume that $pr(A=a|Z=a,U,X)$ and $pr(Z=z|X)$, $\forall (a,z) \in \{0,1\}^2$, are uniformly bounded away from zero almost surely.  Let \(\mathbb{I}(C)\) denote the indicator of event $C$, $W$\(\indep\)$H|R$ denote independence between random variables $W$ and $H$ conditional on $R$.

We consider the following IV assumptions:
\begin{assumption}[Relevance]\label{as:iv1} $A \not\indep Z|X$;\end{assumption}
\begin{assumption}[Independence]\label{as:iv2} $U \indep Z | X$; \end{assumption}
\begin{assumption}[Weak Latent Ignorability \& Exclusion Restriction] \label{as:iv3}$Y^{a=0} \indep (A,Z)|X,U$. \end{assumption}

Assumptions \ref{as:iv1} and \ref{as:iv2} formalize IV relevance and independence conditions, while assumption \ref{as:iv3} formalizes a weak form of both the exclusion restriction condition that the instrument only affects the outcome through the treatment, and of latent ignorability, both of which strictly only involve the treatment-free potential outcome, i.e. $Y^{a=0}=Y^{a=0,z=1}=Y^{a=0,z=0}$, and not the potential outcome under treatment, i.e. $Y^{a=1,z}$ is unrestricted and may neither satisfy the exclusion restriction nor unconfoundedness even after conditioning on $(X,U)$. This asymmetry between $Y^{0}$ and $Y^{1}$ is substantively important. Assumption \ref{as:iv3} only requires that, conditional on $(X,U)$, the joint treatment--instrument assignment mechanism carries no additional information about the outcome that would be observed under no treatment. In some applications, this may be more credible than imposing the analogous restriction on $Y^{1}$. The reason is that once treatment is actually received, the realized treated outcome may depend not only on unmeasured $U$ (e.g., latent health or motivation) but also on treatment pathway features that are themselves affected by the instrument, such as the timing of uptake, the provider or facility accessed, treatment intensity, adherence, or the ability to navigate the system after encouragement. By contrast, for individuals under the no-treatment state, such post-treatment pathway variables are absent by design, making it more plausible that, after conditioning on $(X,U)$, neither $A$ nor $Z$ has residual relevance for $Y^{0}$. We return in Section \ref{sec:compare} to the distinction between this weak latent ignorability condition and the stronger condition that would additionally require the same restriction for $Y^{1}$.

\begin{figure}[b]
\centering  
\includegraphics[width=0.88\textwidth]{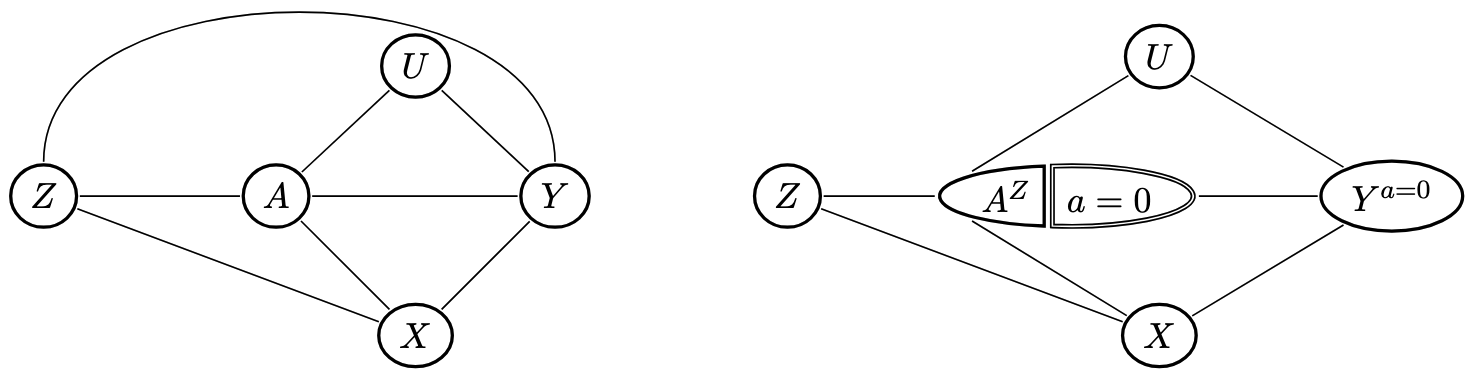}
\caption{A Graphical Representation of the IV model under Assumptions \ref{as:iv1} -- \ref{as:iv3}. The left and right panels present a DAG and a SWIG, respectively.}
\label{fig:D1}
\end{figure}

Figure \ref{fig:D1} presents causal graph representations of the IV model under the above assumptions through (left) a directed acyclic graph (DAG) and (right) a single-world intervention graph (SWIG). The presence of an arrow from \( Z \) to \( Y^{a=1} \) in the DAG illustrates the potential presence of a direct effect of \( Z \) on  \( Y \) even after conditioning on $(U,X,A)$ which assumption \ref{as:iv3} accommodates. The SWIG provides a graphical representation of assumption \ref{as:iv3} \cite{richardson2013single}, corresponding to a world in which one sets \( A = 0 \) via hypothetical intervention. 

The following notations are used throughout the paper:
\begin{align*}
    & \beta^a(X) := E(Y^a \mid A=1, X), 
    && \beta^a := E(Y^a \mid A=1),
    \\
    & \beta(X) := E(Y^1 - Y^0 \mid A=1, X) = \beta^1(X) - \beta^0(X), 
    && \beta := E(Y^1 - Y^0 \mid A=1) = \beta^1 - \beta^0,
    \\
    &p_z(X) := pr(A=1 \mid Z=z, X),
    &&e_z(X) := E\{Y(1-A)|Z=z,X\}, \\
    &\pi_z(X) := f(Z=z|X), 
    && \rho(X) := pr(A=1|X),
    \\
    &\Omega(X) := 1/\{p_1(X) - p_0(X)\},
    &&\delta(X) := \bb{e_1(X) - e_0(X)}/\bb{p_1(X)-p_0(X)},
    \\
    &\psi^1(X) := E\ba{Y|A=1,X}, 
    && \psi^0(X) := -\delta(X), \\
    &\psi(X) := \psi^1(X) - \psi^0(X),
    &&\psi := E\ba{Y|A=1} + E\bb{A\delta(X)/pr(A=1)}.
\end{align*}

The function $\beta^a(X)$ denotes the mean of $Y^a$ conditional on $\{A=1,X\}$, while $\beta(X)$ corresponds to the conditional ATT, defined as the difference between $\beta^1(X)$ and $\beta^0(X)$. Meanwhile, $\beta^a = \int_{\mc{X}} \beta^a(X) dF(X|A=1)$ and $\beta = \int_{\mc{X}} \beta(X) dF(X|A=1)$ denote their marginal counterparts, respectively. $\delta(X)$ is a single-arm version of the so-called conditional Wald ratio, which in its standard form is given by 
$$\delta^*(X) := \frac{E(Y|Z=1,X)-E(Y|Z=0,X)}{pr(A=1|Z=1,X) - pr(A=1|Z=0,X)};$$
therefore, the modified version simply swaps in the outcome for the untreated only $(1-A)Y$ for $Y$ in the reduced form appearing in the numerator, justifying the single-arm Wald ratio designation. This reflects the identification challenge for the ATT which aims to recover the average potential outcome for the treated arm, had the treatment been withheld. Given the definition of $\delta(X)$, the quantity $\psi(X)$ denotes the difference between $\psi^1(X)$, the empirical mean of the outcome for the treated arm, and $\psi^0(X)$, which corresponds to $-\delta(X)$. The marginal  $\psi = \int_{\mc{X}} \psi(X) dF(X|A=1)$ is obtained by averaging $\psi(X)$ over the distribution of covariates for the treated arm. Moreover, for a sequence of random variables \(\{V_N\}\), we define:
\(\mathbb{P}(V) = \sum_{i=1}^N V_i/N\), the empirical mean of the observed data;  \(|\mathcal{I}_k|\), the cardinality of the set \(\mathcal{I}_k\), and
\(\mathbb{P}_{\mathcal{I}_k}(V) = \sum_{i\in \mathcal{I}_k} V_i /|\mathcal{I}_k|\), the empirical mean over the subset \(\mathcal{I}_k\).  
Furthermore, we use the standard notation \(V_N = O_P(r_N)\) to denote that \(V_N / r_N\) is stochastically bounded, \(V_N = o_P(r_N)\) to denote that \(V_N / r_N\) converges to 0 as \(N \to \infty\), and $\longrightarrow_D$ to denote the convergence in distribution.

\subsection{The Multiplicative IV Model}
As discussed in the introduction, the ATT cannot be identified solely on the basis of assumptions \ref{as:iv1} -- \ref{as:iv3}. A fourth IV condition will typically be required. We introduce the following new condition, which we refer to as the \emph{Multiplicative Instrumental Variable (MIV) Model}:  
\begin{assumption}[MIV Model]\label{as:miv}
The latent propensity score for treatment satisfies the following multiplicative model: 
\begin{align*} pr(A=1|Z,X,U)= exp\{\alpha_1(Z,X)+\alpha_2(U,X)\},
\end{align*}
\textit{where $\alpha_1(\cdot,\cdot)$ and $\alpha_2(\cdot,\cdot)$ are arbitrary functions with the sole restriction $\alpha_1(0,X)=0$ almost surely and the natural constraint $pr(A=1|Z,X,U) \in (0,1)$ almost surely.}
\end{assumption}

Assumption \ref{as:miv} effectively rules out the presence of a $U-Z$ interaction on the multiplicative scale for the treatment latent propensity score model conditional on $X$. Therefore, the assumption places the onus on the analyst's ability to measure and condition on a sufficiently rich set of covariates, so that the multiplicative association between the instrument and the treatment remains stable across values of the hidden confounder. The assumption is not empirically testable. We now state a key nonparametric identification result under Assumption \ref{as:miv}.
\begin{theorem}\label{thm1}
    Suppose that assumptions \ref{as:iv1} -- \ref{as:miv} hold for the IV model. Then, $\beta^0(X)$ is identified from the observed data by the following expression:
    \begin{align*}
        &\beta^0(X) = \psi^0(X).
    \end{align*}
Additionally, the counterfactual mean parameter $\beta^0$ is identified by $\psi^0$.    
\end{theorem}

By consistency, $\beta^1(X)$ and $\beta^1$ are  identified by the empirical mean $\psi^1(X)$ and $\psi^1$, respectively. The conditional ATT, $\beta(X)$, and the marginal ATT, $\beta$, are therefore identified by $\psi(X)$ and $\psi$, respectively.

This result appears to be new to the IV literature, although a related result was obtained by \citeA{hernan2006instruments} under stronger conditions; we refer the reader to \Cref{sec:compare} for a detailed discussion of the relationship between our result and theirs. The result can be viewed as providing a multiplicative analog to the result of \citeA{wang2018bounded}, who established the nonparametric identification of the ATE under an additive latent propensity score model $pr(A=1|Z,X,U)=\alpha_1(Z,X)+\alpha_2(U,X)$. It is also worth relating the result to that of \citeA{vytlacil2002independence}, who established an equivalence between the monotone IV framework of \citeA{imbens1994identification} and a certain latent index model. In fact, \Cref{tab:as_lim} provides an instructive comparison of IV models identifying the most prominent causal effects in the literature, through the lens of a so-called generalized latent index model (GLIM), where $X$ is suppressed for ease of exposition. 

\begin{table}[!htp]
\small
\centering
\begin{tabular}{|c|c|c|c|}
\hline
\makecell{Under Assumptions \ref{as:iv1} -- \ref{as:iv3} and\\ the Additional Assumption} & 
\makecell{Generalized Latent Index Model \\$z \in \bb{0,1}$}  & \boldmath{$-\delta$} Identifies  \\ 
\hline
Monotonicity
&
\makecell{
Equivalent to
\\
$A^z = \mb{I}\{g(z) + U \geq \epsilon_z \},$
\\
$g(z) = pr(A=1|Z=z),$
\\
$U \sim Uniform(0,1),$ 
\\
degenerate $\epsilon_z=1$.\\
}
&
\makecell{
$E\ba{Y^0|A^1=1,A^0=0}$
\\ 
\cite{imbens1994identification} \\
\cite{vytlacil2002independence}
}
\\ 
\hline
Additive IV Model
&
\makecell{
Implied by
\\
$A^z = \mb{I}\{g(z)+U \geq \epsilon_z\},$
\\
for some $g(z)$ s.t. $g(z)+U \in (0,1)$,
\\
$\epsilon_z \sim Uniform(0,1)$.
}
&
\makecell{
$E\ba{Y^0}$
\\
\cite{wang2018bounded}
}
\\ 
\hline
Logistic IV Model
&
\makecell{
Implied by
\\
$A^z = \mb{I} \bb{g(z) + U \geq \epsilon_z}$,
\\
$g(z) + U \in [-\infty, \infty]$,
\\
$\epsilon_z \sim \textit{Logistic}(0, 1), \; \epsilon_0 \indep \epsilon_1$.
}
&
\makecell{
$E\ba{Y^0 \mid \mc{N} = 1}$
\\
\cite{tchetgen2024nudge}
}
\\ 
\hline
Multiplicative IV Model 
&
\makecell{
Implied by
\\
$A^z = \mb{I}\{g(z) \times  U \geq \epsilon_z\},$
\\
for some $g(z)$ s.t. $g(z) \times U \in (0,1)$,
\\
$\epsilon_z \sim Uniform(0,1)$.
}
&
\makecell{
$E\ba{Y^0|A=1}$
\\
This Work
}
\\ 
\hline
\end{tabular}
\caption{Causal Effects Identified by a Single-arm Wald Ratio under Different GLIM Setups. 
 $\mc{N}$ denotes an indicator for whether a person's treatment is nudge-able, i.e., $\mc{N} = \mb{I} (A^1 \neq A^0) = \mb{I}(A^1=1, A^0=0)$ $+ \mb{I}(A^1=0, A^0=1)$.
}
\label{tab:as_lim}
\end{table}

The table reveals that the four parameters, the LATE, ATE, NATE and ATT, can be obtained as different special cases of a treatment selection IV model, which we refer to as a generalized latent index model (GLIM). The model specifies the treatment as $ A^z=\mb{I}\{h(z,U) \geq \epsilon_z \}$, $z \in \bb{0,1}$, where $h(z,u)$ is an index function from a specific class, defined on the instrument and an unmeasured confounder, representing an individual’s net utility, and $(\epsilon_0,\epsilon_1)$ is a vector of random thresholds reflecting residual independent causes of the potential treatments $A^0$ and $A^1$ respectively. Treatment uptake occurs when a person’s net utility (\emph{baseline readiness / self-assessed ability / motivation}) exceeds the corresponding threshold (\emph{cost/constraint}). Notably, the GLIM allows the distribution of the treatment-free potential outcome $Y^0$, conditional on $U$ to remain unrestricted. However, the monotonicity condition restricts two potential sources of heterogeneity of the GLIM, mainly by additive separability of the index function, $U$ and $z$ do not interact, i.e.  $h(z,U)=g(z) + U $, and by $\epsilon_0=\epsilon_1=1$ almost surely, the joint distribution of $A^0$ and $A^1$ is degenerate given $U$ and $Z$; i.e. other than $Z$, any other cause of the treatment, is captured by $U$ and thus necessarily a hidden confounder. In contrast, while the additive IV model likewise admits an additively separable index function, the model does not further restrict the dependence between the potential treatments $A^0$ and $A^1$ conditional on $U$ by allowing for independent residual causes of treatment as heterogeneous thresholds $\epsilon_0\neq\epsilon_1$.  Similarly, \citeA{tchetgen2024nudge} recently showed that the NATE is identified under an analogous GLIM with an additively separable utility function; however, with independent logistic thresholds $(\epsilon_0,\epsilon_1)$. 

Finally, the GLIM motivating the MIV also accommodates heterogeneous thresholds through a multiplicative form on the latent index, namely \(h(z,U)=g(z)U\). In particular, after taking logs, it admits an additively separable structure but under a non-uniform threshold
$$\log g(z)+\log U \ge \log \epsilon_z.$$
After reparameterization, this can be written as
$$h(U)-c(z)\ge e_z,$$
where \(h(U)\) summarizes latent motivation or readiness, \(c(z)\) is an instrument-dependent cost, and \(e_z\) is an idiosyncratic shock or cost, which follows an \emph{exponential(1)} distribution. This representation is helpful for clarifying the substantive content of multiplicative separability: the instrument need not equally shift everyone’s latent utility by the same additive amount irrespective of latent propensity for treatment uptake. Instead, it operates through a threshold mechanism in which changes induced by \(Z\) interact with individuals’ latent baseline propensity \(U\). On the log additive scale, the model allows the instrument’s effect on treatment uptake to be heterogeneous and to scale with latent propensity, which best captures settings in which it may be difficult for a given instrument to overcome a person's inherent lack of preference for the treatment in view. Importantly, while the LIM is exactly equivalent to monotonicity under the IV model of \citeA{imbens1994identification}, the additive, logistic and multiplicative IV models are implied by but not sufficient for the GLIMs described in the table.

\section{A Semiparametric Efficient Estimator}
\label{sec:semi}
\subsection{Semiparametric Efficiency Bound}
In this section, we derive the efficient influence function (EIF) of the identifying functional $\psi$, under a nonparametric model denoted $\mc{M}$ 
that places no restriction on the observed data distribution. 
The following theorem states the main result. 
\begin{theorem}
\label{thm2}
\textbf{(1)} The EIF of $\psi$ in model $\mc{M}$ is:
    \begin{align*}
        &EIF(O;\psi) = \frac{1}{pr(A=1)} \left[ A\left\{ Y+\delta(X)- \psi \right\} +  
         \theta(O) \right], \\
        &where \ \theta(O) := \rho(X)\frac{2Z-1}{\pi_Z(X)} \Omega(X)\left[Y(1-A) - e_Z(X) - \{A - p_Z(X)\} \delta(X)\right].
    \end{align*}
    Hence, the corresponding semiparametric efficiency bound for $\psi$ in model $\mc{M}$ is $var\{EIF(O;\psi)\}$. \textbf{(2)} The EIF is a multiply robust moment equation for $\psi$ in the sense that it is an unbiased moment equation, i.e. $E\{EIF(O;\psi)\} =0$, under the union of the following three models for $z\in\{0,1\}$: 
\begin{align*}
    &\mc{M}_1: \textit{models for } p_0(X), e_0(X) \textit{ and } \delta(X)  \textit{ are evaluated at their true value;} \\
    &\mc{M}_2:\textit{models for }  p_z(X) \textit{ and }  \pi_z(X) \textit{ are evaluated at their true value;} \\ 
    &\mc{M}_3: \textit{models for } \delta(X) \textit{ and }  \pi_z(X) \textit{ are evaluated at their true value.}     
\end{align*}    
\end{theorem}
The first component of the EIF, i.e., $A \{ Y+\delta(X)- \psi \}/pr(A=1)$, can be interpreted as the moment equation for the marginal $\psi$ were, contrary to fact, the nuisance function $\delta(X)$ known to the analyst, essentially re-expressing the identifying formula in Theorem \ref{thm1}. The second component $\theta(O)/pr(A=1)$  can be seen as a correction term, which accounts for having to estimate the unknown nuisance functions, and delivers the multiple robustness property described in part (2) of the theorem, and, as we show below, paves the way for constructing an estimator of $\psi$ with bias of at most second order.  An immediate implication of multiple robustness entails, letting the superscript $*$ indicate a misspecified model, that for $z\in\{0,1\}$:
\begin{align*}
    &\mc{M}_1: E\{EIF(O;\psi,e_1^*(X),e_0(X),p_1^*(X),p_0(X),\pi_z^*(X),\delta(X))\} =0; \\
    &\mc{M}_2: E\{EIF(O;\psi,e_1^*(X),e_0^*(X),p_1(X),p_0(X),\pi_z^*(X),\delta^*(X))\} =0; \\
    &\mc{M}_3: E\{EIF(O;\psi,e_1^*(X),e_0^*(X),p_1^*(X),p_0^*(X),\pi_z(X),\delta(X))\} =0.
\end{align*} 
\subsection[Estimator]{Semiparametric Efficient Debiased Estimator}
\label{sec:proposed_estimator}
The proposed estimator is based on the EIF and employs the Double/Debiased Machine Learning (DDML) approach to incorporate machine learning estimators of nuisance parameters \cite{chernozhukov2018double}. Specifically, DDML involves cross-fitting which entails partitioning the sample \(\{O_i: i=1,\ldots,N\}\) into non-overlapping folds \(\{\mc{I}_1, \ldots, \mc{I}_K\}\); then estimating nuisance functions over the estimation fold \(\mc{I}_k^c\) (the complement set containing all samples except those in \(\mc{I}_k\)); and estimating \(\psi\) using samples from the evaluation fold \(\mc{I}_k\); and finally averaging the $K$ estimates obtained from $K$ sample splits. Each of the $K$ estimates is obtained based on the following three-step procedure. The first two steps involve estimating nuisance functions over $\mc{I}_k^c$, and the third step targets the marginal $\psi$ over $\mc{I}_k$. 
\\[0.2cm]
\noindent (\textit{Step 1: Estimation of $p_z(X)$, $\pi_z(X)$ and $e_z(X)$}): 
\\[0.1cm]
\indent For a given evaluation fold $\mc{I}_k^{c}$, we first partition its samples into two nonoverlapping folds of the same size, say $\mc{I}_k^{c1}$ and $\mc{I}_k^{c2}$, and obtain $\hat{p}_z^{(-k,c1)}(X)$, $\hat{\pi}_z^{(-k,c1)}(X)$ and $\hat{e}_z^{(-k,c1)}(X)$ 
from the first fold $\mc{I}_k^{c1}$. One can use an ensemble method like the Superlearner, which combines several candidate machine learning estimators up to the user's choice and outputs a certain convex combination of fitted values of user-specified learners. The approach has formally been established to have the key oracle property that the resulting estimator is guaranteed to be asymptotically equivalent to the best candidate estimator \cite{van2007super} with respect to a given risk (e.g. mean squared error (MSE)).
\\[0.2cm]
(\textit{Step 2: Estimation of $\Omega(X)$ and $\delta(X)$}):
\\[0.1cm]
\indent 
Estimating $\Omega(X)$ and $\delta(X)$ on the second fold $\mc{I}_k^{c2}$ requires special care,
as the true functionals of $\Omega(X)$ and $\delta(X)$ are not directly accessible for standard regression analysis. To facilitate the subsequent construction of a multiply robust estimator and ensure both convergence rate and numerical stability, we adopt the Forster-Warmuth (FW) counterfactual regression approach developed by \citeA{yang2023forster}. 
\\[0.1cm]
\indent The implementation of the FW counterfactual regression approach involves two parts:
\begin{enumerate}
    \item Derive the uncentered EIF denoted by $f(O)$, also known as oracle pseudo-outcome,  of the marginal mean of the counterfactual regression of interest. Plug in the estimated nuisance functions from Step 1 into the oracle pseudo-outcome to obtain $\hat{f}(O)$. Use $\hat{f}(O)$ as \emph{feasible} pseudo-outcome.
    \item Fit a regression model using the FW-Learner using a different subset of samples. Regress the pseudo-outcome $\hat{f}(O)$ on basis functions of the covariates $X$.
\end{enumerate}

\indent For part 1, the uncentered EIF $f_{\delta(X)}(O)$ of $E\bb{\delta(X)|A=1}$ is given in \Cref{thm2} up to a constant shift
\begin{align*}
    f_{\delta(X)}(O) = \frac{1}{pr(A=1)}
    \bc{
\begin{aligned}
    &\quad \quad \quad A \frac{e_1(X)-e_0(X)}{p_1(X) - p_0(X)}  +  \\
    &\rho(X)\frac{2Z-1}{\pi_Z(X)} \frac{Y(1-A) - e_Z(X) - \{A - p_Z(X)\} \delta(X) }{p_1(X) - p_0(X)} 
\end{aligned}
} ,
\end{align*}
and the uncentered EIF $f_{\Omega(X)}(O)$ of $E\bb{\Omega(X)|A=1}$ takes the form
$$ f_{\Omega(X)}(O) = \frac{1}{pr(A=1)} 
\bc{\frac{A}{{p_1(X) - p_0(X)}} 
+ \rho(X) \frac{2Z-1}{\pi_Z(X)} \frac{{A - p_Z(X)}}{\bb{p_1(X)-p_0(X)}^2}
}.$$
Of note, if nuisance functions $p_z(X)$, $\pi_z(X)$ and $e_z(X)$ were known, $f_{\delta(X)}(O)$ and $f_{\Omega(X)}(O)$ would be conditionally unbiased for the counterfactual regression, justifying referring to them as pseudo-outcome, i.e, $E\bb{f_{\delta(X)}(O)|X} = E\bb{\delta(X)|A=1,X}$ and $E\bb{f_{\Omega(X)}(O)|X} = E\bb{\Omega(X)|A=1,X}$.

For part 2, we now describe the FW-learner. Consider a complete system of basis functions $\{\phi_j(\cdot)\}_{j=1}^\infty$, and define $\bar{\phi}_J(X) = \bb{ \phi_0(X) := 1,\, \phi_1(X),\, \ldots,\, \phi_J(X)}$
as the vector of the first \( J+1 \) basis functions. Examples of such basis functions with good approximation properties include wavelets, Fourier series, splines and polynomial series. Let \( m^*_{f(O)}(X) = E\bb{f(O) \mid X} \) denote the conditional expectation of $f(O)$, and  \( \hat{m}_{J-f(O)}(X)\) denote the estimator of the conditional expectation of $f(O)$ obtained using basis functions of order $J$. Suppressing the superscript $(-k,c1)$ and using the estimators from Step 1, we obtain $\hat{f}_{\delta(X)}(O)$ and $\hat{f}_{\Omega(X)}(O)$. The FW-Learner of $\delta(X)$ and $\Omega(X)$ are defined as
\begin{align*}
&\hat{\delta}(X)
=
\hat{m}_{J-\delta(X)}(x)  
:= \bb{1 - h_n(x)}
\bar{\phi}_J^T(x) 
\bb{\sum_{i\in \mc{I}_{k}^{c2}} \bar{\phi}_J(X_i) \bar{\phi}_J^T(X_i) + \bar{\phi}_J(x)\bar{\phi}_J^T(x)}^{-1}
\sum_{i\in \mc{I}_{k}^{c2}} \bar{\phi}_J(X_i) \hat{f}_{\delta(X)}(O),
\\
&\widehat{\Omega}(X)
=
\hat{m}_{J-\Omega(X)}(x)  
:= \bb{1 - h_n(x)} 
\bar{\phi}_J^T(x) 
\bb{\sum_{i\in \mc{I}_{k}^{c2}} \bar{\phi}_J(X_i) \bar{\phi}_J^T(X_i) + \bar{\phi}_J(x)\bar{\phi}_J^T(x)}^{-1}
\sum_{i\in \mc{I}_{k}^{c2}} \bar{\phi}_J(X_i) \hat{f}_{\Omega(X)}(O),
\end{align*}
where 
\begin{align*}
h_n(x) 
&:= \bar{\phi}_J^T(x)
\bb{\sum_{i\in \mc{I}_{k}^{c2}} \bar{\phi}_J(X_i)\bar{\phi}_J^T(X_i) + \bar{\phi}_J(x)\bar{\phi}_J^T(x)}^{-1}
\bar{\phi}_J(x) \in [0,1].
\end{align*}

The motivation for using the FW counterfactual regression approach is that, if appropriately tuned (selecting the number of basis functions to include $J$ via cross-validation), the FW-Learner for \(\Omega(X)\) and \(\delta(X)\) can be shown to achieve the oracle minimax rate for the MSE---the rate obtained when the true functions $p_z(X)$, $\pi_z(X)$ and $e_z(X)$ are known---which may potentially be faster than \(o_P\ba{N^{-1/4}}\). \Cref{c:fw-lem}, from \citeA{yang2023forster}, formally establishes the theoretical results of applying the FW-Learner to the pseudo-outcomes $\hat{f}(O)$.
\begin{Lemma}
    \label{c:fw-lem}
    Let \( H_f(x) = E\bb{\hat{f}(O) \mid X = x, \hat{f}} \), and \( \sigma_m^2 \) be an upper bound on \(E\bb{\hat{f}^2(O) \mid X, \hat{f}}\) almost surely \( X \). Suppose that \( X \) has a density with respect to \( \mu \) that is bounded by \( \kappa \). Then, the FW-Learner \( \hat{m}_J \) satisfies
    \begin{align*}
    &
    \bc{ E\bc{\bb{\hat{m}_{J-f(O)}(X) - m^*_{f(O)}(X)}^2 \mid \hat{f} } }^{1/2} 
    \leq M_1 + M_2
    \\
    &
    M_1 = \sqrt{\frac{2\sigma_m^2 J}{|\mc{I}_k^{c2}|}} + \sqrt{2\kappa M_J^\Psi \ba{m^*_{f(O)}}}
    \ , \quad 
    M_2 = \sqrt{6} \bc{ E \bc{\bb{H_f(X) - m^*_{f(O)}(X)}^2 \mid \hat{f}} }^{1/2}  
    \end{align*}    
\end{Lemma}

The first component of the term $M_1$ is a variance term which naturally scales as $J/N$ for a series regression estimator with $J$ basis functions. Thus, as expected, the series estimator is consistent provided that the number of basis functions $J$ scales at rates slower than $N$. The second component of $M_1$, $M_J^\Psi\ba{m^*_{f(O)}}$ corresponds to the approximation error of the function $m^*_{f(O)}$ associated with truncating its series representation to $J$ basis functions (even if as we allow, $J$ grows with sample size). Thus, $M_1$ is an upper bound on the  MSE of the oracle FW-Learner had the pseudo-outcome $f(O)$ been observed for all units. The second term $M_2$ scales as $H_f-m^*_{f(O)}$, which captures the bias incurred from having to estimate the oracle pseudo-outcome $f(O)$ with the empirical pseudo-outcome $\hat{f}(O)$.  According to Theorem 2 of \citeA{yang2023forster}, by virtue of using an EIF to estimate the pseudo-outcome, $M_2$ is guaranteed to be of second-order. Indeed, this second order bias is characterized precisely in the Supplementary Material A.4 for our specific setting. The key takeaway of \Cref{c:fw-lem} is that the FW-Learner of $\delta(X)$ and  $\Omega(X)$ is guaranteed to perform as well as the oracle FW-learner (with access to the oracle pseudo-outcome), provided the bias due to estimation of nuisance functions needed to evaluate the pseudo-outcome can be made sufficiently small such that it is negligible relative to the oracle MSE $M_1$.

Again, to make full use of the observed data, we swap the roles of specific folds $\mc{I}_k^{c1}$ and $\mc{I}_k^{c2}$, repeat Steps 1 and 2, and average two estimates labeled by superscripts $(-k,c1)$ and $(-k,c2)$ to obtain $\hat{p}_z^{(-k)}(X)$, $\hat{\pi}_z^{(-k)}(X)$, $\hat{e}_z^{(-k)}(X)$, $\widehat{\Omega}^{(-k)}(X)$ and $\hat{\delta}^{(-k)}(X)$.

Alternative orthogonal learning approaches may also be considered for counterfactual pseudo-outcome regression, including the framework of \citeA{kennedy2023towards} (see also \citeA{rambachan2022robust}), and one could also potentially adapt the orthogonal learning framework of \citeA{foster2023orthogonal} to construct alternative orthogonal learners in our setting. However, to our knowledge, the oracle and MSE guarantees we rely on here have not yet been established for such alternatives at the same level of generality as for the FW learner of \citeA{yang2023forster}, in order to target the specific nuisance functions arising in our setting. Specifically, the FW learner provides a concrete roadmap for constructing pseudo-outcomes with second-order bias control for a broad class of counterfactual regression functions, including those arising in our IV setting; furthermore, it places essentially no restrictions on the regressor distribution and guarantees a bounded estimated regression function under relatively mild conditions. These properties motivate our focus on the FW learner here. Whether comparable guarantees can be established for alternative orthogonal learners remains unknown and is beyond the scope of the present paper.
\\[0.2cm]
(\textit{Step 3: Cross-Fit Estimator of $\psi$}):
\\[0.1cm]
\indent 
For each $k = 1,\ldots,K$, we use the evaluation fold $\mc{I}_k$ to estimate $\hat{\psi}^{(k)}$ using nuisance functions estimated from the estimation fold $\mc{I}_k^{c}$ to evaluate the EIF; we then average them to obtain the final cross-fit estimator: 
\begin{align*}
    &\hat{\psi}^{EIF-FW}=\frac{1}{K} \sum_{i=1}^K \hat{\psi}^{(k)}, \numberthis \label{psi_esti} \\
    &\textit{where} \\
    &\hat{\psi}^{(k)} =  
    \frac{1}{\mb{P}(A)} 
    \mb{P}_{\mc{I}_k}
    \left(
    \begin{aligned}
      &A\bb{Y+\hat{\delta}^{(-k)}(X)} + \hat{\rho}^{(-k)}(X) \frac{2Z-1}{ \hat{\pi}^{(-k)}_Z(X)}
    \widehat{\Omega}^{(-k)}(X)  \\
      &
        \quad  \quad  \quad 
    \times \bc{
    Y(1-A)-\hat{e}_Z^{(-k)}(X)  - \bb{A-\hat{p}_Z^{(-k)}(X)} \hat{\delta}^{(-k)}(X) 
    }
    \end{aligned}
    \right), \\
    &\hat{\rho}^{(-k)}(X)  = \hat{p}_1^{(-k)}(X) \hat{\pi}_1^{(-k)}(X) + \hat{p}_0^{(-k)}(X)\hat{\pi}_0^{(-k)}(X)
    .
\end{align*}

\subsection[Asymptotic Theory]{Asymptotic Theory for the Proposed Estimator}
\label{stat_property}
Let  $||\cdot||_{P,2}$ denote the $\mc{L}^2(P)$-norm for the arm $A=1$ with respect to the true law $P$ that generates the observed data:
\begin{align*}
    r_{\eta,N}^{(-k)} = \| \hat{\eta}^{(-k)} (X) - \eta(X) \|_{P,2}
    =
    \bigg\{ 
    \int_{\mathcal{X}} 
    \{
    \hat{\eta}^{(-k)} (X) - \eta(X) \}^2 \, dF(X|A=1)
    \bigg]^{1/2}
    \ , 
\end{align*}
for $
    \eta \in \{ p_z,\pi_z,e_z,\Omega,\delta\}$, $z \in \bb{0,1}$ and $k \in \{1,\ldots,K\}$. Suppose the following conditions hold for  estimated nuisance functions for all $k = 1,\ldots,K$: 
\begin{assumption}[Boundedness of the Estimated Nuisance Functions]
\label{as:bounds}

There exist constants $c_1,c_2>0$ such that $\hat{p}_z^{(-k)}(x) \in [c_1,1-c_1]$, $\hat{\pi}^{(-k)}_z(x) \in [c_1,1-c_1]$, $\hat{e}_z^{(-k)}(x)\in [-c_2,c_2]$, $\widehat{\Omega}^{(-k)}(x)\in [-c_2,c_2]$ and $\hat{\delta}^{(-k)}(x)\in [-c_2,c_2]$ for all $z \in \{0,1\}$ and $x \in \mc{X}$.     
\end{assumption}
\begin{assumption}[Consistent Estimation]
\label{as:consis}
$r_{p_z,N}^{(-k)}$, $r_{\pi_z,N}^{(-k)}$, $r_{e_z,N}^{(-k)}$, $r_{\Omega,N}^{(-k)}$ and $r_{\delta,N}^{(-k)}$ are $o_P(1)$  
\end{assumption}
\begin{assumption}[Cross-product Rates] \label{as:rates}
\begin{align*} 
    &r_{\Omega,N}^{(-k)} r_{\delta,N}^{(-k)}
    \textit{ and } \ 
    r_{p_0,N}^{(-k)} r_{\pi_{z},N}^{(-k)}   
    \textit{ and } \    
    \ r_{e_0,N}^{(-k)} r_{\pi_{z},N}^{(-k)} \
    \textit{ and } \        
    r_{\delta,N}^{(-k)} r_{\pi_{z},N}^{(-k)} \
    \textit{ are } o_P\ba{N^{-1/2}}     \textit{ for  $z \in \{0,1\}$}. 
\end{align*}
\end{assumption}

Assumption \ref{as:bounds} requires that the estimators are uniformly bounded, with $p_z(X),\pi_z(X)$ constrained within $(0,1)$, and $e_z(X),$ $\Omega(X),$ $\delta(X)$ within a finite interval for $x \in \mc{X}$. Assumption \ref{as:consis} requires that estimated nuisance functions converge to their true functions as the sample size grows large, which is achieved by many nonparametric machine learning estimators. Assumption \ref{as:rates} requires that the cross-product rates of nuisance function estimators are $o_P\ba{N^{-1/2}}$.  Importantly, if some of these four nuisance functions, $p_z(X)$, ${\pi}_z(X)$, $e_z(X)$, and ${\delta}(X)$, are estimated at sufficiently fast rates, e.g. faster than $o_P\ba{N^{-1/4}}$, the remaining nuisance function is allowed to converge at a slower rate, e.g. slower than  $o_P\ba{N^{-1/4}}$ provided that their cross-product terms remain $o_P\ba{N^{-1/2}}$.  Assumptions \ref{as:consis} and \ref{as:rates} are satisfied with each of nuisance functions estimated at rate $o_P\ba{N^{-1/4}}$, which can be attained by many nonparametric machine learning estimators and the FW counterfactual regression approach under well-established conditions. 

These assumptions are key to studying the asymptotic distribution of the proposed estimator in \eqref{psi_esti} in subsequent \Cref{th:asym}.  Before proceeding, we note that these assumptions align with the conditions for unbiasedness of the EIF proposed in \Cref{thm2}. The corresponding union model $\mc{M}_1\cup\mc{M}_2\cup\mc{M}_3$ with a similar structure and parametric multiply robust estimator has been studied in \citeA{wang2018bounded}'s work on the ATE. Since $\delta(X)$ does not inherently induce a partial likelihood, they developed a method to estimate $\delta(X)$ without requiring the models for $p_Z(X)$ and $e_Z(X)$ to be correct. This motivates implementing Step 2 to estimate the functions $\delta(X)$ and $\Omega(X)$, which may be smoother than their respective components ($e_z(X)$ and $p_z(X)$ for $z\in \bb{0,1}$), directly in a nonparametric manner. The estimation should be conducted with care to retain the potential of multiple robustness.

More specifically, instead of using the FW learner, one could in principle adopt the simpler strategy of substitution estimators $\tilde{\delta}(X) = \{\hat{e}_1(X) -\hat{e}_0(X)\}/\{\hat{p}_1(X) - \hat{p}_0(X)\}$ and $\widetilde{\Omega}(X) = 1/\{\hat{p}_1(X) - \hat{p}_0(X)\}$, which would yield a convergence rate for, say, $\tilde{\delta}(X)$ determined by the worst rate achieved by $\hat{p}_z(X)$ and $\hat{e}_z(X)$ and $\widetilde{\Omega}(X)$ in $\hat{p}_z(X)$ for $z\in \bb{0,1}$. To compare this simpler plug-in strategy to the FW-Learner, denote the estimator of $\psi$ using \(\tilde{\delta}(X)\) and \(\widetilde{\Omega}(X)\) as \(\hat{\psi}^{EIF}\) to distinguish it from \(\hat{\psi}^{EIF-FW}\). The convergence rate requirement for the second-order bias terms in \(\hat{\psi}^{EIF}\) can be derived by decomposing \(r_{\delta,N}^{(-k)}\) in assumption \ref{as:rates} into \(r_{p_z,N}^{(-k)}\) and \(r_{e_z,N}^{(-k)}\), and \(r_{\Omega,N}^{(-k)}\) into \(r_{p_z,N}^{(-k)}\), which leads to a new, more restrictive condition on the convergence rates. In the Supplementary Material S4, we formally show that the upper bound on the convergence rate of the bias term of \(\hat{\psi}^{EIF-FW}\) is no larger than that of \(\hat{\psi}^{EIF}\); that is, the bias of \(\hat{\psi}^{EIF-FW}\) may decay far faster than that of \(\hat{\psi}^{EIF}\) and thus yield improved finite-sample performance.

The following \Cref{th:asym} formally establishes the asymptotic normality of the estimator \(\hat{\psi}^{EIF-FW}\) under the stated assumptions.
\begin{theorem}
\label{th:asym}
    Suppose that assumptions \ref{as:iv1} -- \ref{as:rates} hold. Under model $\mc{M}$, the estimator in \eqref{psi_esti} satisfies
    \begin{align*}
        \sqrt{N}\ba{\hat{\psi}^{EIF-FW}-\psi} \longrightarrow_D N(0,\sigma^2) \  as \  N \rightarrow \infty
        \numberthis \label{asymp-converge}.
    \end{align*}
The variance $\sigma^2$ matches the semiparametric efficiency bound for $\psi$ in the nonparametric model for the observed data, i.e. $\sigma^2=var\{EIF(O;\psi)\}$.
A consistent estimator of $\sigma^2$ is given by
    \begin{align*}
        &\hat{\sigma}^2 = \frac{1}{K}\sum_{i=1}^K \mb{P}_{\mc{I}_k}
        \bc{
        \bc{
        \frac{1}{\mb{P}(A)}
        \bc{
        \begin{aligned}
        & A\bb{Y+\hat{\delta}^{(-k)}(X)-\hat{\psi}^{EIF-FW}}
        + 
        \\
        &
        \hat{\rho}^{(-k)}(X) \frac{2Z-1}{ \hat{\pi}^{(-k)}_Z(X)}
        \widehat{\Omega}^{(-k)}(X)   \bc{
        \begin{aligned}
            &Y(1-A)-\hat{e}_Z^{(-k)}(X) \\
            &- \bb{A-\hat{p}_Z^{(-k)}    (X)}\hat{\delta}^{(-k)}(X)
        \end{aligned}
        }
        \end{aligned}
        }
        }^2
        }
    \end{align*}
\end{theorem}

Of note, assumptions \ref{as:consis} and \ref{as:rates} are conditions where the root $N$ scaled second-order bias becomes $o_P(1)$ for the proposed estimator. Let $z_{1-\alpha/2}$ denote the $(1-\alpha/2)$ quantile of the standard normal distribution. By \Cref{th:asym}, an asymptotic $100(1-\alpha)\%$ confidence interval for $\psi$ is given by 
\[
\left(\hat{\psi}^{EIF-FW} - z_{1-\alpha/2} \frac{\hat{\sigma}}{\sqrt{N}}, \; \hat{\psi}^{EIF-FW} + z_{1-\alpha/2} \frac{\hat{\sigma}}{\sqrt{N}}\right).
\]

\section{Simulation Studies}
\label{sec:sim}
In this section, we evaluate the finite-sample performance of the proposed estimators. We generate i.i.d. $\{Y_i,$ $A_i,$ $Z_i,$ $X_{1i},$ $X_{2i},$ $ U_i\}$ under the following data-generating process (DGP) compatible with assumptions \ref{as:iv1} -- \ref{as:miv}. The marginal ATT $\beta$ under this DGP is 3.164. 
\begin{align*}
    & X_1, X_2 \sim Uniform(0,1);  && U\sim N(4,0.5^2); \\
    & \alpha_1(Z,X) = Z  \bb{0.5 + (X_1 + X_2)/2}; && \alpha_2(U,X) = \ba{-X_1 - X_2 - U/4}; \\
    & pr(Z=1|X) = \frac{exp(-1 + X_1 +X_2)}{1+exp(-1 + X_1 +X_2)}; && pr(A=1|Z,X,U) = exp\{\alpha_1(Z,X) + \alpha_2(U,X)\}; \\  
    & E\ba{Y^0|U,X,Z} = E\ba{Y^0|U,X}= (X_1 + X_2)  exp(U/6); && E\ba{Y^1|U,X,Z} = (X_1 + X_2 + X_1  X_2 + Z)  exp(U/4); \\
    & \epsilon_{Y_1},\epsilon_{Y_0} \sim N(0, 0.5^2); && Y^a = E(Y^a|U,X,Z) + \epsilon_{Y_a},  \textit{ for } a \in\bb{0,1}.
\end{align*}

We consider three competing estimators for $\psi$:
\begin{enumerate}
    \item $\hat{\psi}^{Wald}$: The single-arm Wald ratio-based estimator $\hat{\psi}^{Wald} = \mathbb{P} \big[ A / \mathbb{P}(A)  [ Y + \{\hat{e}_1(X)- \hat{e}_0(X) \} / \{ \hat{p}_1(X) $ $- \hat{p}_0(X)\} ] \big]$, which is a plug-in estimator of the identifying formula in \Cref{thm1}. 
    \item $\hat{\psi}^{EIF}$: The EIF-based estimator, with simple substitution estimators used for $\Omega(X)$ and $\delta(X)$. 
    \item $\hat{\psi}^{EIF-FW}$: The EIF-based estimator $\hat{\psi}^{EIF-FW}$ proposed in \Cref{th:asym}, invoking the FW counterfactual regression approach to estimate $\Omega(X)$ and $\delta(X)$.
\end{enumerate}
\begin{figure}[!htbp]
    \centering
    \begin{minipage}[t]{0.50\textwidth}
    \vspace{0pt}
        \centering
        \includegraphics[width=\linewidth]{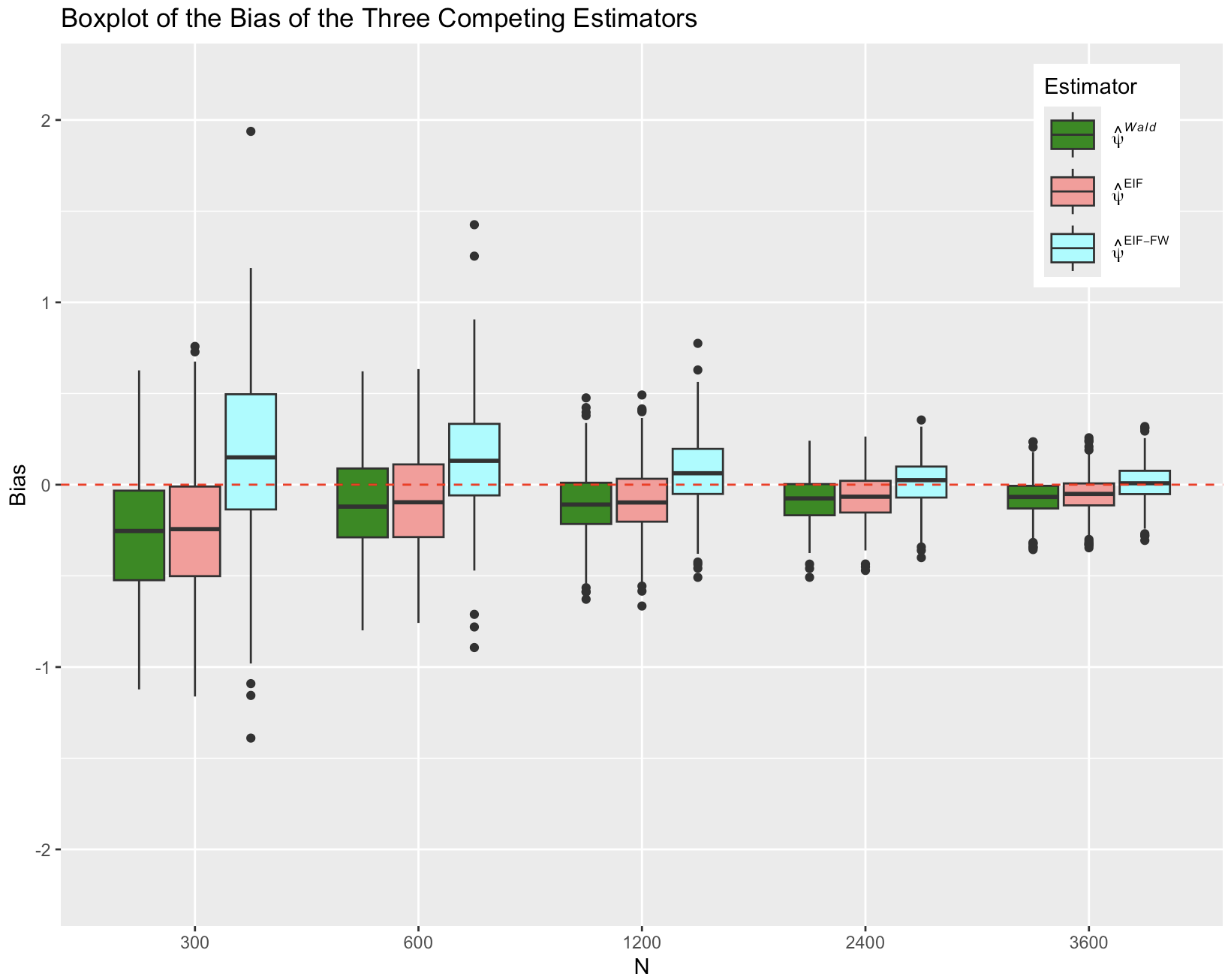}
        \caption{A Graphical Summary of the Simulation Results. Each column gives boxplots of biases of proposed estimators $\hat\psi^{Wald}$, $\hat\psi^{EIF}$ and $\hat\psi^{EIF\text{-}FW}$ at $N=300,600,1200,2400,3600$ respectively.}
        \label{fig:sim_boxplot}
    \end{minipage}
    \hspace{1pt}
    \begin{minipage}[t]{0.48\textwidth}
    \vspace{0pt}
        \centering
        \scriptsize
        \setlength{\tabcolsep}{4pt}
        \renewcommand{\arraystretch}{0.95}
        \begin{tabular}{lcccc}
            \hline
            Sample Size (F-stat) &  & $\hat\psi^{Wald}$ & $\hat\psi^{EIF}$ & $\hat\psi^{EIF\text{-}FW}$ \\
            \hline
            \multirow{4}{*}{300 (22.16)}
            & Bias     & -0.273 & -0.234 & 0.141 \\
            & ASE      & 0.373  & 0.418  & 0.664 \\
            & ESE      & 0.401  & 0.413  & 0.561 \\
            & Coverage & 0.823  & 0.917  & 0.973 \\
            \hline
            \multirow{4}{*}{600 (40.24)}
            & Bias     & -0.110 & -0.090 & 0.126 \\
            & ASE      & 0.273  & 0.289  & 0.367 \\
            & ESE      & 0.279  & 0.286  & 0.326 \\
            & Coverage & 0.883  & 0.933  & 0.957 \\
            \hline
            \multirow{4}{*}{1200 (85.08)}
            & Bias     & -0.102 & -0.091 & 0.066 \\
            & ASE      & 0.193  & 0.204  & 0.220 \\
            & ESE      & 0.185  & 0.186  & 0.199 \\
            & Coverage & 0.903  & 0.943  & 0.947 \\
            \hline
            \multirow{4}{*}{2400 (165.73)}
            & Bias     & -0.081 & -0.067 & 0.015 \\
            & ASE      & 0.137  & 0.143  & 0.146 \\
            & ESE      & 0.126  & 0.127  & 0.130 \\
            & Coverage & 0.897  & 0.947  & 0.973 \\
            \hline
            \multirow{4}{*}{3600 (272.80)}
            & Bias     & -0.068 & -0.054 & 0.006 \\
            & ASE      & 0.112  & 0.117  & 0.116 \\
            & ESE      & 0.108  & 0.110  & 0.109 \\
            & Coverage & 0.887  & 0.937  & 0.957 \\
            \hline
        \end{tabular}
        \caption{Numerical summaries of the performance of each estimator at different sample sizes, with the corresponding first-stage F-statistic reported in parentheses.}
        \label{tab:sim_summary}
    \end{minipage}
\end{figure}
We explore moderate to large sample sizes $N$, taking values in $\{300, 600, 1200, 2400, 3600\}$. Using the simulated data, we implement the estimation procedure and compute the proposed estimator for $\psi$ and $\sigma^2$ outlined in Section \ref{sec:proposed_estimator} and Theorem \ref{th:asym}. 

The procedure begins by splitting the samples into three non-overlapping folds of equal size. The first two folds \( \mathcal{I}_{1} \) and \( \mathcal{I}_{2} \) (referred to as \( \mathcal{I}_{k}^{c1} \) and \( \mathcal{I}_{k}^{c2} \) in Section \ref{sec:proposed_estimator}) are used to estimate nuisance functions \( \hat{p}_z^{(-3)}(X) \), \( \hat{\pi}_z^{(-3)}(X) \), \( \hat{e}_z^{(-3)}(X) \), \( \widehat{\Omega}^{(-3)}(X) \) and \( \hat{\delta}^{(-3)}(X) \) following Steps 1 and 2. The third fold \( \mathcal{I}_{3} \) (referred to as \( \mathcal{I}_{k} \) in Section \ref{sec:proposed_estimator}) is used as the evaluation fold to obtain \( \hat{\psi}^{(3)} \). For the first and the second estimators $\hat{\psi}^{Wald}$ and $\hat{\psi}^{Plug-in}$, folds \( \mathcal{I}_{1} \) and \( \mathcal{I}_{2} \) are merged to directly estimate \( \hat{p}_z^{(-3)}(X) \), \( \hat{\pi}_z^{(-3)}(X) \), and \( \hat{e}_z^{(-3)}(X) \). We choose generalized linear models, Xgboost, and Random Forest as candidate estimators in the SuperLearner algorithm \cite{friedman2010regularization, chen2016xgboost, breiman2001random}. To fully use the data, the estimation and evaluation folds are rotated, and the procedure is repeated three times to produce the final estimate \( \hat{\psi} \) as the average of $\psi^{(1)}$, $\psi^{(2)}$ and $\psi^{(3)}$. We also implement the median adjustment expanded by \citeA{chernozhukov2018double}, splitting the same dataset randomly and repeating the cross-fitting 7 times. This improves the stability by combining estimates of $\psi$ from different partitions of the same dataset, making the final estimator less sensitive to extreme estimates that machine learning methods may produce even at large sample sizes.

Figure \ref{fig:sim_boxplot} visually summarizes the results from 300 estimates at each sample size. In terms of bias, we observe that \(\hat{\psi}^{Wald}\) exhibits substantially greater bias than \(\hat{\psi}^{EIF-FW}\) across all sample sizes, with \(\hat{\psi}^{EIF}\) performing between the two. Figure \ref{tab:sim_summary} provides numerical summaries of these three estimators. As \(N\) increases, the standard errors of all three estimators decrease. Notably, the empirical coverage of the 95\% confidence interval for \(\hat{\psi}^{Wald}\) falls considerably below its nominal rate, whereas the coverage of \(\hat{\psi}^{EIF}\) and \(\hat{\psi}^{EIF-FW}\) aligns with the nominal rate. Overall, \(\hat{\psi}^{EIF-FW}\) significantly outperforms the others as \(N\) grows. This result is expected,  as there are no theoretical guarantees for the asymptotic behavior of $\hat{\psi}^{Wald}$, while the asymptotic normality of $\hat{\psi}^{EIF}$ and $\hat{\psi}^{EIF-FW}$ is formally established in Theorem \ref{th:asym}, with the latter potentially having smaller bias and therefore better coverage.

\section[Relations between Different Identifying Assumptions]{Relations between MIV and Robins' No-Treatment Effect Heterogeneity Assumption}
\label{sec:compare}
For identification of the ATT, our MIV framework makes assumptions \ref{as:iv1} -- \ref{as:miv}. In contrast, as mentioned in \Cref{intro}, the absence of effect heterogeneity of the ATT with respect to the instrument, conditional on covariates, also identifies the ATT via a standard conditional Wald ratio \cite{robins1994correcting}. Specifically, for binary point treatment and instrumental variables, \citeA{robins1994correcting} considered the following formulation of instrumental variable conditions:
\begin{assumption}[Instrument Mean Independence \& Exclusion Restriction \cite{robins1994correcting}]
\label{as:iv-23jr}
$$E\ba{Y^{a=0} | Z,X} \indep Z | X$$
\end{assumption}
\begin{assumption}[No Current Treatment Interaction; NCTI]
\label{as:ncti}
$$E\ba{Y^{a=1}-Y^{a=0} | A=1,Z,X} \indep Z | A=1,X$$
\end{assumption}

Assumption \ref{as:ncti} is a no-treatment effect heterogeneity condition on the additive scale, i.e.\ it states that the conditional ATT given $(Z,X)$ is equal for the $Z=1$ and $Z=0$ arms. Clearly, assumption \ref{as:ncti} restricts the causal effect of primary interest to be constant across strata defined by the instrument, which may be overly restrictive, particularly if inference about such heterogeneity is of scientific interest. At the same time, the assumption is guaranteed to hold under the conditional null ATT and therefore is locally robust, in the sense that the level of a statistical test of the ATT null hypothesis is robust to this assumption. Under these conditions together with instrument relevance (assumption \ref{as:iv1}), \citeA{robins1994correcting} established identification of the conditional ATT $\beta(X)$ with the standard conditional Wald ratio:
$$\beta(X)=\delta^*(X).$$

Interestingly, \citeA{hernan2006instruments} related assumption \ref{as:ncti} and the MIV assumptions under a certain nonparametric structural equation model with independent errors. The corresponding IV model essentially implies a stronger notion of latent ignorability, which we now state:
\begin{assumption}[Strong Latent Ignorability \& Exclusion Restriction]
\label{as:iv3-s}
$$Y^a \indep (A,Z) | X,U,\qquad a\in\{0,1\}.$$
\end{assumption}

Assumption \ref{as:iv3-s} is stronger than assumption \ref{as:iv3} described in \Cref{sec:ns} as the former essentially requires marginal latent ignorability with respect to both potential outcomes, $(Y^{a=0},Y^{a=1})$ together with exclusion restrictions for both potential outcomes, i.e. $Y^{a,z}=Y^{a}$ for $a,z\in\{0,1\}^2$; while the latter only restricts the treatment-free potential outcome $Y^{a=0}$. This distinction is not merely formal. A useful intuition is that weak latent ignorability only concerns the no-treatment potential outcome $Y^{0}$, whereas strong latent ignorability additionally constrains the treated potential outcome $Y^{1}$. In applications where interest centers on the ATT for the treated/exposed group, one may be willing to reason about confounding for the treatment-free counterfactual among the treated, while being less comfortable imposing further restrictions on treated potential outcomes. The latter may fail when the instrument affects not only whether treatment is received, but also how treatment is accessed or experienced among those treated.

Another related intuition arises in encouragement designs. Consider the Oregon Health Insurance Experiment, where $Z$ is lottery assignment, $A$ is Medicaid enrollment, and $Y$ is a health outcome, with latent health need, health literacy, and care-seeking behavior absorbed in $U$ \cite{finkelstein2012oregon}. Under assumption \ref{as:iv3}, one posits that among individuals in the no-insurance state, lottery assignment has no residual bearing on health outcomes once $(X,U)$ are controlled for. This may be scientifically credible, since individuals who do not enroll remain uninsured regardless of lottery assignment. Assumption \ref{as:iv3-s}, however, additionally requires that among those who obtain coverage, the treated potential outcome be unrelated to whether coverage arose through lottery-induced enrollment or through other pathways, conditional on $(X,U)$. That stronger requirement may fail if lottery winners differ from always-takers in how quickly they enroll, which providers they access, how intensively they use care, or how effectively they navigate the Medicaid system after enrollment. This reviewer-suggested example is useful because it illustrates that the stronger restriction concerns features of the treated potential outcome $Y^1$ that are not required for our ATT analysis under MIV.

More broadly, similar considerations arise in other designs, including job-training vouchers, insurance or screening outreach, physician- or facility-preference instruments, and housing voucher lotteries. In such settings, it may be reasonable to require that, after conditioning on $(X,U)$, the instrument has no residual bearing on outcomes under no treatment, while remaining agnostic about whether the instrument also affects outcomes under treatment through timing, provider choice, treatment intensity, adherence, or other pathway-specific features. Thus, the appeal of the MIV framework here is not mainly that it is categorically easier to defend, but that it avoids invoking additional assumptions on $Y^1$ that are not needed to identify the ATT.

Remarkably, \citeA{hernan2006instruments} established that assumptions \ref{as:iv1}, \ref{as:iv2}, \ref{as:iv3-s}, and the MIV model \ref{as:miv} imply assumption \ref{as:ncti} and therefore identify the conditional ATT,
$$\beta(X)=E\bb{E\ba{Y^1-Y^0 | U,X} | A=1,X},$$
by the standard conditional Wald ratio $\delta^*(X)$, without a priori restricting the degree of latent heterogeneity $E\ba{Y^1-Y^0 | U,X}$ in the treated population. Clearly, the causal model of \citeA{hernan2006instruments}, namely assumptions \ref{as:iv1}, \ref{as:iv2}, \ref{as:iv3-s}, and \ref{as:miv}, which we refer to as MIVHR, is a submodel of the MIV causal model considered in this paper. Consequently, under MIVHR the conditional ATT admits dual identifying expressions, namely $\delta^*(X)=\psi(X)$, or equivalently
$$
E\ba{Y^1 | A=1,X}
=
E(Y | A=1,X)
=
\frac{E(YA | Z=1,X)-E(YA | Z=0,X)}
{pr(A=1 | Z=1,X)-pr(A=1 | Z=0,X)}.
$$

This is the treated-arm analog of the single-arm Wald ratio we previously considered only for the treatment-free arm. The above equality is an immediate consequence of the following formal result:
\begin{Lemma}
\label{c:orthogo1}
The MIVHR model defined by assumptions \ref{as:iv1}, \ref{as:iv2}, \ref{as:iv3-s}, and \ref{as:miv} has a testable implication in the observed data in the form of the conditional independence
$$Y \indep Z | A=1,X.$$
\end{Lemma}

Lemma \ref{c:orthogo1} raises an important efficiency question: whether, under MIVHR, explicitly leveraging this additional conditional independence can improve estimation of $E\bb{\psi(X) \mid A=1}$ relative to the estimator previously derived, which is semiparametric efficient in the larger nonparametric model $\mc{M}$ that does not restrict the observed data law. Surprisingly, the answer is negative. Although the extra conditional independence restriction yields the alternative representation of the ATT as $E\bb{\delta^*(X) \mid A=1}$, the efficient influence function for this estimand remains $EIF(O;\psi)$ in the smaller model $\mc{M}_{ind}\subset\mc{M}$. This further underscores an important point: one may prefer not to invoke Assumption \ref{as:iv3-s} at all, since it imposes additional restrictions involving the potential outcome under treatment, yet appears to offer no identification or efficiency gain beyond the MIV assumptions considered here. A formal statement is given below.

\begin{Lemma}
\label{c:orthogo2}
The $EIF(O;\psi)$ in \Cref{thm2}, derived under $\mc{M}$ assuming \ref{as:iv1} -- \ref{as:miv}, is also semiparametric efficient in the smaller semiparametric model $\mc{M}_{ind}$.
\end{Lemma}

At the same time, it is worth emphasizing that the stronger MIVHR model leaves a testable footprint in the observed data and is therefore, at least in principle, falsifiable. The implication of \Cref{c:orthogo1} may in principle be assessed using a conditional independence test. However, such tests target a broad null hypothesis and are well known to be challenging in flexible nonparametric settings \cite{shah2020hardness}, especially when one is interested in departures in a particular direction relevant to the causal estimand. For this reason, it is also natural to consider a more targeted comparison tailored to the ATT. Under MIVHR, the coincidence between the standard conditional Wald estimand and our single-arm Wald estimand motivates a Hausman-type test comparing the two estimands \cite{hausman1978specification}. The two tests play distinct but complementary roles: the conditional independence test assesses a broad observed-data implication of the stronger model, whereas the Hausman comparison asks whether the stronger restrictions imposed by MIVHR matter for inference on the ATT. From this perspective, the Hausman comparison may be especially informative in applications, since rejection directly indicates that the stronger model fails in a way that is reflected in ATT inference. At the same time, the two tests are likely to lead to qualitatively similar conclusions in many settings. In particular, it may be implausible in practice that MIVHR fails while the weaker MIV assumptions continue to hold, so rejection of the Hausman null may also be viewed as circumstantial evidence against the plausibility of MIV itself. Conversely, failure to reject the Hausman null suggests not only that there is no conclusive empirical evidence against the stronger restriction in a direction relevant to the ATT, but also that MIVHR and MIV yield similar ATT-relevant inference in the application at hand.

\section{Application: Job Corps Dataset}
\label{sec:apli}
To illustrate the proposed methods, we analyze data from the National Job Corps Study \cite{schochet2008does}, which evaluated the largest federally funded job training program in the United States for disadvantaged youth ages 16 to 24. Between 1994 and 1996, the study recruited 15,386 eligible applicants through Job Corps outreach and admissions agencies nationwide. Of these, 9,409 applicants (approximately 61\%) were randomly assigned to the program group, while 5,977 applicants (approximately 39\%) were assigned to the control group. Individuals assigned to the program group were allowed to enroll in Job Corps, whereas those assigned to the control group were barred from enrolling in Job Corps for three years after randomization, although they remained free to participate in other education or training programs available in their communities. In our notation, the instrument $Z$ indicates randomized assignment to Job Corps eligibility, while the treatment indicator $A$ records actual enrollment in Job Corps during the follow-up period. The study collected baseline demographic characteristics, including race, gender, and education, together with information on subsequent program enrollment and post-assignment outcomes such as health, employment, and earnings. Due to non-enrollment among some applicants assigned to the program group and loss to follow-up, the final analytic sample used in our study consists of 9,240 individuals. Missing covariates, such as education level and smoking status, were coded using binary missingness indicators. The data are available through the \texttt{R} package \texttt{causalweight} \cite{causalweight}.

Our goal is to estimate the average causal effect of program enrollment on earnings for those who enrolled in at least one year of training. The causal relationship between program enrollment and earnings may be impacted simultaneously by unmeasured confounders, such as career interests, motivation, and personality, which affect both program enrollment and earnings. From the GLIM perspective, the MIV model may be plausible in this context if treatment uptake is governed by a multiplicative latent utility function. In particular, suppose that program enrollment follows a threshold-crossing rule of the form $A^z=\mb{I}\{g(z,X)U \ge \epsilon_z\}$, where $U$ captures unmeasured baseline readiness, motivation, self-assessed ability, or perceived gain from training, and $\epsilon_z$ represents idiosyncratic barriers or shocks affecting take-up. Under the equivalent log-scale representation, this may be written as $h(U,X)-c(z,X)\ge e_z$, where assignment affects enrollment primarily by lowering the effective cost or barrier to participation. In the Job Corps setting, being assigned to the treatment group may reduce informational, administrative, financial, or psychological barriers to enrollment, but such reductions need not have the same effect for everyone on the original scale. Rather, the same assignment may induce a larger change in treatment take-up for individuals with greater latent readiness to invest in their job market skills, while having a smaller effect for individuals with weaker baseline motivation or lower perceived ability to benefit from the program. This type of heterogeneity is accommodated by the multiplicative latent index model.

Program enrollment $A$ is coded as 1 if the participant enrolled in the training program within the first two years after assignment, and 0 otherwise. We consider the fourth-year weekly income, transformed as $log(1 + \textit{Income})$, as the outcome variable $Y$, adjust for all covariates measured before treatment assignment and follow the estimation procedure outlined in Section \ref{sec:proposed_estimator}. Nuisance functions are estimated using the stacked approach of SuperLearner, which includes GLM, Random Forest, and Xgboost as the candidate estimators and the FW counterfactual regression approach. We adopt the median adjustment technique and report results from 100 estimates.

The result from the proposed estimator $\hat{\psi}^{EIF-FW}$ in Theorem \ref{th:asym} is given in the second row of Table \ref{tab:appli}. For comparison, Table \ref{tab:appli} also reports, in the first row, results from the single-arm Wald ratio estimator $\hat{\psi}^{Wald}$ (with confidence intervals obtained by multiplier bootstrap); in the fourth row, the two-stage least squares (2SLS) estimator, which targets the ATT under the MIVHR model and is often viewed as an approximation of the standard Wald ratio estimand in the linear IV setting \cite{angrist2009mostly}; in the third row, the EIF-based estimator under the NCTI framework \cite{levis2024nonparametric}; and in the fifth row, the EIF-based estimator under the no unmeasured confounding assumption.  The first-stage F-statistic is 1202.84, providing strong evidence of instrument relevance in this sample. All estimators give results of a significant positive treatment effect, suggesting that job training enrollment increases long-term earnings for the population represented by the dataset. Although existing literature applies to different subpopulations and relies on various IV assumptions, it also reports the overall positive effect of Job Corps training on long-term earnings \cite{schochet2008does,chen2018going}. In addition, both the conditional independence tests of \citeA{shah2020hardness} and \citeA{scheidegger2022weighted}, as well as the Hausman test for $\hat{\psi}^{EIF-FW}$ and $\hat{\psi}^{EIF-NCTI}$, yield p-values greater than 5\%, so we fail to reject both null hypotheses. As discussed \Cref{sec:compare}, this suggests that there is no conclusive empirical evidence against conditional independence, and that the MIVHR and the MIV deliver consistent inference about the ATT. 

\begin{table}[!htp]
\centering
\small
\begin{tabular}{@{}p{1.8cm}p{6.8cm}p{3.0cm}p{4.2cm}@{}}
\toprule
Estimator & Description & Point Estimate & 95\% Confidence Interval \\ 
\midrule
$\hat{\psi}^{Wald}$ 
& Single-arm Wald ratio-based 
& 0.123 
& $(-0.157,\,0.403)$ \\

$\hat{\psi}^{EIF-FW}$ 
& EIF-based, as given in Theorem \ref{th:asym} 
& 0.312 
& $(0.043,\,0.581)$ \\

$\hat{\psi}^{EIF-NCTI}$ 
& EIF-based, under the NCTI framework 
& 0.431 
& $(0.215,\,0.647)$ \\

$\hat{\psi}^{2SLS}$ 
& Two-stage least squares 
& 0.621 
& $(0.316,\,0.926)$ \\

$\hat{\psi}^{EIF-UC}$ 
& EIF-based, under the unconfoundedness assumption without using $Z$ 
& 0.410 
& $(0.287,\,0.532)$ \\ 
\bottomrule
\end{tabular}

\caption{Numerical summaries of the data analysis.}
\label{tab:appli}
\end{table}
\section{Discussion}
\label{sec:discuss}
In this paper, we studied identification and estimation of the marginal ATT under the MIV model assumption. We derived the EIF and the semiparametric efficiency bound for the marginal ATT within a nonparametric model for the observed data. The proposed estimator for the marginal ATT uses generic machine learning methods and FW counterfactual regression approach to adaptively estimate all nuisance functions. We formally present the conditions on the convergence rates of these nuisance function estimators that are required for the proposed estimator to achieve asymptotic normality. We conducted simulation studies to verify the theoretical properties derived from the proposed estimator and provided a comparison of the MIV model assumption and the previously studied NCTI assumption. Finally, we applied the method to study the ATT of the job training program on subsequent earnings using the Job Corps dataset.

The framework presented in this paper can be extended in several directions. First, unlike NCTI, the proposed MIV framework places no restriction on the outcome model, and thus can potentially be extended to study more general functionals of the treatment-free potential outcome among the treated, such as the median effect of treatment on the treated. This would constitute a worthwhile generalization on which we plan to dedicate a separate manuscript. Second, while this paper focuses on a single binary instrument and a binary exposure, it would be valuable to generalize the proposed method to the setting of more general instruments and exposures, say polytomous or continuous. Third, the identification result established in this paper relies on a key assumption of the MIV model. Future research might consider a sensitivity analysis framework to evaluate the extent to which inference about the ATT is affected when this assumption is relaxed to some extent.

\bibliographystyle{apacite}
\bibliography{feh}

@book{wright1928tariff,
  title={The tariff on animal and vegetable oils},
  author={Wright, Philip Green},
  number={26},
  year={1928},
  publisher={Macmillan}
}

@article{greenland2000introduction,
  title={An introduction to instrumental variables for epidemiologists},
  author={Greenland, Sander},
  journal={International Journal of Epidemiology},
  volume={29},
  number={4},
  pages={722--729},
  year={2000},
  publisher={Oxford University Press}
}

@article{bollen2012instrumental,
  title={Instrumental variables in sociology and the social sciences},
  author={Bollen, Kenneth A},
  journal={Annual Review of Sociology},
  volume={38},
  number={1},
  pages={37--72},
  year={2012},
  publisher={Annual Reviews}
}

@techreport{imbens2014instrumental,
  title={Instrumental variables: An econometrician's perspective},
  author={Imbens, Guido W},
  year={2014},
  institution={National Bureau of Economic Research}
}

@article{dunn2005estimating,
  title={Estimating treatment effects from randomized clinical trials with noncompliance and loss to follow-up: the role of instrumental variable methods},
  author={Dunn, Graham and Maracy, Mohammad and Tomenson, Barbara},
  journal={Statistical Methods in Medical Research},
  volume={14},
  number={4},
  pages={369--395},
  year={2005},
  publisher={Sage Publications Sage CA: Thousand Oaks, CA}
}

@article{baker1994paired,
  title={The paired availability design: a proposal for evaluating epidural analgesia during labor},
  author={Baker, Stuart G and Lindeman, Karen S},
  journal={Statistics in Medicine},
  volume={13},
  number={21},
  pages={2269--2278},
  year={1994},
  publisher={Wiley Online Library}
}

@article{permutt1989simultaneous,
  title={Simultaneous-equation estimation in a clinical trial of the effect of smoking on birth weight},
  author={Permutt, Thomas and Hebel, J Richard},
  journal={Biometrics},
  pages={619--622},
  year={1989},
  publisher={JSTOR}
}

@article{wang2018bounded,
  title={Bounded, efficient and multiply robust estimation of average treatment effects using instrumental variables},
  author={Wang, Linbo and Tchetgen Tchetgen, Eric J},
  journal={Journal of the Royal Statistical Society Series B: Statistical Methodology},
  volume={80},
  number={3},
  pages={531--550},
  year={2018},
  publisher={Oxford University Press}
}

@article{cui2021semiparametric,
  title={A semiparametric instrumental variable approach to optimal treatment regimes under endogeneity},
  author={Cui, Yifan and Tchetgen Tchetgen, Eric J},
  journal={Journal of the American Statistical Association},
  volume={116},
  number={533},
  pages={162--173},
  year={2021},
  publisher={Taylor \& Francis}
}

@article{qiu2021optimal,
  title={Optimal individualized decision rules using instrumental variable methods},
  author={Qiu, Hongxiang and Carone, Marco and Sadikova, Ekaterina and Petukhova, Maria and Kessler, Ronald C and Luedtke, Alex},
  journal={Journal of the American Statistical Association},
  volume={116},
  number={533},
  pages={174--191},
  year={2021},
  publisher={Taylor \& Francis}
}

@article{tchetgen2024nudge,
  title={The Nudge Average Treatment Effect},
  author={Tchetgen Tchetgen, Eric J},
  journal={arXiv preprint arXiv:2410.23590},
  year={2024}
}

@article{liu2020identification,
  title={Identification and inference for marginal average treatment effect on the treated with an instrumental variable},
  author={Liu, Lan and Miao, Wang and Sun, Baoluo and Robins, James and Tchetgen Tchetgen, Eric J},
  journal={Statistica Sinica},
  volume={30},
  number={3},
  pages={1517},
  year={2020},
  publisher={NIH Public Access}
}

@article{robins1994correcting,
  title={Correcting for non-compliance in randomized trials using structural nested mean models},
  author={Robins, James M},
  journal={Communications in Statistics-Theory and methods},
  volume={23},
  number={8},
  pages={2379--2412},
  year={1994},
  publisher={Taylor \& Francis}
}

@article{tchetgen2013alternative,
  title={Alternative identification and inference for the effect of treatment on the treated with an instrumental variable},
  author={Tchetgen Tchetgen, Eric J and Vansteelandt, Stijn},
  year={2013},
  publisher={bepress}
}

@article{yang2023forster,
  title={Forster-Warmuth Counterfactual Regression: A Unified Learning Approach},
  author={Yang, Yachong and Kuchibhotla, Arun Kumar and Tchetgen Tchetgen, Eric J },
  journal={arXiv preprint arXiv:2307.16798},
  year={2023}
}

@article{levis2024nonparametric,
  title={Nonparametric identification and efficient estimation of causal effects with instrumental variables},
  author={Levis, Alexander W and Kennedy, Edward H and Keele, Luke},
  journal={arXiv preprint arXiv:2402.09332},
  year={2024}
}

@article{chernozhukov2018double,
 ISSN = {13684221, 1368423X},
 author = {Victor Chernozhukov and Denis Chetverikov and Mert Demirer and Esther Duflo and Christian Hansen and Whitney Newey and James Robins},
 journal = {The Econometrics Journal},
 number = {1},
 pages = {C1--C68},
 publisher = {[Royal Economic Society, Oxford University Press]},
 title = {Double/debiased machine learning for treatment and structural parameters},
 urldate = {2025-01-08},
 volume = {21},
 year = {2018}
}

@article{schochet2008does,
  title={Does job corps work? Impact findings from the national job corps study},
  author={Schochet, Peter Z and Burghardt, John and McConnell, Sheena},
  journal={American Economic Review},
  volume={98},
  number={5},
  pages={1864--1886},
  year={2008},
  publisher={American Economic Association}
}

@article{chen2018going,
  title={Going beyond LATE: bounding average treatment effects of Job Corps training},
  author={Chen, Xuan and Flores, Carlos A and Flores-Lagunes, Alfonso},
  journal={Journal of Human Resources},
  volume={53},
  number={4},
  pages={1050--1099},
  year={2018},
  publisher={University of Wisconsin Press}
}

@book{angrist2009mostly,
  title={Mostly harmless econometrics: An empiricist's companion},
  author={Angrist, Joshua D and Pischke, J{\"o}rn-Steffen},
  year={2009},
  publisher={Princeton University Press}
}

@article{richardson2013single,
  title={Single world intervention graphs {(SWIGs)}: {A} unification of the counterfactual and graphical approaches to causality},
  author={Richardson, Thomas S and Robins, James M},
  journal={Center for the Statistics and the Social Sciences, University of Washington Series. Working Paper},
  volume={128},
  number={30},
  pages={2013},
  year={2013},
  publisher={Citeseer}
}

@article{hernan2006instruments,
  title={Instruments for causal inference: An epidemiologist's dream?},
  author={Hern{\'a}n, Miguel A and Robins, James M},
  journal={Epidemiology},
  volume={17},
  number={4},
  pages={360--372},
  year={2006},
  publisher={LWW}
}

@article{vytlacil2002independence,
  title={Independence, monotonicity, and latent index models: An equivalence result},
  author={Vytlacil, Edward},
  journal={Econometrica},
  volume={70},
  number={1},
  pages={331--341},
  year={2002},
  publisher={JSTOR}
}

@article{van2007super,
  title={Super learner},
  author={van der Laan, Mark J and Polley, Eric C and Hubbard, Alan E},
  journal={Statistical Applications in Genetics and Molecular Biology},
  volume={6},
  number={1},
  year={2007},
  publisher={De Gruyter}
}

@inproceedings{chen2016xgboost,
  title={Xgboost: A scalable tree boosting system},
  author = {Chen, Tianqi and Guestrin, Carlos},
  booktitle={Proceedings of the 22nd ACM SIGKDD International Conference on Knowledge Discovery and Data Mining},
  pages={785--794},
  year={2016}
}

@article{breiman2001random,
  title={Random forests},
  author={Breiman, Leo},
  journal={Machine Learning},
  volume={45},
  pages={5--32},
  year={2001},
  publisher={Springer}
}

@article{friedman2010regularization,
  title={Regularization paths for generalized linear models via coordinate descent},
  author={Friedman, Jerome and Hastie, Trevor and Tibshirani, Rob},
  journal={Journal of Statistical Software},
  volume={33},
  number={1},
  pages={1},
  year={2010},
  publisher={NIH Public Access}
}

@Manual{causalweight,
    title = {causalweight: Estimation Methods for Causal Inference Based on Inverse
Probability Weighting},
    author = {Bodory, Hugo and  Huber, Martin and Kueck, Jannis},
    year = {2024},
    note = {R package version 1.1.1},
    url = {https://CRAN.R-project.org/package=causalweight},
}

@article{newey1990semiparametric,
  title={Semiparametric efficiency bounds},
  author={Newey, Whitney K},
  journal={Journal of applied econometrics},
  volume={5},
  number={2},
  pages={99--135},
  year={1990},
  publisher={Wiley Online Library}
}

@book{tsiatis2006semiparametric,
  title={Semiparametric theory and missing data},
  author={Tsiatis, Anastasios A},
  volume={4},
  year={2006},
  publisher={Springer}
}

@article{imbens1994identification,
  title={Identification and Estimation of Local Average Treatment Effects},
  author={Imbens, Guido W and Angrist, Joshua D},
  journal={Econometrica},
  volume={62},
  number={2},
  pages={467--475},
  year={1994},
  publisher={Econometric Society}
}

@article{heckman1979sample,
  title={Sample selection bias as a specification error},
  author={Heckman, James J},
  journal={Econometrica: Journal of the econometric society},
  pages={153--161},
  year={1979},
  publisher={JSTOR}
}

@article{finkelstein2012oregon,
  title={The Oregon health insurance experiment: evidence from the first year},
  author={Finkelstein, Amy and Taubman, Sarah and Wright, Bill and Bernstein, Mira and Gruber, Jonathan and Newhouse, Joseph P and Allen, Heidi and Baicker, Katherine and Oregon Health Study Group, the},
  journal={The Quarterly journal of economics},
  volume={127},
  number={3},
  pages={1057--1106},
  year={2012},
  publisher={MIT Press}
}

@article{shah2020hardness,
  title={The hardness of conditional independence testing and the generalised covariance measure},
  author={Shah, Rajen D and Peters, Jonas},
  year={2020}
}

@article{kennedy2023towards,
  title={Towards optimal doubly robust estimation of heterogeneous causal effects},
  author={Kennedy, Edward H},
  journal={Electronic Journal of Statistics},
  volume={17},
  number={2},
  pages={3008--3049},
  year={2023},
  publisher={The Institute of Mathematical Statistics and the Bernoulli Society}
}

@article{rambachan2022robust,
  title={Robust design and evaluation of predictive algorithms under unobserved confounding},
  author={Rambachan, Ashesh and Coston, Amanda and Kennedy, Edward},
  journal={arXiv preprint arXiv:2212.09844},
  year={2022}
}

@article{scheidegger2022weighted,
  title={The weighted generalised covariance measure},
  author={Scheidegger, Cyrill and H{\"o}rrmann, Julia and B{\"u}hlmann, Peter},
  journal={Journal of Machine Learning Research},
  volume={23},
  number={273},
  pages={1--68},
  year={2022}
}

@article{hausman1978specification,
  title={Specification tests in econometrics},
  author={Hausman, Jerry A},
  journal={Econometrica: Journal of the econometric society},
  pages={1251--1271},
  year={1978},
  publisher={JSTOR}
}

@article{foster2023orthogonal,
  title={Orthogonal statistical learning},
  author={Foster, Dylan J and Syrgkanis, Vasilis},
  journal={The Annals of Statistics},
  volume={51},
  number={3},
  pages={879--908},
  year={2023},
  publisher={Institute of Mathematical Statistics}
}

\setcounter{page}{1}
\appendix
\begin{appendices}
\section{Supplementary Material}
\DoToC
\subsection*{Supplementary Material Outline}

We present the proofs of the theorems and lemmas in the order in which they appear in the main text. In Sections \ref{a:iden} and \ref{a:semi}, we provide the proofs of \Cref{thm1}, \Cref{thm2}, and \Cref{th:asym}, respectively. Following the proof of \Cref{th:asym}, in Section \ref{a:fw_discussion}, we elaborate on the FW counterfactual regression approach, providing additional details and contrasting the theoretical properties of $\hat{\psi}^{EIF-FW}$ and $\hat{\psi}^{EIF}$. Finally, in Sections \ref{a:ortho1} and \ref{a:ortho2}, we provide the proofs of the lemmas presented in \Cref{sec:compare}.

\newpage

\allowdisplaybreaks
\newpage
\subsection{Proof of \Cref{thm1} -- Identification of the ATT}
\label{a:iden}
To show $\psi^0(X) = -\delta(X)$, we first note that the MIV assumption \ref{as:miv} introduces the independence condition $U \indep Z|A=1,X$:
\begin{align*}
	&f(U|A=1,Z,X) 
	\stackrel{Bayes' \  Rule}{=}
	\frac{pr(A=1|U,Z,X)f(U|Z,X)}{pr(A=1|Z,X)} \stackrel{A2}{=}\frac{pr(A=1|U,Z,X)f(U|X)}{pr(A=1|Z,X)} \\
	&=
	\frac{pr(A=1|U,Z,X)f(U|X)}{\int pr(A=1|U,Z,X) dU} 
	\stackrel{A4}{=}
	\frac{exp\{\alpha_1(Z,X)+\alpha_2(U,X)\}f(U|X)}{\int exp\{\alpha_1(Z,X)+\alpha_2(U,X)\} dU} 
	=\frac{exp\{\alpha_2(U,X)\}f(U|X)}{\int exp\{\alpha_2(U,X)\} dU}  \ \textit{independent of Z.}
\end{align*}
Then, we have
\begin{align*}
	&\delta(X) \\
	&= \frac{E\{(1-A)Y|Z=1,X\} - E\{(1-A)Y|Z=0,X\}}{pr(A=1|Z=1,X) - pr(A=1|Z=0,X)} \\
	&= \frac{E_U\bc{E\{(1-A)Y|U,Z=1\} - E\{(1-A)Y|U,Z=0\}}}{pr(A=1|Z=1,X) - pr(A=1|Z=0,X)} \\	
	&= \frac{E_U\bb{E(Y^0|U,A=0,Z=1)pr(A=0|U,Z=1,X) - E(Y^0|U,A=0,Z=0)pr(A=0|U,Z=0,X)}}{pr(A=1|Z=1,X) - pr(A=1|Z=0,X)} \\
	&\stackrel{A3}{=} \frac{E_U\bb{E(Y^0|U,X)pr(A=0|U,Z=1,X) - E(Y^0|U,X)pr(A=0|U,Z=0,X)\}}}{pr(A=1|Z=1,X) - pr(A=1|Z=0,X)} \\		
	&= -\frac{E_U\bc{E(Y^0|U,X) \bb{\frac{pr(A=1|U,Z=1,X)}{pr(A=1|U,Z=0,X)} - 1}pr(A=1|U,Z=0,X)}}{pr(A=1|Z=1,X) - pr(A=1|Z=0,X)} \\				
	&\stackrel{A4}{=} -\frac{E_U\bc{E(Y^0|U,X) \bb{\frac{pr(A=1|Z=1,X)}{pr(A=1|Z=0,X)} - 1} pr(A=1|U,Z=0,X)}}{pr(A=1|Z=1,X) - pr(A=1|Z=0,X)} \\			
	&= -\bb{\frac{pr(A=1|Z=1,X)}{pr(A=1|Z=0,X)} - 1} \frac{E_U\{E(Y^0|U,X)pr(A=1|U,Z=0,X)\}}{pr(A=1|Z=1,X) - pr(A=1|Z=0,X)} \\					
	&\stackrel{Bayes' \  Rule}{=} - \bb{ \frac{pr(A=1|Z=1,X)}{pr(A=1|Z=0,X)} - 1} 
	\frac{ \int E(Y^0|U,X) f(U|X) \frac{f(U|A=1,Z=0,X)pr(A=1|Z=0,X)}{f(U|Z=0,X)} dU }{pr(A=1|Z=1,X) - pr(A=1|Z=0,X)} \\		
	&\stackrel{A2}{=}  -(pr(A=1|Z=1,X) - pr(A=1|Z=0,X)) 	\frac{ \int E(Y^0|U,X) f(U|X) \frac{f(U|A=1,Z=0,X)}{f(U|X)} dU}{pr(A=1|Z=1,X) - p'r(A=1|Z=0,X)} \\
	&=  -\int E(Y^0|U,X)f(U|A=1,Z=0,X)dU \\	
	&\stackrel{U \indep Z|A=1,X}{=}  -\int E(Y^0|U,X) f(U|A=1,X) dU \\
	&= -  \int E(Y^0|U,X) f(U|A=1,X) dU \\
	&= -  E\ba{Y^0|A=1,X}.
\end{align*}

\subsection{Proof of \Cref{thm2}}
\label{a:semi}
In this section, we first prove part \textbf{(1)} and then part \textbf{(2)} of \Cref{thm2}.

\textbf{\textit{Deriving the EIF}}

We assume regular parametric submodels $\mc{P}_v = \{\mc{P}_v : v \in \mathbb{R}\}$, indexed by a real-valued parameter \(v\), within the nonparametric model \(\mc{M}\) of interest \cite{newey1990semiparametric, tsiatis2006semiparametric}. Let \(P_v\) denote the probability distribution under parameter value \(v\), with \(P = P_{v=0}\) representing the true data-generating distribution. Denote the observed data by \(O = (Y, A, Z, X)\), and let \(f_v\) be the density (or probability mass) function of \(P_v\). We write \(\nabla_v = \frac{\partial}{\partial v}\), and denote the pathwise derivative of \(f_v\) by \(f'_v = \nabla_v f_v\), and likewise for the probability mass function: \(pr'_v = \nabla_v pr_v\). The score function for the submodel is given by \(S_v = f'_v / f_v\) (or \(S_v = pr'_v / pr_v\) in the discrete case). We omit the subscript $v$ when $v=0$, meaning the quantity is defined under the true law $P$. Let $\mathcal{L}_{2}(W) = \{ h(W) | E\{h^2(W) \}< \infty \}$ denote the Hilbert space of square-integrable functions of a random variable $W$, equipped with the inner product $\langle h_1,h_2 \rangle = E\{ h_1(W) h_2(W) \}$. 

The following properties T1, T2 and T3 are used for the following EIF derivation:  
\begin{align*} 
&\text{T1: }  E\{h(X) S(Y|X)\} =0 
\ , \quad \forall h(X) \in \mc{L}_2(X);
\quad 
\textit{(score has expectation 0 and by iterated expectation;)}\\
&\text{T2: }  E[h(X) \{ Y - E(Y|X)\}] =0 \ , \quad \forall h(X) \in \mc{L}_2(X);
\quad 
\textit{(by iterated expectation;)} \\
&\text{T3: }  S(O) = S(Y,A,Z,X) = S(Y|A,Z,X) + S(A|Z,X) + S(Z|X) + S(X) \ . 
\quad 
\textit{(by the chain rule.)}
\end{align*}

We begin deriving the EIF for $-\psi^0 = E\bb{A\delta(X)/pr(A=1)}$. The canonical gradient for $-\psi_v^0 = E_v\bb{A\delta_v(X)/pr_v(A=1)}$ in the parametric submodel $\mc{M}_v$ is given by
\begin{align*}
	&-\frac{\partial}{\partial v} \psi_v^0 
 \\
	&=\frac{\partial}{\partial v} \int\delta_v(X) \frac{A}{pr_v(A=1)} f_v(A,X) d(A,X)
 \\
	&= \int \frac{\partial}{\partial v} \left\{ \delta_v(X) \frac{A}{pr_v(A=1)} f_v(A,X) \right\} d(A,X)\\
	&= \int \delta_v(X) \frac{A}{pr_v(A=1)} \frac{f'_v(A,X)}{f_v(A,X)} f_v(A,X)  d(A,X) + \int  \{\nabla_{v} \delta_v(X) \}\frac{A}{pr_v(A=1)} f_v(A,X)  d(A,X) 
 \notag  
 \\ 
	&\quad \quad \quad +\int \delta_v(X) \left\{-\frac{A}{pr_v(A=1)}\frac{pr'_v(A=1)}{pr_v(A=1)} \right\} f_v(A,X)  d(A,X) 
 \\
	&= 
	\underbrace{E_v\left\{\delta_v(X) \frac{A}{pr_v(A=1)} S_v(A,X)\right\}}_{\text{Part 1}} + 
	\underbrace{E_v\left\{ (\nabla_{v}\delta_v(X)) \frac{A}{pr_v(A=1)} \right\}}_{\text{Part 2}} - 
	\underbrace{E_v\left\{\delta_v(X) \frac{A}{pr_v(A=1)} S_v(A=1)\right\}}_{\text{Part 3}} 
\end{align*}
The second-to-last equation above holds by the chain rule of differentiation.  We find alternative representations of Parts 1-3 as follows: 

Part 1:

By T1, one has
\begin{align*}
	& E_v\left\{\delta_v(X) \frac{A}{pr_v(A=1)} S_v(A,X)\right\} = 
	E_v\left\{\delta_v(X) \frac{A}{pr_v(A=1)} S_v(Y,A,Z,X)\right\} = 
	E_v\left\{\delta_v(X) \frac{A}{pr_v(A=1)} S_v(O)\right\}.
\end{align*}

Part 3:
\begin{align*}
	&E_v\left\{\delta_v(X) \frac{A}{pr_v(A=1)} S_v(A=1)\right\} \\
	&= \int \delta_v(X)  \frac{A}{pr_v(A=1)} S_v(A=1) f_v(A,X) d(A,X) \\
	&= \int \delta_v(X)  \frac{A}{pr_v(A=1)} S_v(A=1) f_v(A=1,X) d(A,X) \\		
	&= \int \delta_v(X)  S_v(A=1) f_v(X|A=1) d(A,X) \\
	&= S_v(A=1) E_{v}\{\delta_v(X)|A=1\}\\
	&= -S_v(A=1) \psi_v^0\\
	&= -E_v\left\{\psi_v^0 \frac{A}{pr_v(A=1)} S_v(A)\right\} \\
	&= -E_v\left\{\psi_v^{0} \frac{A}{pr_v(A=1)} S_v(Y,A,Z,X)\right\} \\
	&= -E_v\left\{\psi_v^{0} \frac{A}{pr_v(A=1)} S_v(O)\right\}
\end{align*}
The third-to-last equality follows by noting that $E_v\left\{A/pr_v(A=1) S_v(A)\right\}=S_v(A=1)$, and the second-to-last equality holds by T1.

Part 2:
\begin{align*}
	&E_v\left[\{\nabla_{v}\delta_v(X)\} \frac{A}{pr_v(A=1)} \right] \\
	&=\int \{\nabla_{v}\delta_v(X)\} \frac{A}{pr_v(A=1)} f_v(X,A) d(X,A)
\end{align*}
Note that $\nabla_{v}\delta_v(X)$ can be expressed as follows by the chain rule:
\begin{align}
	&\nabla_{v}\delta_v(X) 
	= \frac{
	\{\nabla_{v} \delta_{Yv}(X)\} \delta_{Av}(X) - \delta_{Yv}(X) 
	\{\nabla_{v}\delta_{Av}(X)\} ]
	}
	{\{\delta_{Av}(X)\}^2},
    \label{eq-delta}
\end{align}
where
\begin{align*}
	&\delta_{Yv}(X) := E_v\{(1-A)Y|Z=1,X\} - E_v\{(1-A)Y|Z=0,X\}, \\
	&\delta_{Av}(X) := pr_v(A=1|Z=1,X)-pr_v(A=1|Z=0,X).
\end{align*}
Hence, it suffices to find expressions of $\nabla_{v}\delta_{Yv}(X) = \nabla_{v}E_v\{(1-A)Y|Z=1,X\} - \nabla_{v}E_v\{(1-A)Y|Z=0,X\}$ and $\nabla_{v}\delta_{Av}(X) = \nabla_{v}pr_v(A-1|Z=1,X) - \nabla_{v}pr_v(A=1|Z=0,X)$. We first derive an expression of $\nabla_{v}E_v\{(1-A)Y|Z=1,X\}$:
\begin{align*}
	&\nabla_{v} E_v\{(1-A)Y|Z=z,X\} \\
	&= \frac{\partial}{\partial v} \int Y (1-A)f_v(Y,A|Z=z,X)d(Y,A) \\
	&= \int Y (1-A)\frac{f'_v(Y,A|Z=z,X)}{f_v(Y,A|Z=z,X)} f_v(Y,A|Z=z,X) d(Y,A)\\ 
	&=E_v\{Y(1-A)S_v(Y,A|Z=z,X)|Z=z,X\} \\
	&=E_v 
	\bc{ \bc{Y(1-A) - E_v\{Y(1-A)|Z=z,X\}}
	S_v(Y,A|Z=z,X)
	|Z=z,X }
\end{align*}
The last equality holds by T2.

$\nabla_{v}\delta_{Yv}(X)$ has the following representation:
\begin{align}	
	& \nabla_{v} \delta_{Yv}(X)
    \nonumber
    \\
	&= \nabla_{v}E_v\{(1-A)Y|Z=1,X\} - \nabla_{v}E_v\{(1-A)Y|Z=0,X\}
    \nonumber
    \\
	&= E_v 
	\Big[ [Y(1-A) - E_v\{Y(1-A)|Z=1,X\}] 
	S_v(Y,A|Z=1,X)
	|Z=1,X \Big]
	  \nonumber
    \\
	& \qq - E_v 
	\Big[ [Y(1-A) - E_v\{Y(1-A)|Z=0,X\}] 
	S_v(Y,A|Z=0,X)
	|Z=0,X \Big]   \nonumber
    \\
	&= 
	E_v \bc{\frac{2Z-1}{f_v(Z|X)} [Y(1-A) - E_v\{Y(1-A)|Z,X\} ]S_v(Y,A|Z,X) \Big|X } \nonumber
    \\
	&= 
	E_v \bc{\frac{2Z-1}{f_v(Z|X)} [Y(1-A) - E_v\{Y(1-A)|Z,X\} ]S_v(Y,A,Z|X) \Big|X }.
    \label{eq-deltaY}
\end{align}
The last equality holds by T1. 

Based on analogous algebra, one can show that:
\begin{align}
	&\nabla_{v} \delta_{Av}(X) = E_v\bc{\frac{2Z-1}{f_v(Z|X)} \{A - pr_v(A=1|Z,X)\}S_v(Y,A,Z|X) \Big|X }
    \label{eq-deltaA}
\end{align}

Combining \eqref{eq-delta}-\eqref{eq-deltaA}, we have
\begin{align*}
	&\nabla_{v}\delta_v(X) \\
	&= \frac{1}{\delta_{Av}(X)^2}\Big[
	\{\nabla_{v} \delta_{Yv}(X)\} \delta_{Av}(X) - \delta_{Yv}(X) \{\nabla_{v} \delta_{Av}(X)\}	
	\Big]\\
	&= \frac{1}{\delta_{Av}(X)^2}
	\left[ 
    \begin{array}{l} 
    E_v\bc{\frac{2Z-1}{f_v(Z|X)} [Y(1-A) - E_v\{Y(1-A)|Z,X\}]S_v(Y,A,Z|X) \Big|X }
	\delta_{Av}(X)   \\ 
	  \hspace*{1cm} -  \delta_{Yv}(X) E_v \bc{\frac{2Z-1}{f(Z|X)} \{A - pr_v(A=1|Z,X)\}S_v(Y,A,Z|X)\} \Big|X} 
    \end{array}\right]\\
	&=E_v\left[ 
    \left.
    \begin{array}{l}
    \frac{1}{\delta_{Av}(X)^2}\frac{2Z-1}{f_v(Z|X)} [Y(1-A) - E_v\{Y(1-A)|Z,X\} ]S_v(Y,A,Z|X)  \delta_{Av}(X) 
    \\
    \hspace*{1cm}  - \frac{2Z-1}{f(Z|X)} \{A - pr_v(A=1|Z,X)\}S_v(Y,A,Z|X) \delta_{Yv}(X) \}
    \end{array}
    \right| \, X \right] \\
	&=E_v\left[ \frac{2Z-1}{f_v(Z|X)} \frac{1}{\delta_{Av}(X)}[Y(1-A) - E_v\{Y(1-A)|Z,X\}   -  \{A - pr_v(A=1|Z,X)\} \delta_{v}(X)] S_v(Y,A,Z|X) \Big|X \right].
\end{align*}

Based on this result, Part 2 is expressed as follows:
\begin{align*}
	&E_v \Big[ \{\nabla_{v}\delta_v(X)\} \frac{A}{pr_v(A=1)} \Big] \\
	&=\int \{\nabla_{v}\delta_v(X)\} f_v(X|A=1)dX  \\
	&=\int  \frac{ pr_v(A=1|X)}{pr_v(A=1)}  \{\nabla_{v}\delta_v(X)\} f_v(X) dX \\
	&=\int \int \frac{ pr_v(A=1|X)}{pr_v(A=1)} \frac{2Z-1}{f_v(Z|X)} \frac{1}{\delta_{Av}(X)}[Y(1-A) - E_v\{Y(1-A)|Z,X\}   -  \{A - pr_v(A=1|Z,X)\} \delta_{v}(X) ] \\ 
	&\qq \qq  \times S_v(Y,A,Z|X) f_v(Y,A,Z|X) d(Y,A,Z) \ f_v(X) dX\\
	&=\int \frac{ pr_v(A=1|X)}{pr_v(A=1)}  \frac{2Z-1}{f_v(Z|X)} \frac{1}{\delta_{Av}(X)}[Y(1-A) - E_v\{Y(1-A)|Z,X\}   -  \{A - pr_v(A=1|Z,X)\} \delta_{v}(X) ] \\ 
	&\qq \qq \times S_v(Y,A,Z,X)  f_v(Y,A,Z,X) d(Y,A,Z,X) 
    \\	
	&=\int \frac{pr_v(A=1|X)}{pr_v(A=1)} 
	 \frac{2Z-1}{f_v(Z|X)} \frac{1}{\delta_{Av}(X)}[Y(1-A) - E_v\{Y(1-A)|Z,X\}   -  \{A - pr_v(A=1|Z,X)\} \delta_{v}(X) ]  S_v(O) f_v(O) dO \\ 	
	 	&=E_v\left[
	 	\frac{pr_v(A=1|X)}{pr_v(A=1)}  \frac{2Z-1}{f_v(Z|X)} \frac{1}{\delta_{Av}(X)}
	 	[
	 	Y(1-A) - E_v\{Y(1-A)|Z,X\} - \{A - pr_v(A=1|Z,X)\}\delta_v(X)
	 	]
	 	S_v(O)\right]
\end{align*}
The third-to-last equality holds by T2.

Define $\theta_v(O)$ as
$$\theta_v(O) := \frac{pr_v(A=1|X)}{pr_v(A=1)}  \frac{2Z-1}{f_v(Z|X)} \frac{1}{\delta_{Av}(X)}
	 	[
	 	Y(1-A) - E_v\{Y(1-A)|Z,X\} - \{A - pr_v(A=1|Z,X)\}\delta_v(X)
	 	].
	 	$$
Combining the derived expressions for Part 1, Part 2 and Part 3, we establish the following result:
\begin{align*}
	-&\frac{\partial}{\partial v} \psi_v^0 \\
	&= \text{Part 1 + Part 2 - Part 3} \\
	&= E_v\left[\delta_v(X) \frac{A}{pr_v(A=1)} S_v(O) + \theta_v(O)  S_v(O) +\psi_v^0 \frac{A}{pr_v(A=1)} S_v(O) \right] \\
	&= E_v\left[ \bc{ \frac{A}{pr_v(A=1)} \{ \delta_v(X) + \psi_v^0\} + \theta_v(O) } S_v(O)\right]
\end{align*}
This result implies that $\psi^0$ is pathwise differentiable \cite{newey1990semiparametric} with an influence function $EIF(O;-\psi^0)$ defined as follows:
\begin{align*}
	&EIF(O;-\psi^0) = \frac{A}{pr(A=1)} \{ \delta(X) + \psi^0\} + \theta(O) , \\
	&\theta(O) := \frac{pr(A=1|X)}{pr(A=1)}  \frac{2Z-1}{f(Z|X)} \frac{1}{\delta_A(X)}
	\bc{ Y(1-A) - E\bb{Y(1-A)|Z,X} - \bb{A - pr(A=1|Z,X)} \delta(X)}.
\end{align*}

It is well known that an influence function for $\psi^1$ is given by:
\begin{align*}
	&EIF(O;\psi^1) = \frac{A}{pr(A=1)} \bb{Y(1-A)-\psi^1}.
\end{align*}
Combining these influence functions for $\psi^0$ and $-\psi^0$, an influence for $\psi$ is given by 
\begin{align*}
	&EIF(O;\psi) \\
	&=  \frac{A}{pr(A=1)} \{ Y+\delta(X) - (\psi^1-\psi^0) \} + \theta(O) \\
	&=  \frac{A}{pr(A=1)} \{ Y+\delta(X) - \psi \} + \theta(O) \numberthis
    \label{eq-EIF}.
\end{align*}
Of note, $EIF(O;\psi)$ derived above satisfies the following equation:
\begin{equation*}
 \frac{\partial}{\partial v} \psi_v \bigg|_{v=0}
 = E\bb{EIF(O;\psi) S(O)}.
\end{equation*}
In other words, $\psi$ is pathwise differentiable 
\cite{newey1990semiparametric}, with an influence function $EIF(O;\psi)$. Since the model $\mathcal{M}$ is nonparametric, $EIF(O;\psi)$ is the unique influence function satisfying \eqref{eq-EIF}. Consequently, $EIF(O;\psi)$ is the EIF for $\psi$. This completes the proof.

\textbf{\textit{Multiple Robustness Property}}

We verify that the EIF is an unbiased moment equation under each of the models $\mc{M}_1$, $\mc{M}_2$ and $\mc{M}_3$. Again, we let the superscript $*$ indicate a misspecified model and use the reparameterization that 
\begin{align*}
	&e_Z(X) = e_0(X) + \bb{e_1(X)-e_0(X)}Z, \\
	&p_Z(X) = p_0(X) + \bb{p_1(X)-p_0(X)}Z,\\
	&Y(1-A) - e_Z(X) - \bb{A - p_Z(X)}\delta(X)\\
	&= Y(1-A) - e_0(X) - \bb{e_1(X)-e_0(X)}Z - \bc{A - p_0(X) - \bb{p_1(X)-p_0(X)}Z}\delta(X) \\
	&= Y(1-A) - e_0(X) - \bb{A - p_0(X)} \delta(X)  - \bb{e_1(X)-e_0(X)}Z  + \bb{p_1(X)-p_0(X)}\delta(X)Z \\
	&= Y(1-A) - e_0(X) - \bb{A - p_0(X)} \delta(X)  - \bb{e_1(X)-e_0(X)}Z  + \bb{e_1(X)-e_0(X)}Z \\	
	&= Y(1-A) - e_0(X) - \bb{A - p_0(X)} \delta(X).
\end{align*}
$\textit{Under $\mc{M}_1$, models for } p_0(X), e_0(X) \textit{ and } \delta(X)  \textit{ are evaluated at their true value}$:
\begin{align*}
	& E\{EIF(O;\psi,e_1^*(X),e_0(X),p_1^*(X),p_0(X),\pi_z^*(X),\delta(X))\} \\
	&=
	E\bb{
	\frac{A}{pr(A=1)}\ba{Y + \delta(X) -\psi}
	} +
	E\bc{
	\frac{\rho^*(X)}{pr(A=1)} \frac{2Z-1}{\pi^*_Z(X)}c\Omega^*(X)^{-1} \bc{Y(1-A)-e_0(X) - \bb{A-p_0(X)}\delta(X)}
	}
	\\
	&=
	\frac{pr(A=1)}{pr(A=1)}\ba{\psi^1 + \psi^0 -\psi}
	 +
	E\bc{
	\frac{\rho^*(X)}{pr(A=1)} \frac{2Z-1}{\pi^*_Z(X)}\Omega^*(X)^{-1} \bc{Y(1-A)-e_0(X) - \bb{A-p_0(X)}\delta(X)}
	}	
	\\
	&=
	E\bc{
	\frac{\rho^*(X)}{pr(A=1)} \frac{2Z-1}{\pi^*_Z(X)}\Omega^*(X)^{-1}  \bc{ Y(1-A)-e_0(X) - \bb{A-p_0(X)}\delta(X)}
	}
	\\
	&=	E\bc{
	\frac{\rho^*(X)}{pr(A=1)} \frac{\pi_1(X)}{\pi^*_1(X)} \Omega^*(X)^{-1} \bc{ e_1(X)-e_0(X) - \bb{p_1(X)-p_0(X)}\delta(X)} 
	}
	\\
	&\qq
	-
	E\bc{
	\frac{\rho^*(X)}{pr(A=1)} \frac{\pi_0(X)}{\pi^*_0(X)}\Omega^*(X)^{-1}  \bc{ e_0(X)-e_0(X) - \bb{p_0-p_0(X)}\delta(X)}
	}	\\
	&=	E\bc{
	\frac{\rho^*(X)}{pr(A=1)} \frac{\pi_1(X)}{\pi^*_1(X)} \Omega^*(X)^{-1}  \bc{ e_1(X)-e_0(X) - \bb{e_1(X)-e_0(X)}} - 0	}	=0.
\end{align*}
$\textit{Under $\mc{M}_2$, models for } p_z(X) \textit{ and }  \pi_z(X) \ z\in \bb{0,1} \textit{ are evaluated at their true value:}$
\begin{align*}
	&E\{EIF(O;\psi,e_1^*(X),e_0^*(X),p_1(X),p_0(X),\pi_z(X),\delta^*(X))\} \\
	&=
	E\bb{
	\frac{A}{pr(A=1)}\ba{Y + \delta^*(X) -\psi}
	} +
	E\bc{
	\frac{\rho(X)}{pr(A=1)} \frac{2Z-1}{\pi_Z(X)} \Omega(X)^{-1}\bc{Y(1-A)-e_0^*(X) - \bb{A-p_0^*(X)}\delta^*(X)}
	}
	\\
	&= \psi^1 - \psi + E\bb{ \frac{\rho(X)}{pr(A=1)} \delta^*(X)} + E\bc{
	\frac{\rho(X)}{pr(A=1)} \frac{\pi_1(X)}{\pi_1(X)}\Omega(X)^{-1} \bc{e_1(X)-e_0^*(X) - \bb{p_1(X)-p_0^*(X)}\delta^*(X)}
	}
	\\
	&\qq 	-E\bc{
	\frac{\rho(X)}{pr(A=1)} \frac{\pi_0(X)}{\pi_0(X)}\Omega(X)^{-1} \bc{e_0(X)-e_0^*(X) - \bb{p_0(X)-p_0^*(X)}\delta^*(X)}
	}
	\\
	&=\psi^1 - \psi 
	+
	E\bc{
	\frac{\rho(X)}{pr(A=1)} \bc{
	 \delta^*(X) +
	 \Omega(X)^{-1} \bc{e_1(X)-e_0(X) - \bb{p_1(X)-p_0(X)}\delta^*(X)}
	}
	}	
	\\
	&=\psi^1 - \psi 
	+
	E\bc{
	\frac{\rho(X)}{pr(A=1)} \bc{
	 \delta^*(X) +
	 \bb{\frac{e_1(X)-e_0(X)}{p_1(X) -p_0(X)} - \delta^*(X)}
	}
	}		
	\\
	&=\psi^1 - \psi 
	+
	E\bb{
	\frac{\rho(X)}{pr(A=1)}  
	 \delta(X)
	}	= \psi^1 - \psi^0 -\psi = 0		
	.
\end{align*}
$\textit{Under $\mc{M}_3$, models for } \delta(X) \textit{ and }  \pi_z(X) \ z\in \bb{0,1} \textit{ are evaluated at their true value:} $
\begin{align*}
	&E\{EIF(O;\psi,e_1^*(X),e_0^*(X),p_1^*(X),p_0^*(X),\pi_z(X),\delta(X))\}
	\\
	&=
	E\bb{
	\frac{A}{pr(A=1)}\ba{Y + \delta(X) -\psi}
	} +
	E\bc{
	\frac{\rho^*(X)}{pr(A=1)} \frac{2Z-1}{\pi_Z(X)}\Omega(X)^{-1} \bc{Y(1-A)-e_0^*(X) - \bb{A-p_0^*(X)}\delta(X)}
	} \\
	&=
	0
	+
	E\bc{
	\frac{\rho^*(X)}{pr(A=1)} \frac{\pi_1(X)}{\pi_1(X)}  \Omega(X)^{-1}\bc{ e_1(X)-e_0^*(X) - \bb{p_1(X)-p_0^*(X)}\delta(X)} 
	}
	\\
	&\qq -
	E\bc{
	\frac{\rho^*(X)}{pr(A=1)} \frac{\pi_0(X)}{\pi_0(X)}  \Omega(X)^{-1} \bc{ e_0-e_0^*(X) - \bb{p_0-p_0^*(X)}\delta(X)}
	}
	\\
	&=
	E\bc{
	\frac{\rho^*(X)}{pr(A=1)} \Omega(X)^{-1}  \bc{ e_1(X)-e_0^*(X) - \bb{p_1(X)-p_0^*(X)}\delta(X) - e_0(X)+ e_0^*(X) + \bb{p_0-p_0^*(X)}\delta(X)}
	}	
	\\
	&=
	E\bc{
	\frac{\rho^*(X)}{pr(A=1)} \Omega(X)^{-1}  \bc{ e_1(X)-e_0(X)-  \bb{p_1(X)-p_0(X)}\delta(X)  }
	}=0.
\end{align*}
\subsection{Proof of \Cref{th:asym} -- Asymptotic Normality of the Proposed Estimator}
In this section, we prove \Cref{th:asym} in the main text. For simplicity, we use the shorthand notation $f = f(X)$ when no ambiguity arises ($\psi$ exclusively used to denote the marginal quantity), which applies to functions such as \( p_z = p_z(X) \), \( e_z = e_z(X) \), \( \pi_z = \pi_z(X) \), \( \rho = \rho(X) \), \( \delta = \delta(X) \), and \( \Omega = \Omega(X)\).

If one can show that $\hat{\psi}^{(k)}$ has the asymptotic representation 
\begin{align*}
	\sqrt{|\mc{I}_k|} \ba{\hat{\psi}^{(k)} - \psi} = \frac{1}{\sqrt{|\mc{I}_k|}} \sum_{i \in \mc{I}_k} EIF(O_i;\psi) + o_P(1), 
	\numberthis \label{e-single}
\end{align*}
and one further establishes $\hat{\psi}^{EIF-FW} = K^{-1} \sum_{k=1}^K \hat{\psi}^{(k)}$ has asymptotic representation
\begin{align*}
	\sqrt{N} \ba{\hat{\psi}^{EIF-FW} - \psi} = \frac{1}{\sqrt{N}} \sum_{i=1}^N EIF(O_i;\psi) + o_P(1). 
	\numberthis \label{e-merged}
\end{align*}
Then, the asymptotic normality result holds from a standard central limit theorem. Hence, we begin by proving that \eqref{e-single} holds under assumptions \ref{as:iv1} -- \ref{as:rates}.

Recall 
\begin{align*}
	&\hat{\psi}^{(k)} = \bb{\mb{P}(A)}^{-1}
	\mb{P}_{\mc{I}_k}
	\bc{
	 A\{Y+\hat{\delta}^{(-k)}(X)\} + \hat{\theta}^{(-k)}(X)
     }, \\
	& 
	\hat{\theta}^{(-k)}(O) = \hat{\rho}^{(-k)}(X) \frac{2Z-1}{ \hat{\pi}^{(-k)}_Z(X)}
	\widehat{\Omega}^{(-k)}(X) 
	\bc{
	Y(1-A)-\hat{e}_Z(X)^{(-k)}  - \bb{A-\hat{p}_Z^{(-k)}(X)}\hat{\delta}^{(-k)}(X) 
	}, \\
	&\hat{\rho}^{(-k)}(X)  = \hat{p}_1^{(-k)}(X) \hat{\pi}_1^{(-k)}(X) + \hat{p}_0^{(-k)} (X)\hat{\pi}_0^{(-k)}(X).
\end{align*}

We proceed with the rewritten $\theta$
\begin{align*}
	&\theta  =  \frac{\rho}{pr(A=1)} \frac{2Z-1}{\pi_Z}  \Omega\{ Y(1-A) - e_Z - (A-p_Z)\delta \} =  \frac{\rho}{pr(A=1)}\frac{2Z-1}{\pi_Z} \Omega \{ Y(1-A) - e_0 - (A-p_0)\delta \}
\end{align*}
as
\begin{align*}
	&e_Z = e_0 + (e_1 - e_0)Z \qq p_Z = p_0 + (p_1 - p_0)Z \qq p_Z \delta = \{ p_0 + (p_1 - p_0)Z \} \delta = p_0 \delta + (e_1-e_0)Z \\
	&Y(1-A) - e_Z - (A-p_Z)\delta = Y(1-A) - e_0 - (e_1 - e_0)Z - A\delta + p_0 \delta + (e_1-e_0)Z = Y(1-A) - e_0 - (A-p_0)\delta.
\end{align*}

Let $M_i=  A_i\bb{Y_i+\delta_i} + \theta_i$ and $\widehat{M}_i^{(-k)}=  A_i\bb{Y_i+\hat{\delta}^{(-k)}_i} + \hat{\theta}^{(-k)}_i$. The left-hand side of \eqref{e-single} is given by
\begin{align*}
	&\sqrt{|\mc{I}_k|} \ba{\hat{\psi}^{(k)} - \psi}  \\
	&=\frac{1}{\sqrt{|\mc{I}_k|}} 
	\sum_{i \in \mc{I}_k}
	\bb{\frac{\widehat{M}_i^{(-k)}}{\mb{P}(A)} - \frac{A_i\psi}{\mb{P}_{\mc{I}_k}(A)}} \\
	&=\frac{1}{\sqrt{|\mc{I}_k|}} 
	\sum_{i \in \mc{I}_k}
	\bb{\frac{pr(A=1)}{\mb{P}(A)}\frac{\widehat{M}_i^{(-k)}}{pr(A=1)} - \frac{pr(A=1)}{\mb{P}_{\mc{I}_k}(A)}\frac{A_i\psi}{pr(A=1)}} 
	\\	
	&=\frac{1}{2}\frac{1}{\sqrt{|\mc{I}_k|}} 
	\underbrace{	
	\bb{
	\frac{pr(A=1)}{\mb{P}(A)} - \frac{pr(A=1)}{\mb{P}_{\mc{I}_k}(A)}
	}
	}_{o_p(1)} 
	\underbrace{	
	\sum_{i \in \mc{I}_k} \frac{\widehat{M}_i^{(-k)} + A_i \psi}{pr(A=1)}
	}_{O_p(1)} 	
	+
	\frac{1}{2}\frac{1}{\sqrt{|\mc{I}_k|}}
	\underbrace{		
	\bb{
	\frac{pr(A=1)}{\mb{P}(A)} + \frac{pr(A=1)}{\mb{P}_{\mc{I}_k}(A)}
	}
	}_{2+o_p(1)} 	
	\underbrace{		
	\sum_{i \in \mc{I}_k} \frac{\widehat{M}_i^{(-k)} - A_i \psi}{pr(A=1)}	
	}_{AN: O_p(1)} 		
	\\
	&=\frac{1}{\sqrt{|\mc{I}_k|}} \sum_{i \in \mc{I}_k}\bb{\frac{\widehat{M}_i^{(-k)}-A_i\psi}{pr(A=1)}} + o_p(1)
\end{align*}
The fourth line holds because $\mb{P}_{\mc{I}_k}(A) = pr(A=1)+o_P(1)$ and $\mb{P}(A) = pr(A=1)+o_P(1)$ by the law of large number and $\widehat{M}_i^{(-k)}$ is bounded (each component of $\widehat{M}_i^{(-k)}$ is bounded based on A5).

Let \( \mb{G}_{\mc{I}_k}(V) = |\mc{I}_k|^{-1/2} \sum_{i \in \mc{I}_k} \left\{ V_i - E(V_i) \right\} \) be the empirical process of \( V_i \) centered by \( E(V_i) \).  Similarly, let \( \mb{G}^{(-k)}_{\mc{I}_k}(\widehat{V}^{(-k)}) = |\mc{I}_k|^{-1/2} \sum_{i \in \mc{I}_k} \bb{ \widehat{V}^{(-k)}_i - E^{(-k)}\ba{\widehat{V}^{(-k)}} } \) be the empirical process of \( \widehat{V}^{(-k)} \) centered by \( E^{(-k)}\ba{\widehat{V}^{(-k)}} \),  where \( E^{(-k)}(\cdot) \) is the expectation after considering random functions obtained from \( \mc{I}_k^c \) as fixed functions.  

The empirical process of \( \widehat{M}_i^{(-k)}- A_i\psi \) is
\begin{align}
	&
	|\mc{I}_k|^{-1/2} 
	\sum_{i \in \mc{I}_k}
	\bb{
	\widehat{M}_i^{(-k)}-A_i\psi
	}
	\notag
	\\
	&=
	\mb{G}_{\mc{I}_k}(M_i - A\psi) 
	\label{b-term1}
	\\
	&\quad + 
	 |\mc{I}_k|^{1/2} E^{(-k)} \ba{\widehat{M}_i^{(-k)} - M}
	 \label{b-term2}	
	\\
	&\quad + \mb{G}^{(-k)}_{\mc{I}_k} \ba{ \widehat{M}_i^{(-k)} - M} \label{b-term3}		
\end{align}
From the two subsections below, we show that (\ref{b-term2}) and (\ref{b-term3}) are $o_P(1)$, indicating that (\ref{b-term1}) is asymptotically normal, and consequently, $(AN)$ is $O_p(1)$. Moreover, we establish (\ref{e-single}), concluding the proof as follows:
\begin{align*}
	\sqrt{|\mc{I}_k|} (\hat{\psi}^{(k)} - \psi) = \frac{1}{\sqrt{|\mc{I}_k|}} \sum_{i \in \mc{I}_k}\bb{\frac{\widehat{M}_i^{(-k)}-A_i\psi}{pr(A=1)}} + o_p(1) 
	= \frac{1}{\sqrt{|\mc{I}_k|}} \sum_{i \in \mc{I}_k} EIF(O_i;\psi) + o_p(1).
\end{align*}

\textbf{\textit{Asymptotic Property of (\ref{b-term2})}}
\\

The term (\ref{b-term2})
\begin{align*}
	&|\mc{I}_k|^{1/2} E^{(-k)} \ba{\widehat{M}^{(-k)} - M} \\
	&= 
	|\mc{I}_k|^{1/2} E^{(-k)} \bb{
	A\ba{Y+\hat{\delta}^{(-k)}} + \hat{\theta}^{(-k)} -  A\ba{Y+\delta} - \theta
	}
	\\
	&= 
	|\mc{I}_k|^{1/2} E^{(-k)} \bb{
	A\ba{\hat{\delta}^{(-k)} - \delta} + \hat{\theta}^{(-k)} -  \theta
	}	
	\\
	&= 
	|\mc{I}_k|^{1/2} E^{(-k)} \bb{
	A\ba{\hat{\delta}^{(-k)} - \delta} + \hat{\theta}^{(-k)} -  \theta
	}.
\end{align*}

We note $E\ba{\theta} =0$ and $A$ in $E\bb{A \ba{\hat{\delta}-\delta}}$ contributes $\rho=pr(A=1|X)$.
\begin{align*}
	&E\ba{\theta} = E\bc{ 
	\rho \frac{2Z-1}{\pi_Z} \Omega \bb{
	Y(1-A) - e_0 - (A-p_0) \delta
	}
	}
	=
	E\bc{ 
	\rho \frac{2Z-1}{\pi_Z} \Omega E\bc{\bb{
	Y(1-A) - e_0 - (A-p_0) 
	}
	|X
	}
	\delta
	} \\
	&= 	E\bc{ 
	\rho \delta \frac{\pi_1}{\pi_1} \Omega 
	\underbrace{
	E\bc{\bb{
	e_1- e_0 - (p_1-p_0) \delta}
	|X
	}
	}_{=0}	
	}
	+
	E\bc{ 
	\rho \delta \frac{\pi_0}{\pi_1} \Omega 
	\underbrace{	
	E\bc{\bb{
	e_0- e_0 - (p_0-p_0) \delta}
	|X
	}
	}_{=0}		
	} =0 \\ 
	\\
	&E\bb{A \ba{\hat{\delta}-\delta}} = E\bb{pr(A=1|X)\ba{\hat{\delta}-\delta}} = E\bb{\rho\ba{\hat{\delta}-\delta}}.
\end{align*}

Then, the term (\ref{b-term2}) becomes
\begin{align*}
	&|\mc{I}_k|^{1/2} E^{(-k)} \ba{\widehat{M}^{(-k)} - M} \\
	&= 
	|\mc{I}_k|^{1/2} E^{(-k)} \bb{
	\rho
	\ba{\hat{\delta}^{(-k)} - \delta} + \hat{\theta}^{(-k)} 
	}\\
	&= 
	|\mc{I}_k|^{1/2} E^{(-k)} \bc{
	\rho
	\ba{\hat{\delta}^{(-k)} - \delta} + 
	\hat{\rho}^{(-k)} \frac{2Z-1}{\hat{\pi}_Z^{(-k)}}
	\widehat{\Omega}^{(-k)}
	\bb{
	Y(1-A) - \hat{e}_0^{(-k)} - \ba{A-\hat{p}_0^{(-k)}} \hat{\delta}^{(-k)}
	}
	}\\
	&= 
	|\mc{I}_k|^{1/2} E^{(-k)} 
	\bc{
	\rho
	\ba{\hat{\delta}^{(-k)} - \delta} + 
	\hat{\rho}^{(-k)} \frac{2Z-1}{\hat{\pi}_Z^{(-k)}}
	\widehat{\Omega}^{(-k)}
	\bb{
	\delta \Omega^{-1}Z + e_0  - \hat{e}_0^{(-k)} +\ba{p_0-\hat{p}_0^{(-k)}} \hat{\delta}^{(-k)} + \Omega^{-1}Z\hat{\delta}^{(-k)}
	}
	}\\
	&= 
	|\mc{I}_k|^{1/2} E^{(-k)} 
	\bc{
	\rho
	\ba{\hat{\delta}^{(-k)} - \delta} + 
	\hat{\rho}^{(-k)} \frac{2Z-1}{\hat{\pi}_Z^{(-k)}}
	\widehat{\Omega}^{(-k)}
	\bb{
	\ba{\delta-\hat{\delta}^{(-k)}}
	\Omega^{-1}Z
	+
	\ba{p_0-\hat{p}_0^{(-k)}} \hat{\delta}^{(-k)}
	+ 
	\ba{e_0 - \hat{e}_0^{(-k)} }
	}
	}
	\\
	&=
	|\mc{I}_k|^{1/2} E^{(-k)} \Biggl[
	\rho
	\ba{\hat{\delta}^{(-k)} - \delta} 
	+
	\hat{\rho}^{(-k)} \frac{\pi_1}{\hat{\pi}_Z^{(-k)}}
	\frac{\widehat{\Omega}^{(-k)}}{\widehat{\Omega}} \ba{\hat{\delta}^{(-k)} - \delta} 
	\\
	&\qq \qq
	+\hat{\rho}^{(-k)} \frac{\pi_1}{\hat{\pi}_Z^{(-k)}}
	\widehat{\Omega}^{(-k)}
	\bb{
	\ba{p_0-\hat{p}_0^{(-k)}} \hat{\delta}^{(-k)}
	+ 
	\ba{e_0 - \hat{e}_0^{(-k)} }
	}
	\\
	&\qq \qq
	-\hat{\rho}^{(-k)} \frac{\pi_0}{\hat{\pi}_Z^{(-k)}}
	\widehat{\Omega}^{(-k)}
		\bb{
	\ba{p_0-\hat{p}_0^{(-k)}} \hat{\delta}^{(-k)}
	+ 
	\ba{e_0 - \hat{e}_0^{(-k)} }
	}
	\Biggl].	
\end{align*}

We handle each term of the above expression:
\begin{align*}
	&|\mc{I}_k|^{1/2} E^{(-k)} \ba{\widehat{M}^{(-k)} - M} = |\mc{I}_k|^{1/2} E^{(-k)} \ba{\textrm{T1} + \textrm{T2} + \textrm{T3} + \textrm{T4}}
	\\
	& \rho = p_1 \pi_1 + p_0 \pi_0 \\
	& \text{T1} = (p_1\pi_1 + p_0\pi_0) \ba{\hat{\delta}^{(-k)} - \delta}  \\
	& \text{T2} = 
	\ba{\hat{p}_1^{(-k)}\hat{\pi}_1^{(-k)} + \hat{p}_0^{(-k)}\hat{\pi}_0^{(-k)}}
	 \frac{\pi_1}{\hat{\pi}_1^{(-k)}} \frac{\widehat{\Omega}^{(-k)}}{\Omega} \ba{ \delta - \hat{\delta}^{(-k)}}\\
	& \text{T3} = 
	\ba{\hat{p}_1^{(-k)}\hat{\pi}_1^{(-k)} + \hat{p}_0^{(-k)}\hat{\pi}_0^{(-k)}}
	\frac{\pi_1}{\hat{\pi}_1^{(-k)}} \widehat{\Omega}^{(-k)} \ba{\hat{p}_0^{(-k)} - p_0}  \hat{\delta} - 
	\ba{\hat{p}_1^{(-k)}\hat{\pi}_1^{(-k)} + \hat{p}_0^{(-k)}\hat{\pi}_0^{(-k)}} \frac{\pi_0}{\hat{\pi}_0^{(-k)}} \widehat{\Omega}^{(-k)}
	\ba{\hat{p}_0^{(-k)} - p_0}  \hat{\delta}^{(-k)} \\ 
	&\text{T4} = 
	\ba{\hat{p}_1^{(-k)}\hat{\pi}_1^{(-k)} + \hat{p}_0^{(-k)}\hat{\pi}_0^{(-k)}}
	\frac{\pi_1}{\hat{\pi}_1^{(-k)}} \widehat{\Omega}^{(-k)} \ba{e_0 -  \hat{e}_0^{(-k)}}  - 
	\ba{\hat{p}_1^{(-k)}\hat{\pi}_1^{(-k)} + \hat{p}_0^{(-k)}\hat{\pi}_0^{(-k)}}
	\frac{\pi_0}{\hat{\pi}_0^{(-k)}} \widehat{\Omega}^{(-k)} \ba{e_0 -  \hat{e}_0^{(-k)}} 
\end{align*}

We focus on T1 first.
\begin{align*}
	&\text{T1} \\
	&= (p_1\pi_1 + p_0\pi_0) \ba{\hat{\delta}^{(-k)} - \delta}  \\
	&= p_1 \ba{\pi_1-\hat{\pi}_1^{(-k)}}\ba{\hat{\delta}^{(-k)} - \delta} + p_0 \ba{\pi_0-\hat{\pi}_0^{(-k)}}\ba{\hat{\delta}^{(-k)} - \delta} \\
	&\quad 	+ p_1 \hat{\pi}_1^{(-k)}\ba{\hat{\delta}^{(-k)} - \delta} + p_0\hat{\pi}_0^{(-k)}\ba{\hat{\delta}^{(-k)} - \delta} \\
	&= p_1 \frac{\rho}{\rho}\ba{\pi_1-\hat{\pi}_1^{(-k)}}\ba{\hat{\delta}^{(-k)} - \delta} 
    + p_0 \frac{\rho}{\rho}\ba{\pi_0-\hat{\pi}_0^{(-k)}}\ba{\hat{\delta}^{(-k)} - \delta}
	\tag{T1-1} \\
	&\quad + \hat{\pi}_1^{(-k)} \frac{\rho}{\rho} \ba{p_1 -\hat{p}_1^{(-k)}} \ba{\hat{\delta}^{(-k)} - \delta} + \frac{\rho}{\rho}\hat{\pi}_0 \ba{p_0-\hat{p}_0^{(-k)} } \ba{\hat{\delta}^{(-k)} - \delta} 	\tag{T1-2} \\
	& \quad + \ba{\hat{p}_1^{(-k)}\hat{\pi}_1^{(-k)}  + \hat{p}_0^{(-k)}\hat{\pi}_0^{(-k)}}  \ba{\hat{\delta}^{(-k)} - \delta}	\tag{T1-3} 
	\\ \\
	& E\ba{\text{T1-1}} \lesssim \norm{\hat{\delta}^{(-k)} - \delta} \cdot  \norm{\hat{\pi}_1^{(-k)}  - \pi_1} + \norm{\hat{\delta}^{(-k)} - \delta}  \cdot  \norm{\hat{\pi}_0^{(-k)} - \pi_0} \\
	& E\ba{\text{T1-2}} \lesssim \norm{\hat{\delta}^{(-k)} - \delta} \cdot  \norm{\hat{p}_1^{(-k)} - p_1} + \norm{\hat{\delta}^{(-k)} - \delta} \cdot  \norm{\hat{p}_0^{(-k)} -p_0}
\end{align*}

We consider T1-3 and T2 together.
\begin{align*}
	& \text{T1-3} +\text{T2} \\
	&= \ba{\hat{p}_1^{(-k)}\hat{\pi}_1^{(-k)}  + \hat{p}_0^{(-k)}\hat{\pi}_0^{(-k)}}  
	\ba{\hat{\delta}^{(-k)} - \delta} +
	\ba{\hat{p}_1^{(-k)}\hat{\pi}_1^{(-k)} + \hat{p}_0^{(-k)}\hat{\pi}_0^{(-k)}}
	 \frac{\pi_1}{\hat{\pi}_1^{(-k)}} \frac{\widehat{\Omega}^{(-k)}}{\Omega} \ba{ \delta - \hat{\delta}^{(-k)}} 
	 \\
	&=\hat{\rho}^{(-k)}\frac{\hat{\pi}_1^{(-k)}\Omega}{\hat{\pi}_1^{(-k)}\Omega} \ba{\hat{\delta}^{(-k)} - \delta}  - \hat{\rho}^{(-k)}\frac{\pi_1 \widehat{\Omega}^{(-k)}}{\hat{\pi}_1^{(-k)}\Omega} \ba{\hat{\delta}^{(-k)} - \delta} \\
	&=\frac{\hat{\rho}^{(-k)}}{\hat{\pi}_1^{(-k)}\Omega}
	\ba{
	\hat{\pi}_1^{(-k)}\Omega - \pi_1\widehat{\Omega}^{(-k)}
	}
	\ba{\hat{\delta}^{(-k)} - \delta} \\
	&=\frac{\hat{\rho}^{(-k)}}{\hat{\pi}_1^{(-k)}\Omega}
	\ba{
	\hat{\pi}_1^{(-k)}\Omega - \hat{\pi}_1^{(-k)}\widehat{\Omega}^{(-k)} +  \hat{\pi}_1^{(-k)}\widehat{\Omega}^{(-k)} - \pi_1\widehat{\Omega}^{(-k)}
	}
	\ba{\hat{\delta}^{(-k)} - \delta} \\	
	&=
	\frac{\hat{\rho}^{(-k)}\widehat{\Omega}^{(-k)}}{\rho\hat{\pi}_1^{(-k)}\Omega} 
        \rho
	\ba{
	\hat{\pi}_1^{(-k)} - \pi_1
	}
	\ba{\hat{\delta}^{(-k)} - \delta}
	\tag{T2-1}	 
	\\	
	&\quad + 
	\frac{\hat{\rho}^{(-k)}\hat{\pi}_1^{(-k)}}{\rho\hat{\pi}_1^{(-k)}\Omega}
        \rho
	\ba{
	\Omega - \widehat{\Omega}^{(-k)} 	}
	\ba{\hat{\delta}^{(-k)} - \delta}		
	\tag{T2-2}
	\\
	& E\ba{\text{T2-1}} \lesssim  \norm{\hat{\delta}^{(-k)} -\delta} \cdot \norm{\hat{\pi}_1^{(-k)}  - \pi_1}
	\\
	& E\ba{\text{T2-2}} \lesssim  \norm{\hat{\delta}^{(-k)} -\delta} \cdot \norm{\widehat{\Omega}^{(-k)} -\Omega}
\end{align*}

Next, we handle T3.
\begin{align*}
&\text{T3} \\ 
&= 	\ba{\hat{p}_1^{(-k)}\hat{\pi}_1^{(-k)} + \hat{p}_0^{(-k)}\hat{\pi}_0^{(-k)}}
	\frac{\pi_1}{\hat{\pi}_1^{(-k)}} \widehat{\Omega}^{(-k)} \ba{\hat{p}_0^{(-k)} - p_0}  \hat{\delta} - 
	\ba{\hat{p}_1^{(-k)}\hat{\pi}_1^{(-k)} + \hat{p}_0^{(-k)}\hat{\pi}_0^{(-k)}} \frac{\pi_0}{\hat{\pi}_0^{(-k)}} \widehat{\Omega}^{(-k)}
	\ba{\hat{p}_0^{(-k)} - p_0}  \hat{\delta}^{(-k)}  \\
&= \hat{\rho}\ba{\frac{\pi_1}{\hat{\pi}_1} - \frac{\pi_0}{\hat{\pi}_0}}
  \hat{\delta}^{(-k)}\widehat{\Omega}^{(-k)} \ba{\hat{p}_0^{(-k)} - p_0} \\
&= \hat{\rho}^{(-k)}\hat{\delta}^{(-k)}\widehat{\Omega}^{(-k)}
\ba{\frac{\pi_1 \hat{\pi}_0^{(-k)} - \hat{\pi}_1^{(-k)}  \pi_0}{\hat{\pi}_1^{(-k)}  \hat{\pi}_0}}
\ba{\hat{p}_0^{(-k)} - p_0} \\
&= \frac{\hat{\rho}^{(-k)}\hat{\delta}^{(-k)}\widehat{\Omega}^{(-k)}}{\hat{\pi}_1^{(-k)}  \hat{\pi}_0}\ba{\hat{p}_0^{(-k)} - p_0}
\ba{\pi_1 \hat{\pi}_0^{(-k)} - \hat{\pi}_1^{(-k)}  \pi_0}
 \\
&= \frac{\hat{\rho}^{(-k)}\hat{\delta}^{(-k)}\widehat{\Omega}^{(-k)}}{\hat{\pi}_1^{(-k)}  \hat{\pi}_0} \ba{\hat{p}_0^{(-k)} - p_0}
\bb{ \frac{1}{2}\ba{\pi_1 - \hat{\pi}_1} \ba{\pi_0 + \hat{\pi}_0^{(-k)}} - \frac{1}{2}\ba{\pi_1 + \hat{\pi}_1^{(-k)}} \ba{\pi_0 - \hat{\pi}_0}}
\\
&=  \frac{1}{2}\frac{\hat{\rho}^{(-k)}\hat{\delta}^{(-k)}\widehat{\Omega}^{(-k)}}{\rho\hat{\pi}_1^{(-k)}  \hat{\pi}_0}\ba{\pi_0 + \hat{\pi}_0^{(-k)}} 
\rho 
\ba{\hat{p}_0^{(-k)} - p_0}
\ba{\pi_1 - \hat{\pi}_1}  \tag{T3-1} \\
&\quad - 
	\frac{1}{2}\frac{\hat{\rho}^{(-k)}\hat{\delta}^{(-k)}\widehat{\Omega}^{(-k)}}{\rho\hat{\pi}_1^{(-k)}  \hat{\pi}_0} 
	\ba{\pi_1 + \hat{\pi}_1^{(-k)}} 
    \rho
    \ba{\hat{p}_0^{(-k)} - p_0} 
    \ba{\pi_0 - \hat{\pi}_0} \tag{T3-2} 
\\ \\
& E\ba{\text{T3-1}} \lesssim  \norm{\hat{\pi}_1^{(-k)}  - \pi_1}  \cdot  \norm{\hat{p}_0^{(-k)} -p_0}
	\\
& E\ba{\text{T3-2}} \lesssim  \norm{\hat{\pi}_0^{(-k)} - \pi_0}  \cdot   \norm{\hat{p}_0^{(-k)} -p_0}
\end{align*}

For T4, following similar procedures from handling T3, we have:
\begin{align*}
	&\textrm{T4} \\
	&= 	\ba{\hat{p}_1^{(-k)}\hat{\pi}_1^{(-k)} + \hat{p}_0^{(-k)}\hat{\pi}_0^{(-k)}}
	\frac{\pi_1}{\hat{\pi}_1^{(-k)}} \widehat{\Omega}^{(-k)} \ba{e_0 -  \hat{e}_0^{(-k)}}  - 
	\ba{\hat{p}_1^{(-k)}\hat{\pi}_1^{(-k)} + \hat{p}_0^{(-k)}\hat{\pi}_0^{(-k)}}
	\frac{\pi_0}{\hat{\pi}_0^{(-k)}} \widehat{\Omega}^{(-k)} \ba{e_0 -  \hat{e}_0^{(-k)}} \\
	&= \hat{\rho}^{(-k)}\widehat{\Omega}^{(-k)} \ba{\frac{\pi_1}{\hat{\pi}_1} - \frac{\pi_0}{\hat{\pi}_0}} \ba{e_0 -  \hat{e}_0^{(-k)}}  \\
	&= \hat{\rho}^{(-k)}\widehat{\Omega}^{(-k)} \frac{1}{\hat{\pi}_1^{(-k)}  \hat{\pi}_0} 
	\bb{\frac{1}{2}\ba{\pi_1 - \hat{\pi}_1} \ba{\pi_0 + \hat{\pi}_0^{(-k)}} - \frac{1}{2}\ba{\pi_1 + \hat{\pi}_1^{(-k)}} \ba{\pi_0 - \hat{\pi}_0}} \ba{e_0 -  \hat{e}_0^{(-k)}}  \\
	&= \frac{1}{2} \frac{\hat{\rho}^{(-k)} \widehat{\Omega}^{(-k)} }{\rho\hat{\pi}_1^{(-k)}  \hat{\pi}_0}  \ba{\pi_0 + \hat{\pi}_0^{(-k)}}  
    \rho
    \ba{\pi_1 - \hat{\pi}_1}  \ba{e_0 -  \hat{e}_0^{(-k)}} \tag{T4-1} \\
	&\quad - \frac{1}{2} \frac{\hat{\rho}^{(-k)}\widehat{\Omega}^{(-k)}}{\rho\hat{\pi}_1^{(-k)}  \hat{\pi}_0} \ba{\pi_1 + \hat{\pi}_1^{(-k)}} \rho \ba{\pi_0 - \hat{\pi}_0}  \ba{e_0 -  \hat{e}_0^{(-k)}}  \tag{T4-1} 
	\\ \\
& E\ba{\text{T4-1}} \lesssim  \norm{\hat{\pi}_1^{(-k)} - \pi_1} \cdot \norm{\hat{e}_0^{(-k)} - e_0} 
	\\
& E\ba{\text{T4-2}} \lesssim  \norm{\hat{\pi}_0^{(-k)} - \pi_0} \cdot \norm{\hat{e}_0^{(-k)} - e_0}
\end{align*}

Therefore, we find the term (\ref{b-term2}) admits the following bias structure (using the abbreviated notation $r_f^{(-k)} = \norm{\hat{f}^{(-k)} - f}$) :
\begin{align*}
	&|\mc{I}_k|^{1/2} E^{(-k)} \ba{\widehat{M}^{(-k)} - M} = |\mc{I}_k|^{1/2} E^{(-k)} \ba{\textrm{T1} + \textrm{T2} + \textrm{T3} + \textrm{T4}} \\
	&\lesssim r_{\delta}^{(-k)} \cdot  r_{\pi_1}^{(-k)} + r_{\delta}^{(-k)}  \cdot  r_{\pi_0}^{(-k)} \\
	&\ + r_{\pi_1}^{(-k)}  \cdot  r_{p_0}^{(-k)} + r_{\pi_0}^{(-k)}  \cdot r_{p_0}^{(-k)} \\
	&\ + r_{\pi_1}^{(-k)} \cdot r_{e_0}^{(-k)}  + r_{\pi_0}^{(-k)} \cdot r_{e_0}^{(-k)}\\
	&\ + r_{\delta}^{(-k)} \cdot r_{\Omega}^{(-k)}.
\end{align*}

We further observe that the robustness property of the EIF is reflected in its bias structure; that is, the EIF remains unbiased when certain nuisance functions are evaluated at their true values (colored in red):
\begin{align*}
	&\mc{M}_1: e_0(X), p_0(X) \ and \ \delta(X)  \text{ are correct, then}\\
	&|\mc{I}_k|^{1/2} E^{(-k)} \ba{\widehat{M}^{(-k)} - M}  \\
	&\lesssim {\color{red}r_{\delta}^{(-k)}} \cdot  r_{\pi_1}^{(-k)} + {\color{red}r_{\delta}^{(-k)}}  \cdot r_{\pi_0}^{(-k)} 
	+ r_{\pi_1}^{(-k)}  \cdot {\color{red}r_{p_0}^{(-k)}} + r_{\pi_0}^{(-k)}  \cdot   {\color{red}r_{p_0}^{(-k)}}
	+ r_{\pi_1}^{(-k)} \cdot {\color{red}r_{e_0}^{(-k)}}  + r_{\pi_0}^{(-k)} \cdot {\color{red}r_{e_0}^{(-k)}}
	+ {\color{red}r_{\delta}^{(-k)}} \cdot r_{\Omega}^{(-k)} 	;
	\\ \\
	&\mc{M}_2: \pi_z(X) \ and \  p_z(X) \text{ are correct, then} \\
	&|\mc{I}_k|^{1/2} E^{(-k)} \ba{\widehat{M}^{(-k)} - M}  \\
	&\lesssim r_{\delta}^{(-k)} \cdot  {\color{red}r_{\pi_1}^{(-k)}} + r_{\delta}^{(-k)}  \cdot  {\color{red}r_{\pi_0}^{(-k)}} 
	+ {\color{red}r_{\pi_1}^{(-k)}}  \cdot  {\color{red}r_{p_0}^{(-k)}} + {\color{red}r_{\pi_0}^{(-k)}}  \cdot   {\color{red}r_{p_0}^{(-k)}} 
	+ {\color{red}r_{\pi_1}^{(-k)}} \cdot r_{e_0}^{(-k)}  + {\color{red}r_{\pi_0}^{(-k)}} \cdot r_{e_0}^{(-k)}
	+ r_{\delta}^{(-k)} \cdot {\color{red}r_{\Omega}^{(-k)}}	;
	\\ \\
	&\mc{M}_3: \pi_z(X) \ and \ \delta(X)  \text{ are correct, then} \\
	&|\mc{I}_k|^{1/2} E^{(-k)} \ba{\widehat{M}^{(-k)} - M}   \\
	&\lesssim {\color{red}r_{\delta}^{(-k)}} \cdot {\color{red}r_{\pi_1}^{(-k)}} + {\color{red}r_{\delta}^{(-k)}}  \cdot  {\color{red}r_{\pi_0}^{(-k)} } 
	 + {\color{red}r_{\pi_1}^{(-k)}}  \cdot  r_{p_0}^{(-k)} + {\color{red}r_{\pi_0}^{(-k)}}  \cdot   r_{p_0}^{(-k)} 
	 + {\color{red}r_{\pi_1}^{(-k)}} \cdot r_{e_0}^{(-k)}  + {\color{red}r_{\pi_0}^{(-k)}} \cdot r_{e_0}^{(-k)}
	 + {\color{red}r_{\delta}^{(-k)}} \cdot r_{\Omega}^{(-k)}.
\end{align*}
\textbf{\textit{Asymptotic Property of (\ref{b-term3})}}

The expectation of the term (\ref{b-term3}) 
\begin{align*}
	&\mb{G}^{(-k)}_{\mc{I}_k} \ba{ \widehat{M}_i^{(-k)} - M} 
\end{align*}
has mean 0 conditional on $\mc{I}_k^{(-k)}$. And, the variance is 
\begin{align*}
	var^{(-k)}\bb{\mb{G}^{(-k)}_{\mc{I}_k} \ba{ \widehat{M}_i^{(-k)} - M}}
	\leq 
	var^{(-k)}\bb{\ba{ \widehat{M}_i^{(-k)} - M} }
	\leq
	E^{(-k)}\bb{\ba{\widehat{M}_i^{(-k)} - M}^2}
\end{align*}

Then, it suffices to find the rate of $E^{(-k)}\bc{\bb{\widehat{M}_i^{(-k)} - M}^2}$. Each element of $\ba{\widehat{M}_i^{(-k)} - M}$ is
\begin{align*}
	&\ba{\widehat{M}_i^{(-k)} - M}
	\\
	&
	=A\ba{\hat{\delta}^{(-k)} - \delta} + \hat{\theta}^{(-k)} -  \theta	
	\\
	&=
	\rho
	\ba{\hat{\delta}^{(-k)} - \delta}+ 
	\hat{\rho}^{(-k)} \frac{2Z-1}{\hat{\pi}_Z^{(-k)}}
	\widehat{\Omega}^{(-k)}
	\bb{
	Y(1-A) - \hat{e}_0^{(-k)} - \ba{A-\hat{p}_0^{(-k)}} \hat{\delta}^{(-k)}
	}
	-
	\rho\frac{2Z-1}{\pi_Z}\Omega\bb{Y(1-A) - e_0 - (A-p_0) \delta}
	\\
	&=
	\rho\ba{\hat{\delta}^{(-k)} - \delta}
	-
	\rho\frac{2Z-1}{\pi_Z}\Omega\bb{Y(1-A) - e_0 - (A-p_0) \delta}
	\\ & \ \
	+
	\ba{\hat{\rho}^{(-k)} - \rho} \frac{2Z-1}{\hat{\pi}_Z^{(-k)}}
	\widehat{\Omega}^{(-k)}
	\bb{
	Y(1-A) - \hat{e}_0^{(-k)} - \ba{A-\hat{p}_0^{(-k)}} \hat{\delta}^{(-k)}
	}
	\\ & \ \	
	+
	\rho
	\frac{2Z-1}{\hat{\pi}_Z^{(-k)}}
	\ba{\widehat{\Omega}^{(-k)} -\Omega}
	\bb{
	Y(1-A) - \hat{e}_0^{(-k)} - \ba{A-\hat{p}_0^{(-k)}} \hat{\delta}^{(-k)}
	}	
	\\ & \ \	
	+
	\rho
	\frac{2Z-1}{\hat{\pi}_Z^{(-k)}}
	\Omega
	\bb{
	- \ba{\hat{e}_0^{(-k)} - e_0} - A\ba{\hat{\delta}^{(-k)} - \delta} + \ba{\hat{p}_0^{(-k)} \hat{\delta}^{(-k)} - p_0 \delta }
	}
	\\ & \ \	
	+
	\rho
	\frac{2Z-1}{\hat{\pi}_Z^{(-k)}}
	\Omega\bb{Y(1-A) -e_0 -A\delta + p_0 \delta}	 
	\\
	&=
	\rho\ba{\hat{\delta}^{(-k)} - \delta} \label{JJ1} \tag{J1}
	\\ & \ \	
	+
	\rho\frac{2Z-1}{\pi_Z\hat{\pi}_Z^{(-k)}}\ba{\pi_Z - \hat{\pi}_Z^{(-k)}}\Omega\bb{Y(1-A) - e_0 - (A-p_0) \delta}
	\label{JJ2} \tag{J2}	
	\\ & \ \
	+\ba{\hat{\rho}^{(-k)} - \rho} \frac{2Z-1}{\hat{\pi}_Z^{(-k)}}
	\widehat{\Omega}^{(-k)}
	\bb{
	Y(1-A) - \hat{e}_0^{(-k)} - \ba{A-\hat{p}_0^{(-k)}} \hat{\delta}^{(-k)}
	}
	\label{JJ3} \tag{J3}
	\\ & \ \	
	+
	\rho
	\frac{2Z-1}{\hat{\pi}_Z^{(-k)}}
	\ba{\widehat{\Omega}^{(-k)} -\Omega}
	\bb{
	Y(1-A) - \hat{e}_0^{(-k)} - \ba{A-\hat{p}_0^{(-k)}} \hat{\delta}^{(-k)}
	}	
	\label{JJ4} \tag{J4}	
	\\ & \ \	
	+
	\rho
	\frac{2Z-1}{\hat{\pi}_Z^{(-k)}}
	\Omega
	\bb{
	- \ba{\hat{e}_0^{(-k)} - e_0} - A\ba{\hat{\delta}^{(-k)} - \delta} + \ba{\hat{p}_0^{(-k)} \hat{\delta}^{(-k)} - p_0 \delta }
	}
	\label{JJ5} \tag{J5}			
	.	
\end{align*}

For a finite number of random variables $\{W_1, \ldots, W_K\}$, there exists a constant $C$ satisfying 
\[
E\left\{\left(\sum_{j=1}^{K} W_j\right)^2\right\} \leq C \cdot E(W_j^2).
\]

Thus, it suffices to study the rate of the $\mc{L}(P)$-norm of each term, which are given in (\ref{JJ1})-(\ref{JJ5}) below.
\begin{align*}
	&(\text{\ref{JJ1}}): E\bc{ \bb{\rho\ba{\hat{\delta}^{(-k)} - \delta}}^2} \lesssim \norm{\hat{\delta}^{(-k)} - \delta}^2 = r_{\delta}^2;\\
	&(\text{\ref{JJ2}}): 
	E\bc{\bb{\rho\frac{2Z-1}{\pi_Z\hat{\pi}_Z^{(-k)}}\ba{\pi_Z - \hat{\pi}_Z^{(-k)}}\Omega\bb{Y(1-A) - e_0 - (A-p_0) \delta}}^2} \\
	&\quad \quad \lesssim \norm{\hat{\pi}_1^{(-k)}- \pi_1}^2  +  \norm{\hat{\pi}_0^{(-k)}- \pi_0}^2 = r_{\pi_1} + r_{\pi_0}; \\
	&(\text{\ref{JJ3}}): 
	E\bc{\bb{\ba{\hat{\rho}^{(-k)} - \rho} \frac{2Z-1}{\hat{\pi}_Z^{(-k)}}
	\widehat{\Omega}^{(-k)}
	\bb{
	Y(1-A) - \hat{e}_0^{(-k)} - \ba{A-\hat{p}_0^{(-k)}} \hat{\delta}^{(-k)}
	}}^2} \\
	&\quad \quad \lesssim  \norm{\hat{p}_1^{(-k)}- p_1}^2  +  \norm{\hat{p}_0^{(-k)}- p_0}^2+\norm{\hat{\pi}_1^{(-k)}- \pi_1}^2  +  \norm{\hat{\pi}_0^{(-k)}- \pi_0}^2 = r_{p_1} + r_{p_0} + r_{\pi_1} + r_{\pi_0} \\
	& \text{ using the decoposition $ab - cd = \frac{1}{2}(a-c)(b+d) + \frac{1}{2}(a+c)(b-d)$ for $\ba{\hat{\rho}^{(-k)} - \rho}$;} \\
	&(\text{\ref{JJ4}}): 
	E\bc{\bb{	\rho
	\frac{2Z-1}{\hat{\pi}_Z^{(-k)}}
	\ba{\widehat{\Omega}^{(-k)} -\Omega}
	\bb{
	Y(1-A) - \hat{e}_0^{(-k)} - \ba{A-\hat{p}_0^{(-k)}} \hat{\delta}^{(-k)}
	}}^2} \lesssim \norm{\hat{\Omega}_1^{(-k)}- \Omega}^2 = r_{\Omega}
	\\
	&(\text{\ref{JJ5}}):
	E\bc{\bb{	\rho
	\frac{2Z-1}{\hat{\pi}_Z^{(-k)}}
	\Omega
	\bb{
	- \ba{\hat{e}_0^{(-k)} - e_0} - A\ba{\hat{\delta}^{(-k)} - \delta} + \ba{\hat{p}_0^{(-k)} \hat{\delta}^{(-k)} - p_0 \delta }
	}}^2}\\
	&\quad \quad \lesssim \norm{\hat{e}_0^{(-k)} - e_0}^2  +  \norm{\hat{p}_0^{(-k)} - p_0}^2= r_{e_0} + r_{p_0}.
\end{align*}

Combining these results, we find
\begin{align*}
	var^{(-k)}\bb{\mb{G}^{(-k)}_{\mc{I}_k} \ba{ \widehat{M}_i^{(-k)} - M}}
	\lesssim r_{\delta} + r_{\Omega} + r_{p_1} + r_{p_0} + r_{\pi_1} + r_{\pi_0} + r_{e_0}.
\end{align*}

Under assumption \ref{as:consis} that $r_{p_z,N}^{(-k)}$, $r_{\pi_z,N}^{(-k)}$, $r_{e_z,N}^{(-k)}$, $r_{\Omega,N}^{(-k)}$ and $r_{\delta,N}^{(-k)}$ are $o_P(1)$ for $z\in \bb{0,1}$, $var^{(-k)}\bb{\mb{G}^{(-k)}_{\mc{I}_k} \ba{ \widehat{M}_i^{(-k)} - M}} = o_P(1)$, indicating $(\ref{b-term3})$ is $o_P(1)$ by Chebyshev's inequality (by showing it has $0$ expectation and shrinking variance).
\\

\textbf{\textit{Consistent Variance Estimation}}

Using the the shorthand notation $f = f(X)$, the consistent estimator of $\sigma^2$ proposed in \Cref{th:asym} is
\begin{align*}
	&\hat{\sigma}^2 = K^{-1} \sum_{i=1}^K \mb{P}_{\mc{I}_k}
		\bc{
		\bb{
		\hat{\gamma}^{(-k)}(O) -\hat{\psi}^{EIF-FW}
		}^2
		}
		\\ 
        &where 
        \\
	&\hat{\gamma}^{(-k)}(O):= 	
	\bb{\mb{P}(A)}^{-1}
    \bc{
 	A\{Y+\hat{\delta}^{(-k)}\}  + \hat{\theta}^{(-k)}(O)
	} \\
	&\hat{\theta}(O) := 
	\hat{\rho}^{(-k)}(X) \frac{2Z-1}{\hat{\pi}^{(-k)}_Z(X)}
	\widehat{\Omega}^{(-k)}(X) 
	\bc{
	Y(1-A)-\hat{e}_Z^{(-k)}(X)  - \bb{A-\hat{p}_Z^{(-k)}(X)}\hat{\delta}^{(-k)}(X) 
	}
	,
\end{align*}
which can also be written as
\begin{align*}
	&\hat{\sigma}^2 = K^{-1} \sum_{i=1}^K \hat{\sigma}^{2,(k)}, \quad 
	\hat{\sigma}^{2,(k)} =		\mb{P}_{\mc{I}_k}
	\bc{
	\bb{ A\{Y+\hat{\delta}^{(-k)}\}  + \hat{\theta}^{(-k)}(O) -\hat{\psi}^{EIF-FW}}^2}.
\end{align*}

Therefore, it suffices to show that \(\hat{\sigma}^{2,k} - \sigma^2 = o_P(1)\), which is represented as follows:
\begin{align*}
	&\sigma^2  = \bb{\mb{P}\ba{A}}^{-2} \mb{P}_{\mc{I}_k}
	\bc{
	\bb{
	{
	A(Y+\delta) + \theta(O) -A\psi
	}
	}^2
	}
	\\
	&\hat{\sigma}^{2,k} =\bb{\mb{P}\ba{A}}^{-2} \mb{P}_{\mc{I}_k}
	\bc{
	\bb{
	{
	A\ba{Y+\delta^{(-k)}} + \theta^{(-k)}(O) -A\hat{\psi}^{EIF-FW}
	}
	}^2
	}
	\\
	&\hat{\sigma}^{2,k} - \sigma^2 
	=
	\bb{pr(A=1)}^{-2} 
	\mb{P}_{\mc{I}_k}
	\bc{
	\bb{
	{
	A\ba{Y+\delta^{(-k)}} + \theta^{(-k)}(O) -A\hat{\psi}^{EIF-FW}
	}
	}^2
	-
	\bb{
	{
	A(Y+\delta) + \theta(O) -A\psi
	}
	}^2	
	} + o_P(1)
\end{align*}
The last equation holds by the law of large numbers. From some algebra, we find the term on the RHS of the last equation is
\begin{align*}
	&\mb{P}_{\mc{I}_k}
	\bc{
	\bb{
	{
	A\ba{Y+\delta^{(-k)}} + \theta^{(-k)}(O) -A\hat{\psi}^{EIF-FW}
	}
	}^2
	-
	\bb{
	{
	A(Y+\delta) + \theta(O) -A\psi
	}
	}^2
	} \\
	&
	=\mb{P}_{\mc{I}_k}	
	\biggl[
	\bb{
	A\ba{Y+\delta^{(-k)}} + \theta^{(-k)}(O) -A\hat{\psi}^{EIF-FW}
	+
	A(Y+\delta) + \theta(O) -A\psi
	}
	\\
	& \quad \quad \quad \cdot	
	\bb{
	A\ba{Y+\delta^{(-k)}} + \theta^{(-k)}(O) -A\hat{\psi}^{EIF-FW}
	-
	A(Y+\delta) - \theta(O) + A\psi
	}	
	\biggl]	\\
	&=\mb{P}_{\mc{I}_k}	
	\biggl[
	\bb{
	A\ba{\delta^{(-k)} - \delta + \delta + Y} + \bb{\theta^{(-k)}(O)-\theta(O)+\theta(O)} + A\ba{\psi - \psi - \hat{\psi}^{EIF-FW}}
	+
	A(Y+\delta) + \theta(O) -A\psi
	}
	\\
	& \quad \quad \quad \cdot	
	\bb{
	A\ba{\delta^{(-k)} - \delta} + \bb{\theta^{(-k)}(O) -\theta(O)} + A\ba{\psi - \hat{\psi}^{EIF-FW}}
	}	
	\biggl]	
	\\
	&=\mb{P}_{\mc{I}_k}	
	\biggl[
	\bb{
	A\ba{\delta^{(-k)} - \delta} + \bb{\theta^{(-k)}(O)-\theta(O)} + A\ba{\psi - \hat{\psi}^{EIF-FW}}
	+
	2\bb{A(Y+\delta) + \theta(O) -A\psi}
	}
	\\
	& \quad \quad \quad \cdot	
	\bb{
	A\ba{\delta^{(-k)} - \delta} + \bb{\theta^{(-k)}(O) -\theta(O)} + A\ba{\psi - \hat{\psi}^{EIF-FW}}
	}	
	\biggl]	
	\\
	&=\mb{P}_{\mc{I}_k}	
	\bc{
	\bb{
	A\ba{\delta^{(-k)} - \delta} + \bb{\theta^{(-k)}(O)-\theta(O)} + A\ba{\psi - \hat{\psi}^{EIF-FW}}
	}^2
	} \label{DD1} \tag{D1}
	\\
	& \quad \quad \quad +	2 \mb{P}_{\mc{I}_k}	
	\bc{
	\bb{
	A\ba{\delta^{(-k)} - \delta} + \bb{\theta^{(-k)}(O) -\theta(O)} + A\ba{\psi - \hat{\psi}^{EIF-FW}}
	}
	\bb{A(Y+\delta) + \theta(O) -A\psi}
	}	
	\label{DD2} \tag{D2}	
\end{align*}

Let $H ^{(-k)} = 	\bb{
	A\ba{\delta^{(-k)} - \delta} + \bb{\theta^{(-k)}(O) -\theta(O)} + A\ba{\psi - \hat{\psi}^{EIF-FW}}
	}$. By the H\"older's inequality:
\begin{align*}
	&\bb{(\text{D1}) + (\text{D2})}^2
	\\
	&=
	\bc{
	\mb{P}_{\mc{I}_k} \bb{H^{(-k)}} + 2 \mb{P}_{\mc{I}_k} \bb{H^{(-k)}} \mb{P}_{\mc{I}_k} \bb{A(Y+\delta) + \theta(O) -A\psi}
	}^2
	\\
	&\lesssim \mb{P}_{\mc{I}_k}	 \bc{ \bb{H^{(-k)}}^2} + 2\mb{P}_{\mc{I}_k}	 \bc{ \bb{H^{(-k)}}^2} 
	\mb{P}_{\mc{I}_k}	 \bc{ \bb{A(Y+\delta) + \theta(O) -A\psi}^2}.
\end{align*}
Since $\mb{P}_{\mc{I}_k}	 \bc{ \bb{A(Y+\delta) + \theta(O) -A\psi}^2} = pr(A=1) \sigma^2 + o_P(1) = O_p(1)$, $(\text{D1}) + (\text{D2})$ is $o_P(1)$ if 
$\mb{P}_{\mc{I}_k} \bc{ \bb{H^{(-k)}}^2}$ is $o_P(1)$. From some algebra, we find
\begin{align*}
	&\mb{P}_{\mc{I}_k} \bc{ \bb{H^{(-k)}}^2} \\
	&=\mb{P}_{\mc{I}_k} \bc{
	\bb{
	A\ba{\delta^{(-k)} - \delta} + \bb{\theta^{(-k)}(O) -\theta(O)} + A\ba{\psi - \hat{\psi}^{EIF-FW}}
	}^2
	} \\
	&\leq 3\mb{P}_{\mc{I}_k}\bc{\bb{\theta^{(-k)}(O) -\theta(O)}^2} + 3\mb{P}_{\mc{I}_k}\bb{\ba{\delta^{(-k)} - \delta}^2} + 3\mb{P}_{\mc{I}_k} \bb{\ba{\psi - \hat{\psi}^{EIF-FW}}^2}
	\\
	&= 3E^{(-k)}\bc{\bb{\theta^{(-k)}(O) -\theta(O)}^2} + o_P(1) + 3E^{(-k)}\bb{\ba{\delta^{(-k)} - \delta}^2}+ o_P(1)+3\ba{\psi - \hat{\psi}^{EIF-FW}}^2 \\
	&=o_p(1)
	.
\end{align*}
Therefore, the proposed estimator $\hat{\sigma}^2$ is consistent for $\sigma^2$
\begin{align*}
	&\hat{\sigma}^{2,k} - \sigma^2  =
	\bb{pr(A=1)}^{-2} 
	\bb{(\text{D1}) + (\text{D2})} + o_P(1) = o_P(1).
\end{align*}

\subsection{Details FW Counterfactual Regression Approach}
\label{a:fw_discussion}
In this section, we detail the theoretical results of applying the FW counterfactual regression approach to construct the proposed estimator. Specifically, we first \textbf{(1)} examine the bias structure of the estimator $\hat{\psi}^{EIF}$, which is based on $\tilde{\delta}(X)$ and $\widetilde{\Omega}(X)$, and \textbf{(2)} show that the estimator $\hat{\psi}^{EIF-FW}$ proposed in \Cref{th:asym}, which is based on $\hat{\delta}(X)$ and $\widehat{\Omega}(X)$, achieves an upper bound on the convergence rate no greater than that of the former. For simplicity, we use the shorthand notation $f = f(X)$ when no ambiguity arises ($\psi$ exclusively used to denote the marginal quantity) and omit the superscript $(-k)$ throughout this section.

\textbf{\textit{Bias Structure of $\hat{\psi}^{EIF}$}}

Using the substitution estimators $\tilde{\delta}$ and $\tilde{\Omega}$ introduces a different bias structure. This can be seen from the fact that the convergence rate of $\tilde{\delta}$ depends on the slowest convergence rate among $\hat{e}_1$, $\hat{e}_0$, $\hat{p}_1$, and $\hat{p}_0$, and $\tilde{\Omega}$ depends on the slowest convergence rate among $\hat{p}_1$ and $\hat{p}_0$.
\begin{align*}
	&\widetilde{\Omega} = \frac{1}{\hat{p}_1 - \hat{p}_0}, \quad 
	\Omega = \frac{1}{p_1-p_0}, \\
	&r_{\Omega} = \norm{\widetilde{\Omega}  - {\Omega}} 
	= \norm{\frac{
 	p_1 - p_0 - \hat{p}_1 + \hat{p}_0
	}{
	\ba{\hat{p}_1 - \hat{p}_0} \ba{p_1 - p_0}
	} };
	\\
	&\tilde{\delta} = \frac{\hat{e}_1 - \hat{e}_0}{\hat{p}_1 - \hat{p}_0}, \quad 
	\delta = \frac{e_1-e_0}{p_1-p_0}, \\
	&r_{\delta} = \norm{\tilde{\delta}  - {\delta}} 
	= \norm{ \frac{
 	\hat{e}_1 p_1 - \hat{e}_1 p_0 - \hat{e}_0 p_1 + \hat{e}_0 p_0
 	- e_1 \hat{p}_1 + e_1 \hat{p}_0 + e_0 \hat{p}_1 - e_0 \hat{p}_0
	}{
	\ba{\hat{p}_1 - \hat{p}_0} \ba{p_1 - p_0}
	}}.
\end{align*}

If $\ba{\hat{p}_1 - \hat{p}_0}\ba{p_1 - p_0} > \epsilon $ for some $\epsilon > 0$ with probability one, then one has
\begin{align*}
	&\norm{\widetilde{\Omega}  - {\Omega}} 
	\leq \frac{1}{\epsilon} \norm{\hat{p}_1 - p_1 + \hat{p}_0 - p_0} 
	\lesssim \norm{\hat{p}_1 - p_1}  + \norm{\hat{p}_0 - p_0} = r_{p_1} + r_{p_0},
	\\ 
	&\norm{\tilde{\delta}  - {\delta}} 
	\leq \frac{1}{\epsilon} \norm{ 	
	\underbrace{\hat{e}_1 p_1}_{(t1)} 
	\underbrace{- \hat{e}_1 p_0}_{(t2)}
	\underbrace{- \hat{e}_0 p_1}_{(t3)}
	\underbrace{+ \hat{e}_0 p_0}_{(t4)}
	\underbrace{- e_1 \hat{p}_1}_{(t1)}
	\underbrace{+ e_1 \hat{p}_0}_{(t2)}
	\underbrace{+ e_0 \hat{p}_1}_{(t3)}
	\underbrace{- e_0 \hat{p}_0}_{(t4)}
	}
 	\\
 	&=\frac{1}{\epsilon} 
	\left\lVert
 	\underbrace{\bb{\hat{e}_1 - e_1} p_1 + \bb{p_1 - \hat{p}_1} e_1}_{(t1)}
 	+
 	\underbrace{\bb{e_1 - \hat{e}_1} \hat{p}_0 + \bb{\hat{p}_0 - p_0} \hat{e}_1}_{(t2)}
 	\right.
	\left.
	\underbrace{\bb{e_0 - \hat{e}_0} \hat{p}_1 + \bb{\hat{p}_1 - p_1} \hat{e}_0}_{(t3)}
	+
 	\underbrace{\bb{\hat{e}_0 - e_0} p_1+\bb{p_0 - \hat{p}_0} e_0}_{(t4)}
	\right\rVert 
	\\
	&\lesssim \norm{\hat{e}_1 - e_1} + \norm{\hat{e}_0 - e_0} + \norm{\hat{p}_1 - p_1} + \norm{\hat{p}_0 - p_0} \\
	&= r_{e_1} + r_{e_0} + r_{p_1} + r_{p_0}.
\end{align*}

To obtain the bias structure of $\hat{\psi}^{EIF}$, one can replace  $r_{\delta}$ and $r_{\Omega}$ in the previously derived bias structure for $\hat{\psi}^{EIF-FW}$ with the above expansion:
\begin{align*}
	&|\mc{I}_k|^{1/2} E \ba{\widehat{M}^{EIF} - M}  \\
	&\lesssim  
	r_{\pi_1} \cdot r_{e_1}  + r_{\pi_1} \cdot r_{e_0} + r_{\pi_0} \cdot r_{e_1} + r_{\pi_0} \cdot r_{e_0}
	\\
	&\ +  	
	r_{\pi_1} \cdot r_{p_1}  + r_{\pi_1} \cdot r_{p_0} + r_{\pi_0} \cdot r_{p_1} + r_{\pi_0} \cdot r_{p_0} 
	\\
	&\ + 
	r_{p_1} \cdot r_{e_1}  + r_{p_1} \cdot r_{e_0} + r_{p_0} \cdot r_{e_1} + r_{p_0} \cdot r_{e_0}
	\\
	&\ + 
	r_{p_1}^2  + r_{p_0}^2 + r_{p_1} \cdot r_{p_0}		
	.
\end{align*}

Here, we note that the quadratic term $r_{p_1}^2=\norm{\hat{p}_z - p_z}^2$ for $z = 0,1$ appears in the bias structure. This implies that using the plug-in one $\tilde{\delta}$ in estimating $\psi$, if $\hat{p}_z$ converges slowly (e.g, $\norm{\hat{p}_z - p_z}^2 = O_p(N^{-k})$ for $0<k<1/2$), the desired $\sqrt{N}$ convergence cannot be achieved, even if the other nuisance components converge sufficiently fast.

Moreover, even if all cross-product terms are $O_p(N^{-k})$ for some $k > 1/2$, an alternative estimation strategy, including the FW counterfactual regression approach that directly targets $\delta$ and $\Omega$, may let the bias shrinks more quickly and thus yield improved finite-sample performance. This is because the functionals $\delta$ and $\Omega$ may be smoother than the nuisance components $e_1$, $e_0$, $p_1$, and $p_0$. Next, we demonstrate that the FW counterfactual regression approach realizes it by targeting directly $\delta$ and $\Omega$.

\textbf{\textit{Properties of FW Counterfactual Regression Approach}}

We have shown
\begin{align*}
	&|\mc{I}_k|^{1/2} E \ba{\widehat{M} - M} \\
	&\lesssim r_{\delta} \cdot  r_{\pi_1} + r_{\delta}  \cdot  r_{\pi_0} \\
	&\ + r_{\pi_1}  \cdot  r_{p_0} + r_{\pi_0}  \cdot r_{p_0} \\
	&\ + r_{\pi_1} \cdot r_{e_0}  + r_{\pi_0} \cdot r_{e_0}  \\
	&\ + r_{\delta} \cdot r_{\Omega}	 = Bound-1;
	  \\
	&|\mc{I}_k|^{1/2} E \ba{\widehat{M}^{EIF}  - M} \\
	&\lesssim  
	r_{\pi_1} \cdot r_{e_1}  + r_{\pi_1} \cdot r_{e_0} + r_{\pi_0} \cdot r_{e_1} + r_{\pi_0} \cdot r_{e_0}
	\\
	&\ +  	
	r_{\pi_1} \cdot r_{p_1}  + r_{\pi_1} \cdot r_{p_0} + r_{\pi_0} \cdot r_{p_1} + r_{\pi_0} \cdot r_{p_0} 
	\\
	&\ + 
	r_{p_1} \cdot r_{e_1}  + r_{p_1} \cdot r_{e_0} + r_{p_0} \cdot r_{e_1} + r_{p_0} \cdot r_{e_0}
		\\
	&\ + 
	r_{p_1}^2  + r_{p_0}^2 + r_{p_1} \cdot r_{p_0}
	 = Bound-2.
\end{align*}

We will show the FW counterfactual regression approach achieves the inequality $$Bound-1 \lesssim Bound-2,$$ which means the upper bound on the convergence rate of the bias term of $\hat{\psi}^{EIF-FW}$ is no greater than that of $\hat{\psi}^{EIF}$. I.e, the bias term of $\hat{\psi}^{EIF-FW}$ can shrink at a rate faster than or equal to that of $\hat{\psi}^{EIF}$. 

Let $v_f$ denote the oracle minimax rate of the functional $f$. the inequality in \Cref{c:fw-lem} gives that
\begin{align*}
	&
	r_{\delta} = \norm{\hat{\delta} - \delta} 
	\lesssim  
	v_\delta + E_{2-\delta}
	, &&
	r_{\Omega} = \norm{\widehat{\Omega} - \Omega}^2 
	\lesssim 
	v_{\Omega} + E_{2-\Omega}, 
	\\
	&where \\
	&E_{2-\delta} \asymp Bound-2
	\quad
	&&E_{2-\Omega} \asymp 
	r_{p_1}^2 + r_{p_0}^2  + 
	r_{\pi_1} \cdot r_{p_1}  + 	r_{\pi_0} \cdot r_{p_0}
	\tag{EQ1}
	\end{align*}
For $\hat{\delta}$, its $E_{2-\delta}$ contains exactly the same terms in the bias structure of $\hat{\psi}^{EIF}$. For $\widehat{\Omega}$, its bias structure includes terms $r_{p_z}^2=\norm{\hat{p}_z-p_z}^2$ and $r_{\pi_z} \cdot r_{p_z}=\norm{\hat{p}_z - p_z} \cdot \norm{\hat{\pi}_z - \pi_z}$ for $z=0,1$. By the definition of $\delta$ and $\Omega$ and the fundamental connection between minimax rates and the smoothness of the nuisance functions, the oracle minimax rate of $\delta$ and $\Omega$ follows the following inequalities:
\begin{align*}
	&v_{\delta} \leq \max\bb{v_{e_1},v_{e_0},v_{p_1},v_{p_0}} 
	\leq 
	\max\bb{r_{e_1},r_{e_0},r_{p_1},r_{p_0}}
	, \quad v_{\Omega} \leq \max\bb{v_{p_1},v_{p_0}}
	\leq
	\max\bb{r_{p_1},r_{p_0}}
	.
\end{align*}

Since $E_{2-\delta}$ and $E_{2-\Omega}$ include the second-order terms, which are ignorable as compared to the upper bound of $v_{\delta}$ and $v_{\Omega}$, then one has:
\begin{align*}
	&r_{\delta} 
	\lesssim 
	r_{e_1} + r_{e_0} + r_{p_1} + r_{p_0} 
	,
	\quad 
        r_{\Omega} \lesssim  r_{p_1}+r_{p_0}.
        \tag{EQ2}
\end{align*}
Then, one can combine (EQ1) and (EQ2) and obtain
\begin{align*}
	&|\mc{I}_k|^{1/2} E \ba{\widehat{M} - M} 
	\\
	&\lesssim r_{\delta} \cdot  r_{\pi_1} + r_{\delta}  \cdot  r_{\pi_0} \\
	&\ + r_{\pi_1}  \cdot  r_{p_0} + r_{\pi_0}  \cdot r_{p_0} \\
	&\ + r_{\pi_1} \cdot r_{e_0}  + r_{\pi_0} \cdot r_{e_0} \\
	&\ + r_{\delta} \cdot r_{\Omega}  = Bound-1
	\\
	&\lesssim \bb{r_{e_1} + r_{e_0} + r_{p_1} + r_{p_0}} \cdot  r_{\pi_1} + \bb{r_{e_1} + r_{e_0} + r_{p_1} + r_{p_0}} \cdot  r_{\pi_0} \\
	&\ + r_{\pi_1}  \cdot  r_{p_0} + r_{\pi_0}  \cdot r_{p_0} \\
	&\ + r_{\pi_1} \cdot r_{e_0}  + r_{\pi_0} \cdot r_{e_0} \\
	&\ + \bb{r_{e_1} + r_{e_0} + r_{p_1} + r_{p_0}} \cdot \bb{r_{p_1} + r_{p_0}}
	\\
	&\lesssim  
	r_{\pi_1} \cdot r_{e_1}  + r_{\pi_1} \cdot r_{e_0} + r_{\pi_0} \cdot r_{e_1} + r_{\pi_0} \cdot r_{e_0}
	\\
	&\ +  	
	r_{\pi_1} \cdot r_{p_1}  + r_{\pi_1} \cdot r_{p_0} + r_{\pi_0} \cdot r_{p_1} + r_{\pi_0} \cdot r_{p_0} 
	\\
	&\ + 
	r_{p_1} \cdot r_{e_1}  + r_{p_1} \cdot r_{e_0} + r_{p_0} \cdot r_{e_1} + r_{p_0} \cdot r_{e_0}
		\\
	&\ + 
	r_{p_1}^2  + r_{p_0}^2 + r_{p_1} \cdot r_{p_0}
	 = Bound-2.	
\end{align*}

\subsection{Proof of \Cref{c:orthogo1}}
\label{a:ortho1}
In Section \ref{a:iden}, we have shown $U \indep Z | A=1,X$ under the MIV model. One also has $Y \indep Z | A=1,X$ by noting that 
\begin{align*}
	& f(Y^a|U,X) =   f(Y^a|U,X,A=1,Z) \quad \textit{(by assumptuion \ref{as:iv3})}\\ 
	& f(U|A=1) = f(U|A=1,Z) \quad \textit{(by $U \indep Z | A=1,X$)}\\
	& f(Y^a|U,A=1) f(U|A=1) =  f(Y^a|U,A=1,Z) f(U|A=1,Z) \\
	& \int f(Y^a|U,A=1) f(U|A=1) dU = \int f(Y^a|U,A=1,Z) f(U|A=1,Z) dU \\
	& \implies f(Y^a|A=1) = f(Y^a|A=1,Z).
\end{align*}
Plug-in $a=1$, and by consistency, one has
\begin{align*}
	& f\ba{Y^1|A=1} = f(Y|A=1) = f(Y|A=1,Z).
\end{align*}
Hence,
\begin{align*}
	& Y \indep Z | A=1.
\end{align*}

\subsection{Proof of \Cref{c:orthogo2}}
\label{a:ortho2}
To derive the EIF under the intersection model (assuming \ref{as:iv1}, \ref{as:iv2}, \ref{as:iv3-s} and \ref{as:miv}), we can project the EIF under the MIV model (assuming \ref{as:iv1}, \ref{as:iv2}, \ref{as:iv3} and \ref{as:miv}) onto the tangent space $\Lambda$ under the intersection model. We first characterize the orthogonal complement of the tangent space under the intersection model, and then do the projection.

\textit{\textbf{Characterization of the Orthogonal Complement of the Tangent Space}}

For the following derivation, we again suppress $X$. Under the intersection model, the joint probability density function of the observed data factors:
\begin{align*}
	&f(Y,A,Z) \\
	&=f(Y,Z|A) f(A) \\
	&=f(Y,Z|A=1)^A f(Y,Z|A=0)^{1-A} f(A) \\	
	&=\{f(Y|A=1)f(Z|A=1)\}^A f(Y,Z|A=0)^{1-A} f(A).
\end{align*}

The above factorization implies the tangent space is characterized by
\begin{align*}
	\Lambda = 
	\bb{
	A S(Y) + AS(Z) + (1-A)S(Y,Z) + S(A): 
	\begin{array}{l}
	E\bb{S(Y)|A=1} = 0 \\  E\bb{S(Z)|A=1} = 0 \\ E\bb{S(Y,Z)|A=0} = 0 \\ E\bb{S(A)} = 0
	\end{array} 
	}
	\cap \mathcal{L}_2(P).
\end{align*}

We will show that the orthogonal complement to the tangent space is
\begin{align*}
	&U_1 = 	\left\{
	A\bc{d_1(Y,Z) - E\bb{d_1(Y,Z)|A=1}}
	: d_1 \ unrestricted
	\right\}
	\\
	&U_2 =
	\left\{
	A\bb{Z - E\ba{Z|A=1}} \bc{d_2(Y) - E\bb{d_2(Y)|A=1}}
	: d_2 \ unrestricted	
	\right\}
	\\
	&\Lambda^\perp= U_1
	\cap \mathcal{L}_2(P) =
	U_2
	\cap \mathcal{L}_2(P).
\end{align*}

Proof:

We first verify that $\Lambda^\perp \supseteq U_2 \cap \mathcal{L}_2(P)$ by showing the inner product equals 0. Take any function $u \in \Lambda$ and $v \in U_2
	\cap \mathcal{L}_2(P)$ such that for some $\{S^*(Y),S^*(Z),S^*(Y,Z),S^*(A),d_1^*(Y,Z)\}$ satisfying their constraints.
\begin{align*}
	&u=u(Y,Z,A) = \underbrace{A \bb{S^*(Y) + S^*(Z)}}_{u_1} + \underbrace{(1-A)S^*(Y,Z)}_{u_2} + \underbrace{S^*(A)}_{u_3} \\
	&v=v(Y,Z,A) = A\bb{d_1^*(Y,Z) - E\bb{d_1^*(Y,Z)|A=1}}
\end{align*}

The inner product:
\begin{align*}
	&E\ba{v u_1} \\
	&=E\bc{A\bb{d_1^*(Y,Z) - E\bb{d_1^*(Y,Z)|A=1}}\bb{S^*(Y) + S^*(Z)}} \\
	&=E\bc{A\bb{d_1^*(Y,Z) - E\bb{d_1^*(Y,Z)|A=1}}S^*(Y) + A\bb{d_1^*(Y,Z) - E\bb{d_1^*(Y,Z)|A=1}} S^*(Z)} \\
	&=E\bc{AE\bb{d_1^*(Y,Z) - E\bb{d_1^*(Y,Z)|A=1}|Y,A=1}S^*(Y) + AE\bb{d_1^*(Y,Z) - E\bb{d_1^*(Y,Z)|A=1}|Z,A=1} S^*(Z)} \\	
	&=E\Big[AE\bc{
	E\bb{d_1^*(Y,Z)|Z,A=1} - E\bb{E\bb{d_1^*(Y,Z)|A=1}|A=1} \big|Y,A=1
	}S^*(Y) \\
	& \quad \quad \quad \quad \quad \quad + 
	AE\bc{
	 E\bb{d_1^*(Y,Z)|Y,A=1} - E\bb{E\bb{d_1^*(Y,Z)|A=1}|A=1} 
	 \big|Z,A=1
	 } S^*(Z)\Big] \\		
	&=E\left[
	AE\bc{ -\int d_1^*(Y,Z) 
	\underbrace{f(Y|A=1) f(Y|A=1)}_{=f(Y,Z|A=1) \ by \ independence }dY  + E[E\bb{d_1^*(Y,Z)|A=1}|A=1] \big|Y,A=1}S^*(Y) 
	\right. 
	\\
	&
	\left.
	\quad \quad \quad \quad \quad \quad 
	- AE\bc{ 
	\int d_1^*(Y,Z) \underbrace{f(Y|Z,A=1) f(Y|Z,A=1)}_{=f(Y,Z|A=1) \ by \ independence}dY 
	+ 
	E\bc{E\bb{d_1^*(Y,Z)|A=1}|A=1} \Big|Z,A=1} S^*(Z)
	\right] \\		
	&=
	E\Big[
	A \bc{-E\bb{E\bb{d_1^*(Y,Z)|A=1}|A=1} + E\bb{E\bb{d_1^*(Y,Z)|A=1}|A=1} } S^*(Y) 
	\\
	&\quad \quad \quad \quad
	+ 
	A \bc{ E\bb{E\bb{d_1^*(Y,Z)|A=1}|A=1}  - E\bb{E\bb{d_1^*(Y,Z)|A=1}|A=1} } S^*(Z)
	\Big] \\	
	&=0 
	\\ \\
	&E\ba{v u_2} = E\bc{(1-A)AS^*(Y,Z) \bb{d_1^*(Y,Z) - E\bb{d_1^*(Y,Z)|A=1}}} = 0 \\ 
	\\
	&E\ba{v u_3} = E\bc{S^*(A) \bb{ d_1^*(Y,Z) - E\bb{d_1^*(Y,Z)|A=1}}}
	= E\bc{S^*(A) E\bc{d_1^*(Y,Z) - E\bb{d_1^*(Y,Z)|A=1}|A=1} } 
	=0
	\\ \\
	&E(vu) = E(v u_1) + E(v u_2) + E(v u_3) = 0.
\end{align*}

Next, we will show the equivalence $U_1\cap \mathcal{L}_2(P) = U_2\cap \mathcal{L}_2(P)$ and $\Lambda^\perp \subseteq U_1\cap \mathcal{L}_2(P)$. Take any function $q(Y,Z,A) \in \Lambda^\perp$. Its inner product with the element from $\Lambda$ shall satisfy the following constraints:
\begin{align*}
	&E\bc{A q(Y,Z,A) \bb{S(Y)+S(Z)}} = E\bc{A q(Y,Z,A=1) \bb{S(Y)+S(Z)}}= 0 \tag{EQ1} \\
	&E\bb{(1-A) q(Y,Z,A) S(Y,Z)} = E\bb{(1-A) q(Y,Z,A=0) S(Y,Z)} =0	\tag{EQ2} \\
	&E\bb{q(Y,Z,A) S(A)} = 0 \tag{EQ3}.
\end{align*}

The first equation (EQ1) means that we only need to focus on $q(Y,Z,A=1)$. For the second equation (EQ2) to hold, $q(Y,A,Z)$ takes the form of $A r(Y,Z)$ for some $r(Y,Z)$. The third equation (EQ3) requires $E\bb{q(Y,Z,A)|A=1}$=0. Hence, $q(Y,Z,A)$ satisfies the constraint
$$q(Y,Z,A) \in \{A r(Y,Z): E\bb{r(Y,Z)|A=1}= 0 \} $$

For any given $Ar(Y,Z)$, we can set $d_1(Y,Z) = r(Y,Z)$ and note $d_1^{\dagger*}=0$. Hence, $\Lambda^\perp \subseteq U_1\cap \mathcal{L}_2(P)$.  The equivalence of $U_1$ and $U_2$ can be seen by noting the first equation (EQ1)  also imposes the constraint that:
\begin{align*}
	&E\bb{A q(Y,Z,A=1) S(Z)} = 0 \ and \ E\bb{A q(Y,Z,A=1) S(Y)} = 0
\end{align*}
This implies the conditional expectation of $q(Y,Z,A=1)$ need to be $0$ and thus $q(Y,Z,A=1)$ can be represented as:
$$q(Y,Z,A=1) = \bb{Z-E\ba{Z|A=1}} \bc{d_2(Y)-E\bb{d_2(Y)|A=1}}$$
for some $d_2(Y)$. As we have shown that the equivalence $U_1\cap \mathcal{L}_2(P) = U_2\cap \mathcal{L}_2(P)$, and $\Lambda^\perp \subseteq U_1\cap \mathcal{L}_2(P)$ and $\Lambda^\perp \supseteq U_2 \cap \mathcal{L}_2(P)$, we conclude that $\Lambda^\perp= U_1
	\cap \mathcal{L}_2(P) =
	U_2
	\cap \mathcal{L}_2(P)$.
\\

\textit{\textbf{Projection onto the the Orthogonal Complement of the Tangent Space}}
\\

For the following derivation, we again suppress $X$. For a given function $c(Y,A,Z)$, we define $c^{\perp}(Y,A,Z)$ as its projection onto the orthogonal complement of the tangent space $\Lambda^\perp$
$$c^{\perp}(Y,A,Z) := \prod \{ c(Y,A,Z) |\Lambda^{\perp} \} = A\bb{Z-E(Z|A=1)}\bc{c^*(Y)-E\bb{c^*(Y)|A=1}}$$
for some $c^*(Y)$.
\\ \\
One can show the function $c^*(Y)$ takes the form:
\begin{align*}
	c^*(Y) = \frac{E\bc{c(Y,A,Z)\bb{Z-E\ba{Z|A=1}}|Y,A=1}}{E\bc{\bb{Z-E\ba{Z|A=1}}^2|A=1}}.
\end{align*}

Proof:

We need to show $c(Y,A,Z) - c^\perp (Y,A,Z)$ satisfies the following equation for any $d(Y)$
\begin{align*}
	&E\bc{\bb{c(Y,A,Z) - c^\perp(Y,A,Z)}A\bb{Z-E\ba{Z|A=1}}\bc{d(Y)-\bb{d(Y)|A=1}}
	} \\
	&=E\big[
	\bc{c(Y,A,Z) -  A\bb{Z-E\ba{Z|A=1}}\bc{c^*(Y)-E\bb{c^*(Y)|A=1}}}
	A\bb{Z-E\ba{Z|A=1}}\bc{d(Y)-\bb{d(Y)|A=1}}
	\big] \\
	&=\underbrace{E
	\bc{
	c(Y,A,Z)A\bb{Z-E\ba{Z|A=1}}\bc{d(Y)-E\bb{d(Y)|A=1}}-c^*(Y)A\bb{Z-E\ba{Z|A=1}}^2\bc{d(Y)-E\bb{d(Y)|A=1}}
	}
	}_{(Q1)} \\
	&\quad \quad + \underbrace{E
	\bc{
	E\bb{c^*(Y)|A=1}A\bb{Z-E\ba{Z|A=1}}^2\bc{d(Y)-E\bb{d(Y)|A=1}}
	}
	}_{(Q2)} \\
	&= Q1 + Q2 \\
	&=0 \\
	&\implies c^*(Y) = \frac{E\bc{c(Y,A,Z)\bb{Z-E\ba{Z|A=1}}|Y,A=1}}{E\bc{\bb{Z-E\ba{Z|A=1}}^2|Y,A=1}}
\end{align*}

In $Q2$, we notice that $\bc{d(Y)-E\bb{d(Y)|A=1}}$ is the only part involving $Y$ and $E\bb{d(Y)|Z,A=1} = E\bb{d(Y)|A=1}$
\begin{align*}
	&Q2 \\
	&= E\bc{E\bb{c^*(Y)|A=1}A\bb{Z-E\ba{Z|A=1}}^2\bc{d(Y)-E\bb{d(Y)|A=1}}} \\
	&= E\bc{E\bb{c^*(Y)|A=1}A\bb{Z-E\ba{Z|A=1}}^2 E\bc{
	\bc{d(Y)-E\bb{d(Y)|A=1}}
	|Z,A=1}
	} \\
	&= E\bc{E\bb{c^*(Y)|A=1}A\bb{Z-E\ba{Z|A=1}}^2 
	\bc{E\bb{d(Y)|A=1} -E\bb{d(Y)|A=1}}} \\	
	&= 0.
\end{align*}

Hence, for the equation $Q1+Q2=0$ to hold, we have $Q1=0$, which is:
\begin{align*}
    & E\bc{c(Y,A,Z)A\bb{Z-E\ba{Z|A=1}}\bc{d(Y)-E\bb{d(Y)|A=1}}} = E\bc{c^*(Y)A\bb{Z-E\ba{Z|A=1}}^2\bc{d(Y)-E\bb{d(Y)|A=1}}} \\
    & \textit{LHS} 
    =E\bc{
	 AE\bb{c(Y,A=1,Z)|Y,A=1} \bb{Z-E\ba{Z|A=1}}|Y,A=1]\bc{d(Y)-E\bb{d(Y)|A=1}}
	 } \\
    & \textit{RHS} 
    =
    E[ Ac^*(Y)E[\bb{Z-E\ba{Z|A=1}}^2|A=1]\bc{d(Y)-E\bb{d(Y)|A=1}}] 
\end{align*}

The RHS equals LHS regardless of $d(Y)$ if and only if
\begin{align*}
	c^*(Y)  =  \frac{E\bc{
	c(Y,A,Z)\bb{Z-E\ba{Z|A=1}}|Y,A=1
	}
	}
	{E\bc{\bb{Z-E\ba{Z|A=1}}^2|A=1}}
\end{align*}

\textbf{\textit{Projection of the EIF under the MIV Model onto the Tangent Space of the Intersection Model}}
\\ 

Upon re-incorporating $X$, one has
\begin{align*}
	&c^{\perp}(Y,A,Z,X) := \prod \{ c(Y,A,Z,X) |\Lambda^{\perp} \} = A\bb{Z-E\ba{Z|A=1,X}}\bc{c^*(Y,X)-E\bb{c^*(Y,X)|A=1,X}}
	\\
	&c^*(Y,X)  =  \frac{E[c(Y,A,Z,X)\bb{Z-E\ba{Z|A=1,X}}|Y,A=1,X]}{
	E\bc{\bb{Z-E\ba{Z|A=1,X}}^2|A=1,X}}.
\end{align*}

The EIF under the MIV model is 
\begin{align*}
	&EIF_{MIV}(O;\psi) \\
	&= 
	\frac{A}{pr(A=1)} \{Y + \delta(X) - \psi\}
	\\ & \qquad + 
	\frac{pr(A=1|X)}{pr(A=1)}  \frac{2Z-1}{f(Z|X)} \frac{1}{\delta^A(X)}\bc{Y(1-A) - E\bb{Y(1-A)|Z,X} - \bb{A - pr(A=1|Z,X)} \delta(X) }
	\\
	&EIF_{MIV}(Y,A=1,Z,X) \\
	&= \underbrace{\frac{1}{pr(A=1)} \{Y + \delta(X) - \psi\}
	}_{C_1(Y,X)} +
	\underbrace{\frac{pr(A=1|X)}{pr(A=1)}  \frac{2Z-1}{f(Z|X)} \frac{1}{\delta^A(X)}\bc{ -1 + pr(A=1|Z,X) \delta(X)}
	}_{C_2(Z,X)}.
\end{align*}

Let $c(Y,A,Z,X) = EIF_{MIV}(O;\psi)$. We derive the projection of $EIF_{MIV}(O;\psi)$ onto $\Lambda^\perp$ by each part $C_1(Y,X)$ and $C_2(Z,X)$. The projected $C_1(Y,X)$ is 0 because the numerator of $c^*(Y,X) $ is 0:
\begin{align*}
	&  E\bc{C_1(Y,X) \bb{Z-E\ba{Z|A=1,X}}|Y,A=1,X} \\ 
	&= E\bc{ E\bc{\bb{Z-E\ba{Z|A=1,X}}|Y,A=1,X} C_1(Y,X)}  \\
	&= E\bc{ E\bc{\bb{Z-E\ba{Z|A=1,X}}|A=1,X} C_1(Y,X)}  \\	
	&= E\bb{ 0 \cdot C_1(X)} \\
	&= 0.
\end{align*}

The numerator of the projected $C_2(Z,X)$ is:
\begin{align*}
	& E\bc{C_2(Z,X) \bb{Z-E\ba{Z|A=1,X}}|Y,A=1,X} \\ 
	&=E\bc{\frac{pr(A=1|X)}{pr(A=1)}  \frac{2Z-1}{f(Z|X)} \frac{\delta(X)}{\delta^A(X)}
	\bc{- \bb{1 - pr(A=1|Z,X)}} 
	\bb{Z-E\ba{Z|A=1,X}} 
	\Big|Y,A=1,X} \\ 
	&=-E\bc{\frac{pr(A=1|X)}{pr(A=1)}  \frac{2Z-1}{f(Z|X)} \frac{\delta(X)}{\delta^A(X)} pr(A=0|Z,X)  \bb{Z-E\ba{Z|A=1,X}}
	\Big|Y,A=1,X} \\ 	
	&=-E\bc{\frac{pr(A=1|X)}{pr(A=1)}  \frac{2Z-1}{f(Z|X)} \frac{\delta(X)}{\delta^A(X)} pr(A=0|Z,X)  \bb{Z-pr(Z=1|A=1,X)}
	\Big|A=1,X} \\ 		
	&=-\frac{pr(A=1|X)}{pr(A=1)}  \frac{1}{pr(Z=1|X)} \frac{\delta(X)}{\delta^A(X)} pr(A=0|Z=1,X)  \bb{1-pr(Z=1|A=1,X)}pr(Z=1|A=1,X) \\
	&\qq -\frac{pr(A=1|X)}{pr(A=1)}  \frac{-1}{pr(Z=0|X)} \frac{\delta(X)}{\delta^A(X)} pr(A=0|Z=0,X)  \bb{0-pr(Z=1|A=1,X)}pr(Z=0|A=1,X) \\ 	
	&=-\frac{pr(A=1|X)}{pr(A=1)}  \frac{1}{pr(Z=1|X)} \frac{\delta(X)}{\delta^A(X)} pr(A=0|Z=1,X)  pr(Z=0|A=1,X)pr(Z=1|A=1,X) \\
	&\qq -\frac{pr(A=1|X)}{pr(A=1)}  \frac{1}{pr(Z=0|X)} \frac{\delta(X)}{\delta^A(X)} pr(A=0|Z=0,X)  pr(Z=1|A=1,X)pr(Z=0|A=1,X) \\ 
	&=  -\frac{pr(A=1|X)}{pr(A=1)}  \frac{\delta(X)}{\delta^A(X)} \bb{\frac{pr(A=0|Z=0,X)}{pr(Z=0|X)} + \frac{pr(A=0|Z=1,X)}{pr(Z=1|X)}}  pr(Z=1|A=1,X)pr(Z=0|A=1,X).
\end{align*}

The denominator can be rewritten as
\begin{align*}
	&E[\bb{Z-E\ba{Z|A=1,X}}^2|A=1,X] \\
	&= pr(Z=0|A=1,X)^2 pr(Z=1|A=1,X) + pr(Z=0|A=1,X) pr(Z=1|A=1,X)^2 \\
	&= pr(Z=0|A=1,X) pr(Z=1|A=1,X).
\end{align*}
Therefore, the corresponding $c^*(Y,X)$ for the EIF under the MIV model is
\begin{align*}
	c^*(Y,X) = -\frac{pr(A=1|X)}{pr(A=1)}  \frac{\delta(X)}{\delta^A(X)} \bb{\frac{pr(A=0|Z=0,X)}{pr(Z=0|X)} + \frac{pr(A=0|Z=1,X)}{pr(Z=1|X)}}.
\end{align*}
Of note, $c^*(Y,X)$ is not a function of $Y$. The projection of $EIF_{MIV}(O;\psi)$ onto $\Lambda^\perp$ gives 
\begin{align*}
	&c^{\perp}(Y,A,Z,X) \\
	&= A\bb{Z-E\ba{Z|A=1,X}} \bc{c^*(Y,X) - E\bb{c^*(Y,X)|A=1,X}} \\
	&= A\bb{Z-E\ba{Z|A=1,X}} \bc{c^*(Y,X) - c^*(Y,X)} 	\\
	&= 0.
\end{align*}
Therefore, we conclude that the EIF under the intersection model is
\begin{align*}
	&EIF_{ind}(O;\psi) \\
	&= EIF_{MIV}(O;\psi) - \prod\bb{EIF_{MIV}(O;\psi)|\Lambda^{\perp}} 
	\\
	&= c(Y,A,Z,X) - c^{\perp}(Y,A,Z,X) \\
	&= EIF_{MIV}(O;\psi) - 0 \\
	&= EIF_{MIV}(O;\psi).
\end{align*}
\end{appendices}
\end{document}